\documentclass[twoside]{article}

\usepackage{url,mathrsfs}
\usepackage{amsmath,wrapfig,color,bigfoot}
\usepackage{amsfonts}
\usepackage{graphicx}
\usepackage{subfigure}
\newtheorem{theorem}{Theorem}

\newtheorem{corollary}{Corollary}

\usepackage[accepted]{aistats2014}
\begin{document}

%

%

\twocolumn[

\aistatstitle{\vspace{-0.15in}In Defense of MinHash Over SimHash\vspace{-0in}}

\aistatsauthor{Anshumali Shrivastava\And Ping Li }

\aistatsaddress{ Department of Computer Science\\
Computing and Information Science\\
Cornell University, Ithaca, NY, USA
\And
Department of Statistics and Biostatistics\\
Department of Computer Science\\
Rutgers University, Piscataway, NJ, USA
 }
 ]

\begin{abstract}\vspace{-0.2in}
MinHash and SimHash are the two widely adopted Locality Sensitive Hashing (LSH) algorithms for large-scale data processing applications. Deciding which LSH to use for a particular problem at hand is an important question, which has no clear answer in the existing  literature. In this study, we provide a theoretical answer (validated by experiments) that MinHash virtually always outperforms  SimHash when the data are binary, as common in practice such as search.

\vspace{0.03in}

The collision probability of MinHash is a function of  {\em resemblance} similarity ($\mathcal{R}$), while the collision probability of SimHash is a function of  {\em cosine}  similarity ($\mathcal{S}$). To provide a common basis for  comparison,  we evaluate  retrieval results in terms of $\mathcal{S}$ for both MinHash and SimHash. This evaluation is valid  as we can prove that MinHash is  a valid LSH with respect to $\mathcal{S}$, by using a  general inequality  $\mathcal{S}^2\leq \mathcal{R}\leq \frac{\mathcal{S}}{2-\mathcal{S}}$. Our \textbf{worst case}  analysis  can show that MinHash significantly outperforms SimHash in  \textbf{high similarity} region.

\vspace{0.03in}

Interestingly, our intensive experiments reveal that MinHash is also substantially better than SimHash even in datasets where most of the data points are not too similar to each other. This is partly because, in practical data, often $\mathcal{R}\geq \frac{\mathcal{S}}{z-\mathcal{S}}$ holds where $z$ is only slightly larger than 2 (e.g., $z\leq 2.1$). Our \textbf{restricted worst case} analysis by assuming $\frac{\mathcal{S}}{z-\mathcal{S}}\leq \mathcal{R}\leq \frac{\mathcal{S}}{2-\mathcal{S}}$ shows that MinHash indeed significantly outperforms SimHash even in \textbf{low similarity} region.

\vspace{0.03in}

We believe the results in this paper  will provide valuable guidelines for search in practice,  especially when the data are sparse.

\end{abstract}

\section{Introduction}
\vspace{-0.1in}

The advent of the Internet has led to generation of massive  and inherently high dimensional data. In many industrial applications, the size of the datasets has long exceeded the memory capacity of a single machine. In web domains, it is not difficult to find datasets with the number of instances and the number of dimensions  going into billions~\cite{Report:TeraLarning11,Report:Sibyl,GoogleBlog}.

The reality that  web data  are typically  sparse and  high dimensional is   due to the wide adoption of the ``Bag of Words'' (BoW) representations for  documents and images. In BoW  representations, it is known that the word frequency within a document follows power law.  Most of the words occur rarely in a document  and most of the higher order shingles in the document occur only once.  It is often the case that just the presence or absence information suffices in practice~\cite{Article:Chapelle_99,Proc:Hein_AISTATS05,Proc:Jiang_CIVR07,Proc:HashLearning_NIPS11}.  Leading search companies routinely use  sparse binary representations in their large data systems~\cite{Report:Sibyl}.

{\bf Locality sensitive hashing (LSH)}~\cite{Proc:Indyk_STOC98} is a general framework of  indexing technique,  devised for efficiently solving the approximate near neighbor search problem~\cite{Article:Friedman_75}. The performance of LSH largely depends on the underlying  particular hashing methods. Two popular hashing algorithms are \textbf{MinHash}~\cite{Proc:Broder} and \textbf{SimHash} (sign normal random projections)~\cite{Proc:Charikar}.

MinHash is an LSH for \textbf{resemblance similarity} which is  defined over binary vectors, while SimHash is an LSH for \textbf{cosine similarity} which works for general real-valued  data. With the abundance of binary data over the web, it has become a practically important question: {\em which LSH should be preferred in binary data?}. This question has not been adequately answered in  existing literature. There were prior attempts to address this problem from various aspects. For example, the paper on {\em Conditional Random Sampling (CRS)}~\cite{Proc:Li_Church_Hastie_NIPS06} showed that random projections can be very inaccurate especially in binary data, for the task of inner product estimation (which is not the same as near neighbor search). A more recent paper~\cite{Proc:Shrivastava_ECML12}  empirically demonstrated that $b$-bit minwise hashing~\cite{Proc:Li_Internetware13} outperformed SimHash and spectral hashing~\cite{Proc:Weiss_NIPS08}.

\textbf{Our contribution}: Our paper  provides an essentially conclusive answer that MinHash should  be used for near neighbor search in binary data, both theoretically and empirically.  To favor SimHash, our theoretical analysis and experiments  evaluate the retrieval results of MinHash in terms of  cosine similarity (instead of resemblance). This is possible because we are able to show that MinHash can  be proved to be  an LSH for  cosine similarity by establishing an inequality which  bounds  resemblance by purely functions of  cosine. 

Because we evaluate MinHash (which was designed for resemblance) in terms of cosine, we will first illustrate the close connection between these two similarities.

\vspace{-0.15in}
\section{Cosine Versus Resemblance}
\vspace{-0.15in}

We focus on binary data, which can be viewed as sets (locations of  nonzeros). Consider two sets $W_1, W_2 \subseteq\Omega = \{1, 2, ..., D\}$. The cosine similarity ($\mathcal{S}$) is
\begin{align}
&\mathcal{S} =  \frac{a}{\sqrt{f_1f_2}}, \ \ \ \ \text{where} \\
&f_1 = |W_1|,\ \ f_2 = |W_2|,\ \ a = |W_1 \cap W_2|
\end{align}
The resemblance similarity, denoted by $\mathcal{R}$, is
\begin{align}
&\mathcal{R} =\mathcal{R}(W_1, W_2)= \frac{|W_1 \cap W_2|}{| W_1 \cup W_2|} = \frac{a}{f_1+f_2 -a}
\end{align}
Clearly these two similarities are closely related. To better illustrate the connection, we re-write $\mathcal{R}$ as
\begin{align}
&\mathcal{R}  = \frac{a/\sqrt{f_1f_2}}{\sqrt{f_1/f_2}+\sqrt{f_2/f_1}-a/\sqrt{f_1f_2}} =\frac{\mathcal{S}}{z-\mathcal{S}}\\\label{eqn_z}
&z = z(r) = \sqrt{r}+\frac{1}{\sqrt{r}}\geq 2\\
&r = \frac{f_2}{f_1}= \frac{f_1f_2}{f_1^2} \leq \frac{f_1f_2}{a^2} = \frac{1}{\mathcal{S}^2}\vspace{-0.2in}
\end{align}
There are two degrees of freedom: $f_2/f_1$, $a/f_2$, which make it  inconvenient for analysis. Fortunately, in  Theorem~\ref{thm_ineq}, we can bound $\mathcal{R}$ by purely  functions of $\mathcal{S}$.

\begin{theorem}\label{thm_ineq}
\begin{align}\vspace{-0.2in}
\mathcal{S}^2 \le \mathcal{R} \le \frac{\mathcal{S}}{2-\mathcal{S}}
\end{align}
\textbf{Tightness} Without making assumptions   on the data, neither the lower bound $\mathcal{S}^2$  or the upper bound $\frac{\mathcal{S}}{2-\mathcal{S}}$ can  be improved in the domain of continuous functions.

\textbf{Data dependent bound} If the data satisfy $z\leq z^*$, where $z$ is defined in (\ref{eqn_z}), then
\begin{align}
\frac{\mathcal{S}}{z^*-\mathcal{S}} \le \mathcal{R} \le \frac{\mathcal{S}}{2-\mathcal{S}}
\end{align}
\textbf{Proof:}\ \ See Appendix A. $\hfill\Box$
\end{theorem}

Figure~\ref{fig:bounds} illustrates that in high similarity region, the upper and lower bounds essentially overlap. Note that, in order to obtain $S\approx1$, we need $f_1\approx f_2$ (i.e., $z \approx 2$).
 \begin{figure}[h!]
\begin{center}
\mbox{
\includegraphics[width=2in]{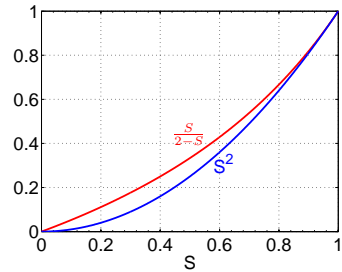} }
\end{center}
\vspace{-0.2in}
\caption{Upper (in red) and lower (in blue) bounds in Theorem~\ref{thm_ineq}, which  overlap in high similarity region.}\label{fig:bounds}\vspace{-0.1in}
\end{figure}

\begin{figure}[h!]
\begin{center}
\mbox{
\includegraphics[width=1.7in]{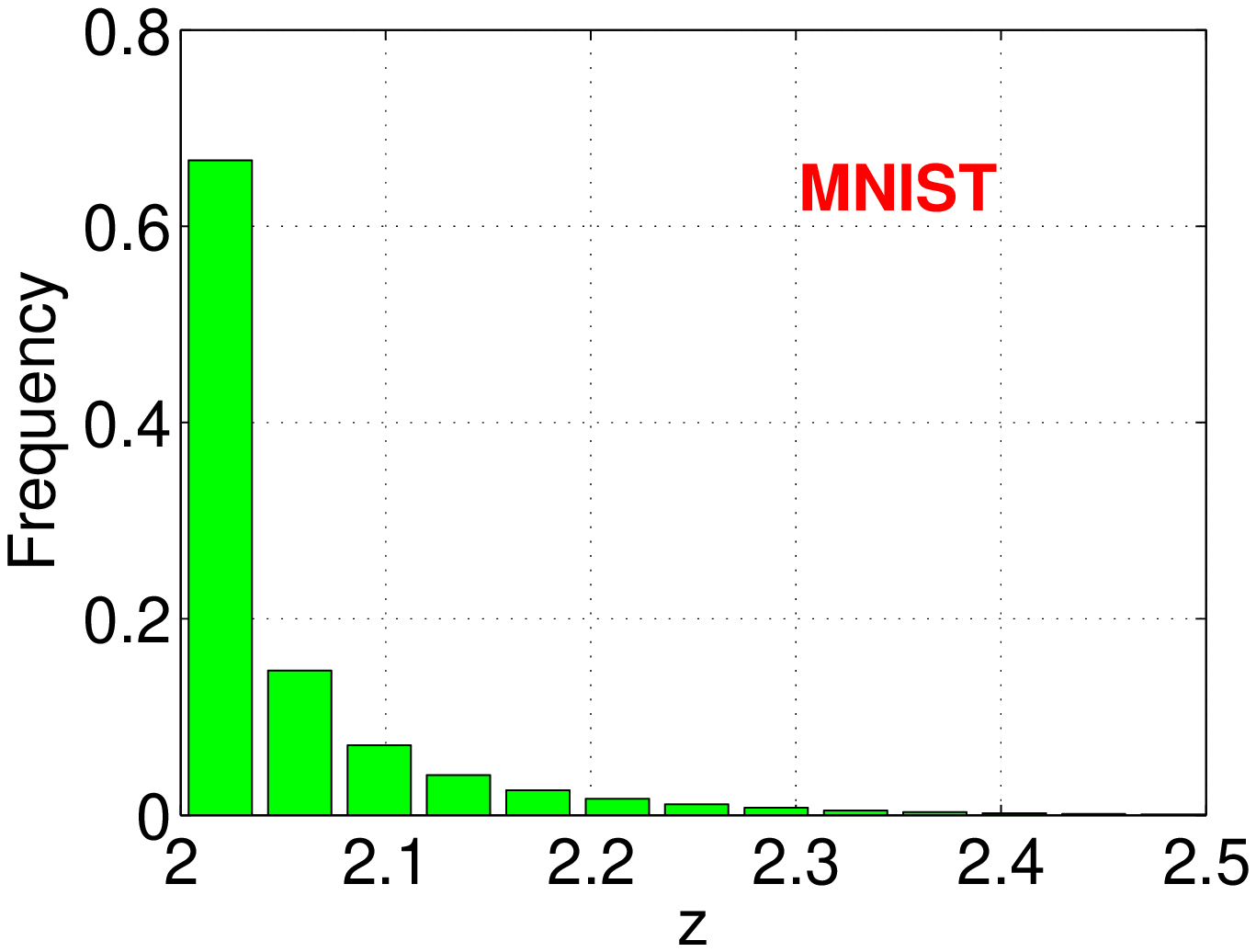}\hspace{-0.15in}
\includegraphics[width=1.7in]{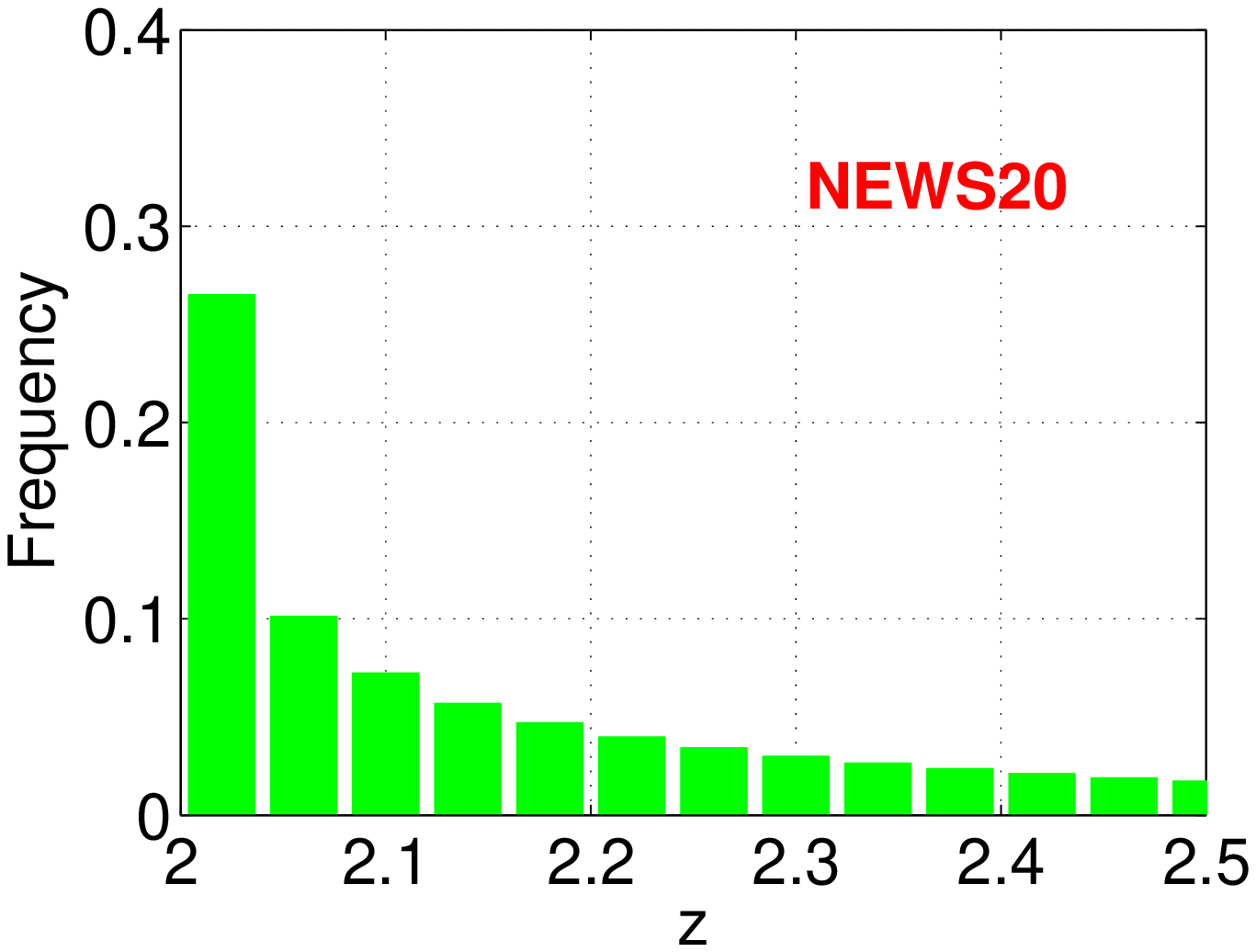}
}
\mbox{
\includegraphics[width=1.7in]{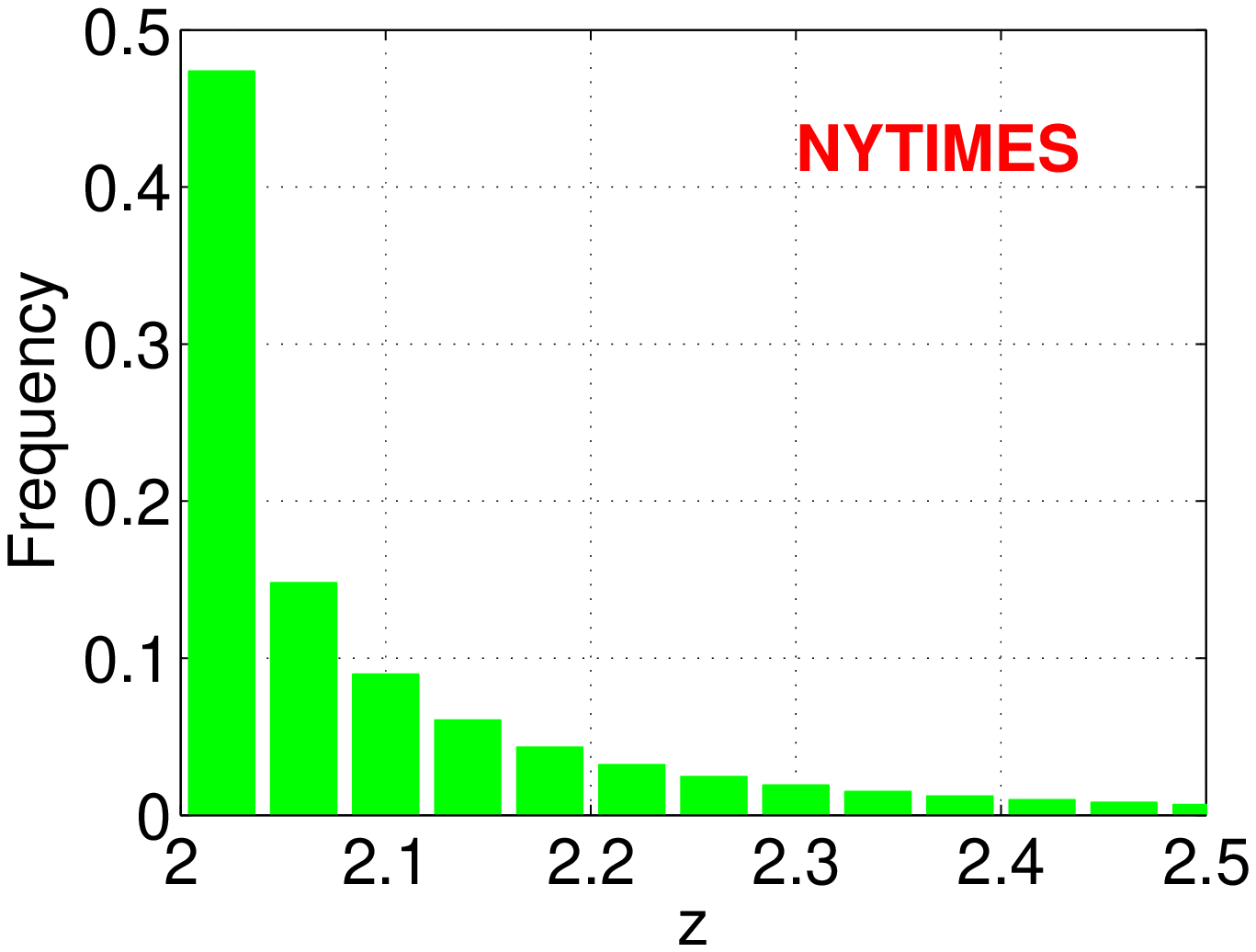}\hspace{-0.15in}
\includegraphics[width=1.7in]{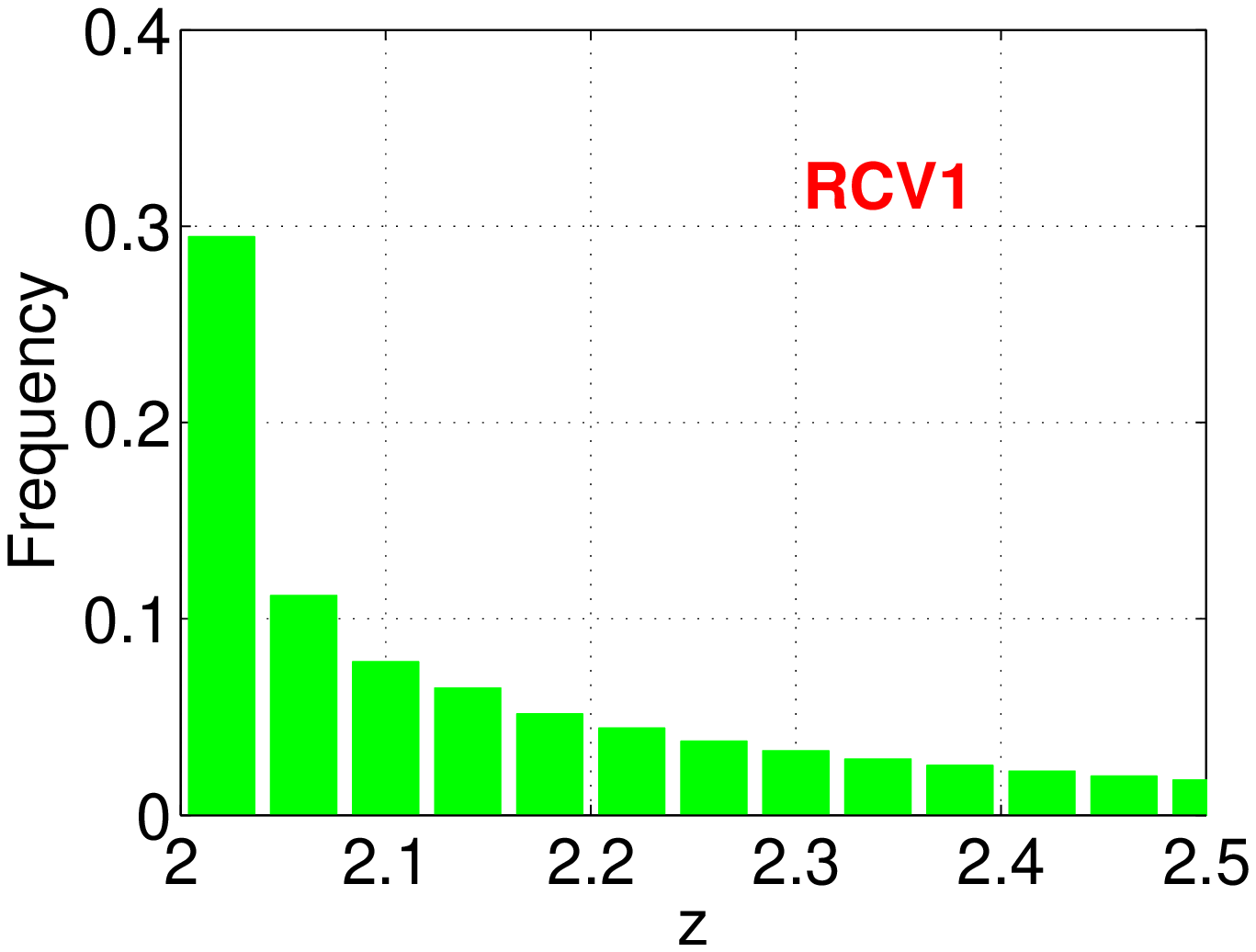}
}
\mbox{
\includegraphics[width=1.7in]{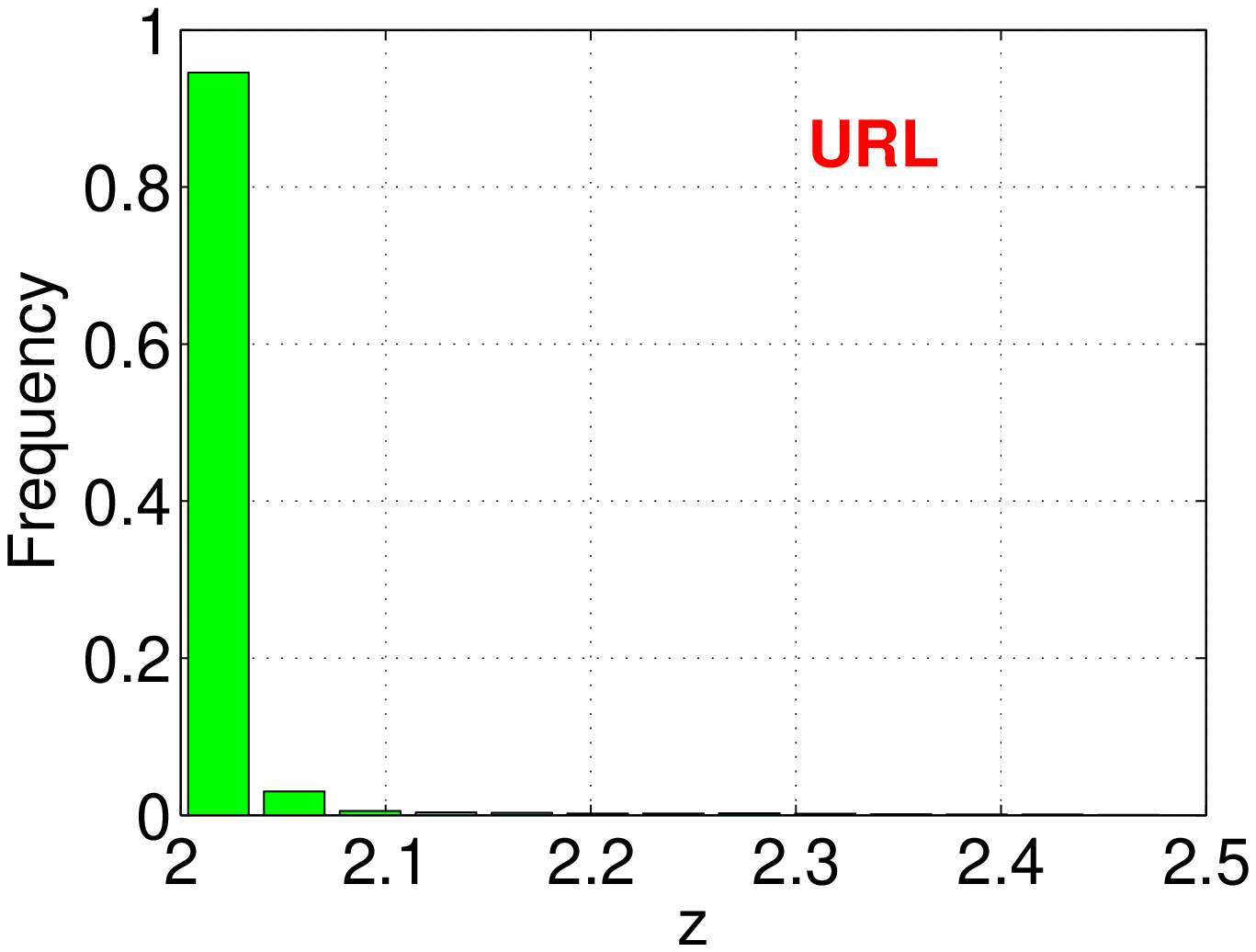}\hspace{-0.15in}
\includegraphics[width=1.7in]{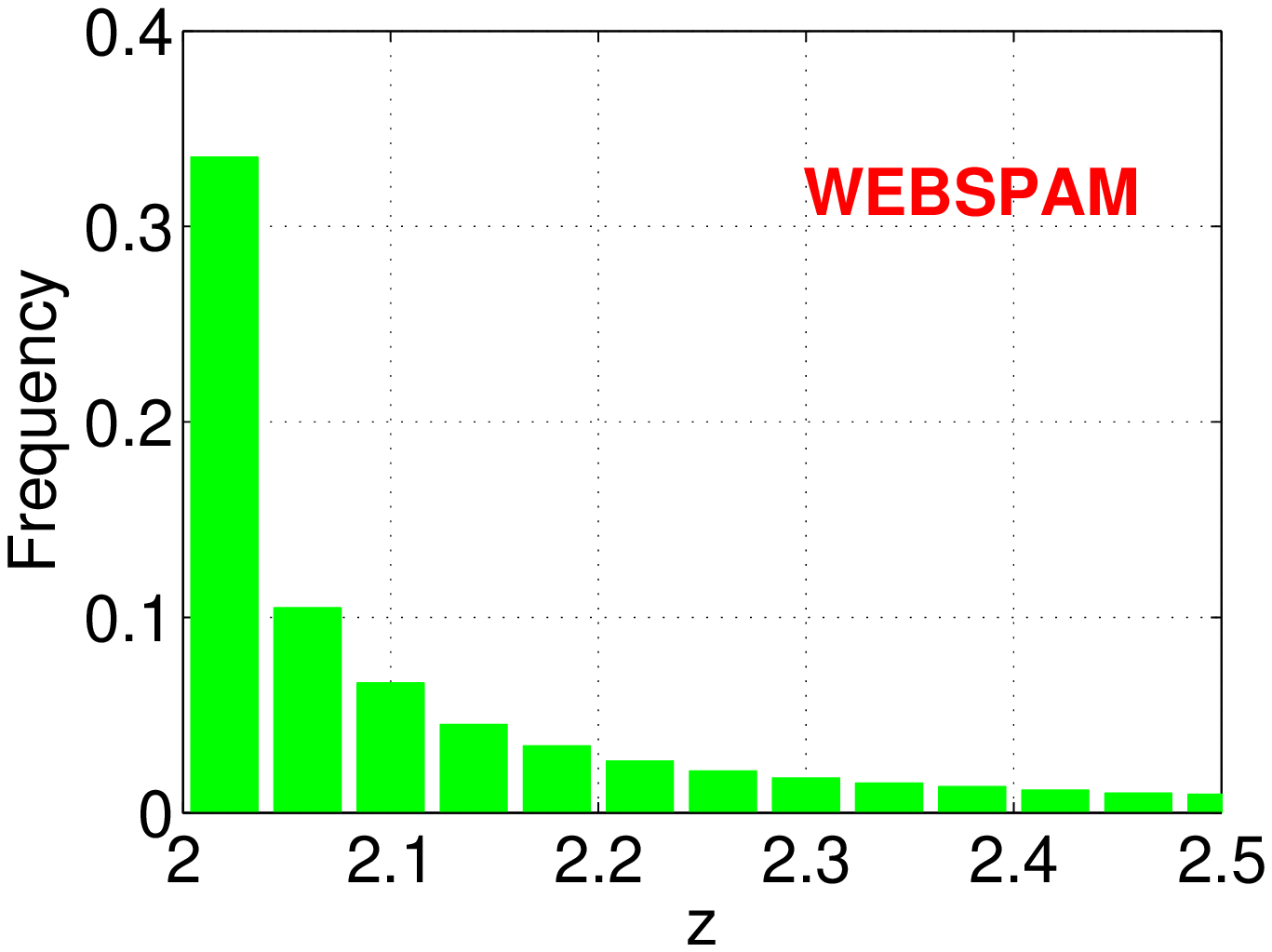}
}
\end{center}
\vspace{-0.2in}
\caption{Frequencies of the $z$ values for all six datasets in Table~\ref{tab_data}, where $z$ is defined in (\ref{eqn_z}). We compute $z$ for every query-train pair of data points. }\label{fig_z}
\end{figure}

While the high similarity region is often of interest, we must also handle data in the low similarity region, because in a realistic dataset, the majority of the pairs are usually not  similar. Interestingly, we observe that for the six datasets in Table~\ref{tab_data}, we often have $\mathcal{R} = \frac{\mathcal{S}}{z-\mathcal{S}}$ with $z$ only being slightly larger than 2; see Figure~\ref{fig_z}.

\begin{table}[h!]
\caption{Datasets}
\begin{center}{\small
{\begin{tabular}{l r r r}
\hline \hline
Dataset        &\# Query    &\# Train  &\# Dim\\
\hline
MNIST      &10,000 &60,000 &784  \\
NEWS20     &2,000  &18,000 &1,355,191 \\
NYTIMES    &5,000 &100,000 &102,660 \\
RCV1       &5,000 &100,000 &47,236 \\
URL        &5,000 &90,000 &3,231,958 \\
WEBSPAM    &5,000 &100,000 &16,609,143
\\\hline\hline
\end{tabular}}
}
\end{center}\label{tab_data}\vspace{-0.2in}
\end{table}

\begin{figure}[h!]
\begin{center}
\mbox{
\includegraphics[width=1.63in]{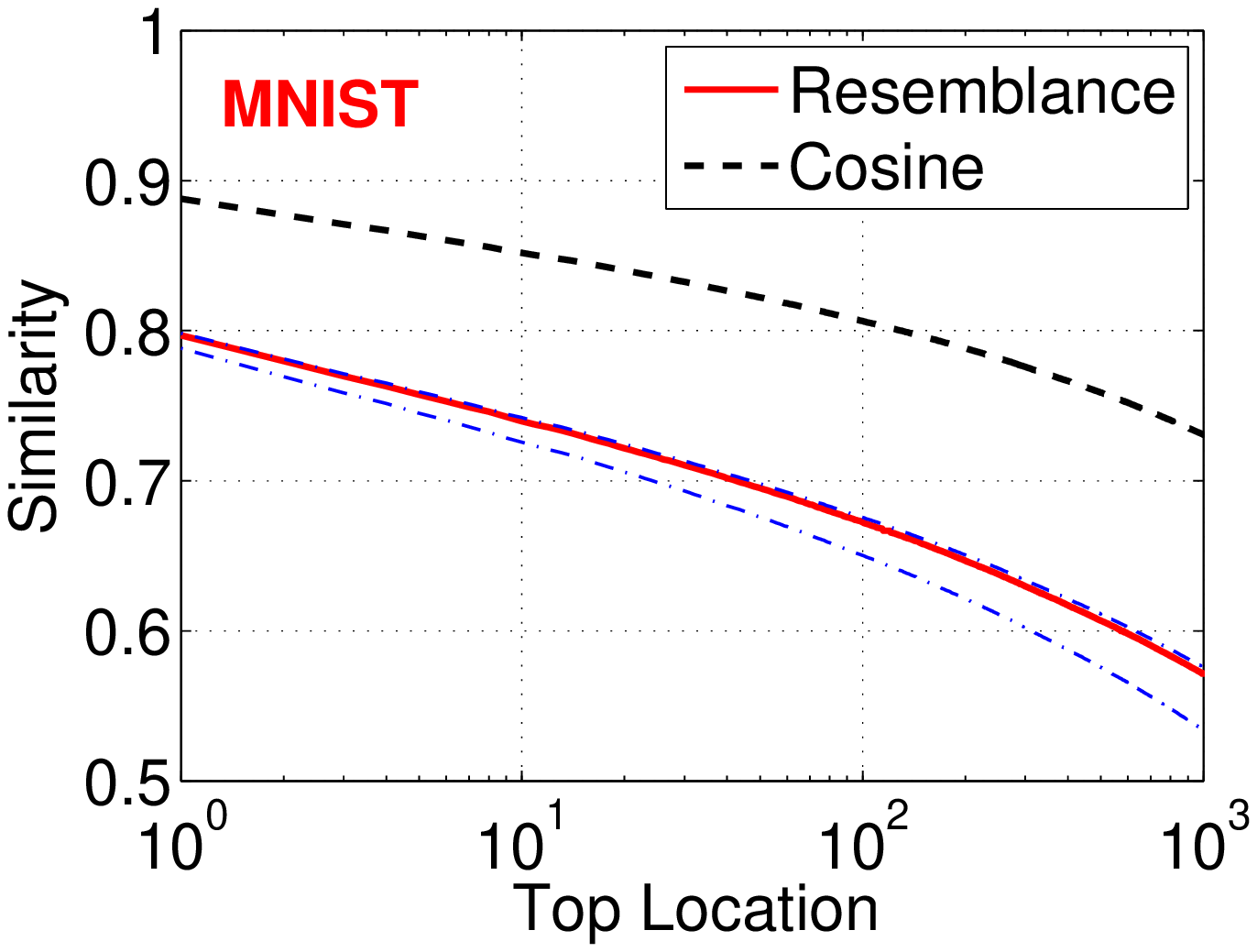}\hspace{-0.13in}
\includegraphics[width=1.63in]{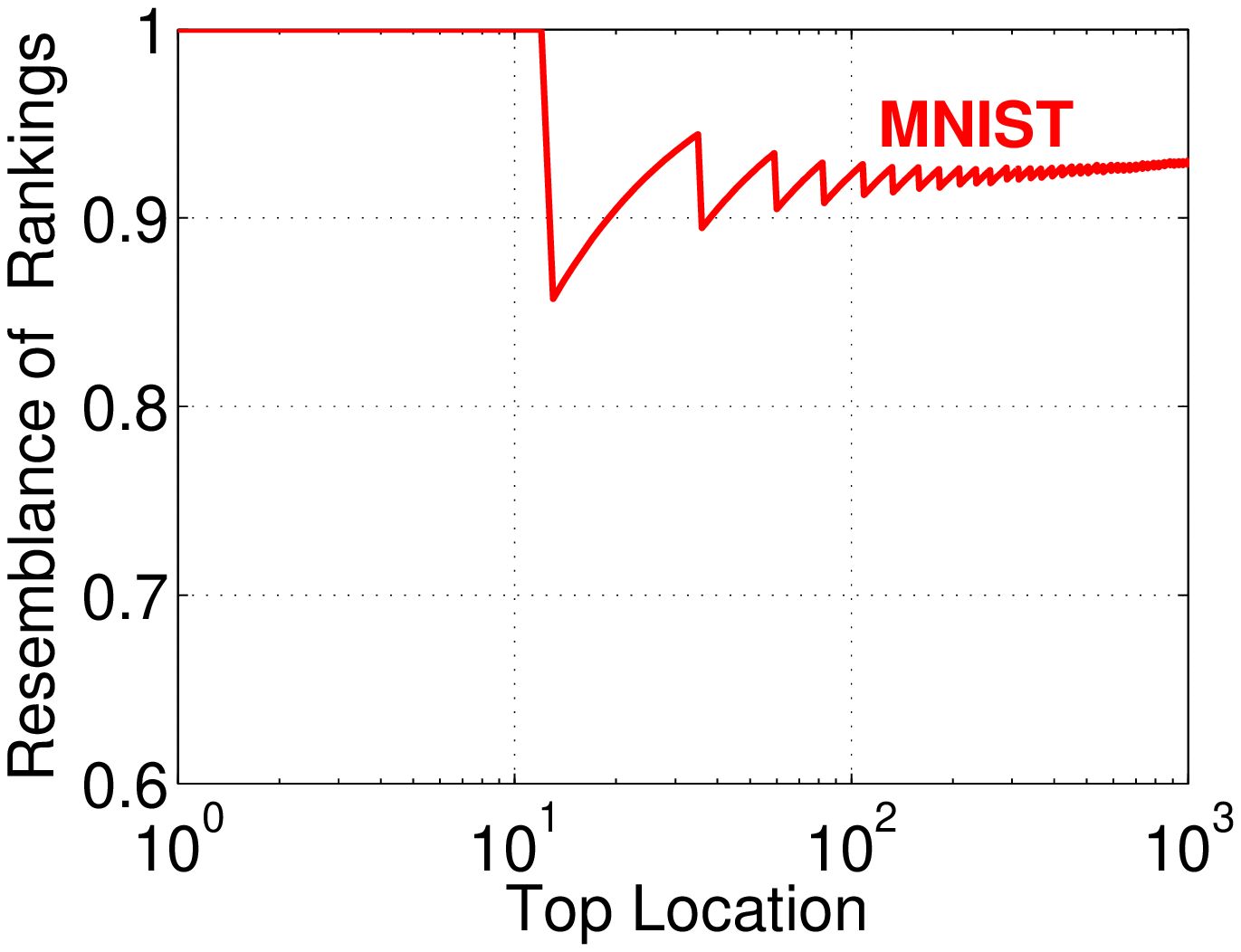}
}

\vspace{-0.03in}

\mbox{
\includegraphics[width=1.63in]{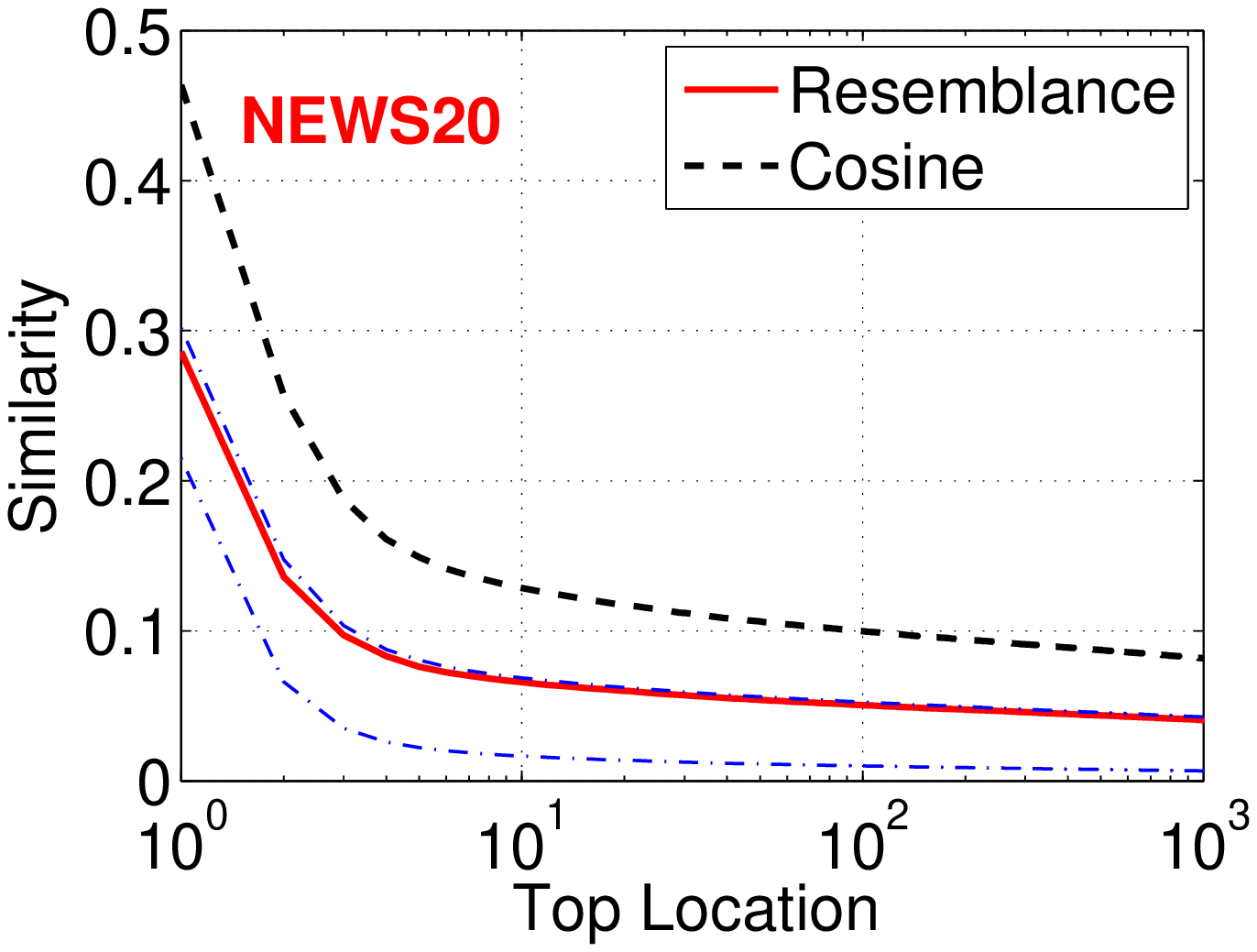}\hspace{-0.13in}
\includegraphics[width=1.63in]{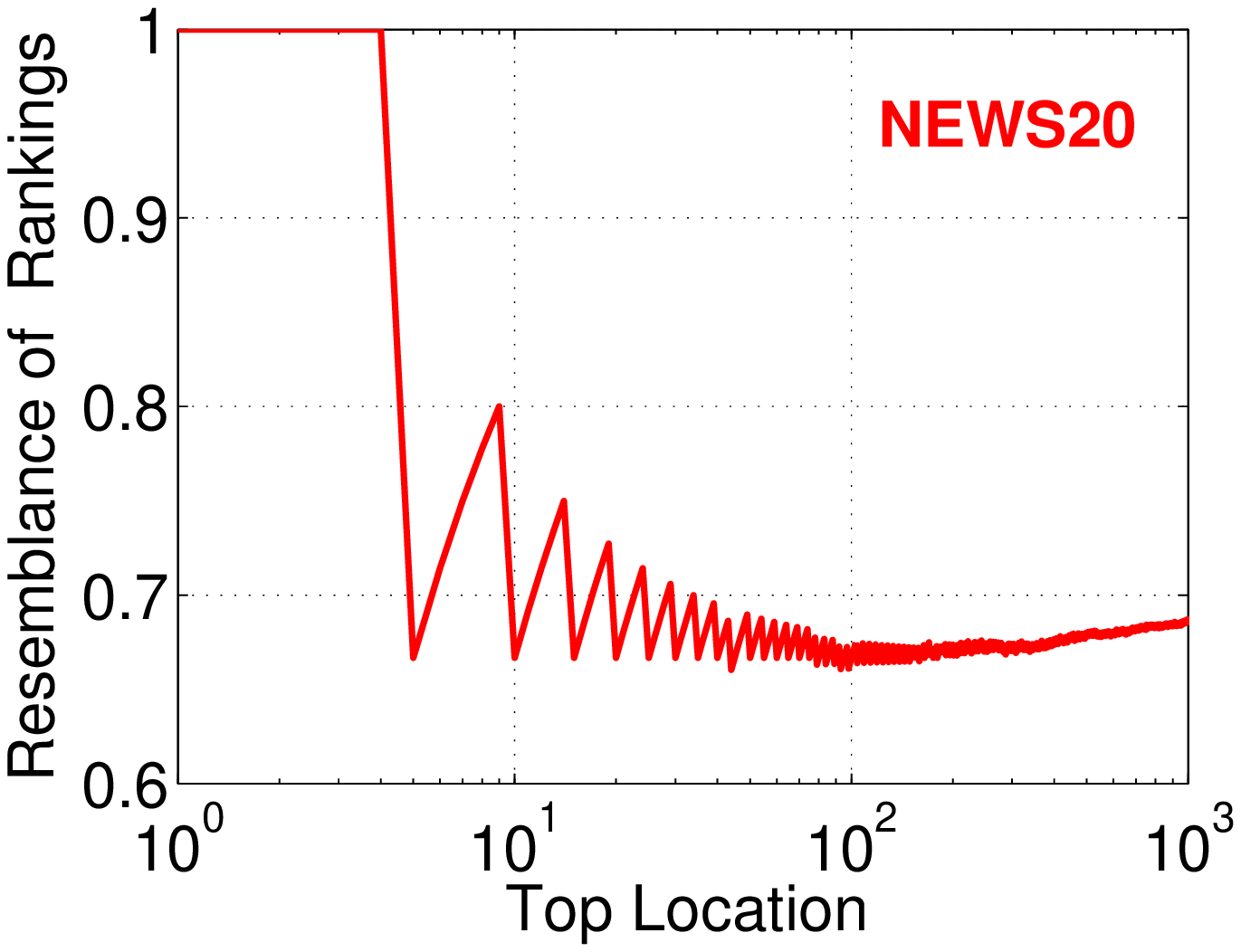}
}

\vspace{-0.03in}

\mbox{
\includegraphics[width=1.63in]{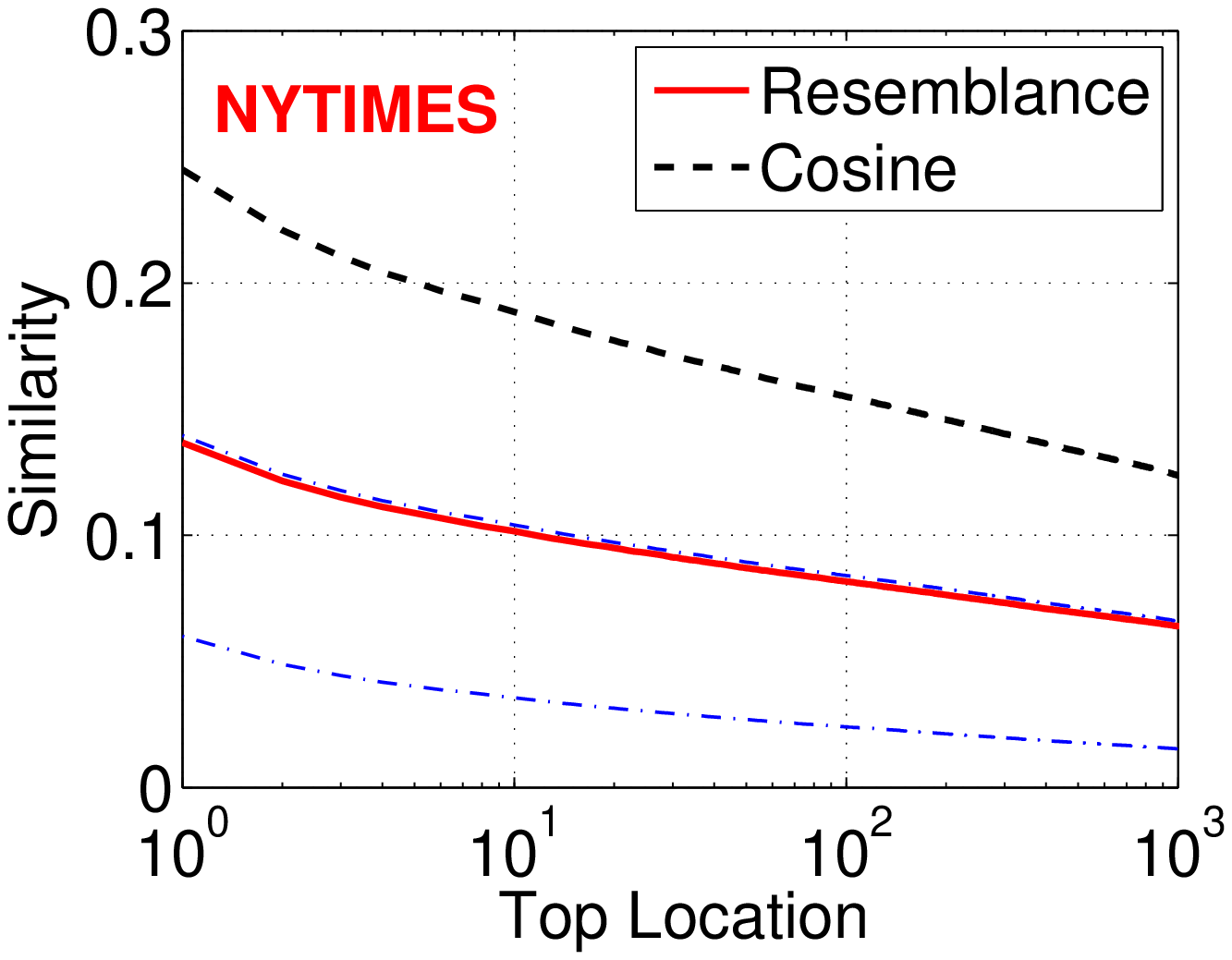}\hspace{-0.13in}
\includegraphics[width=1.63in]{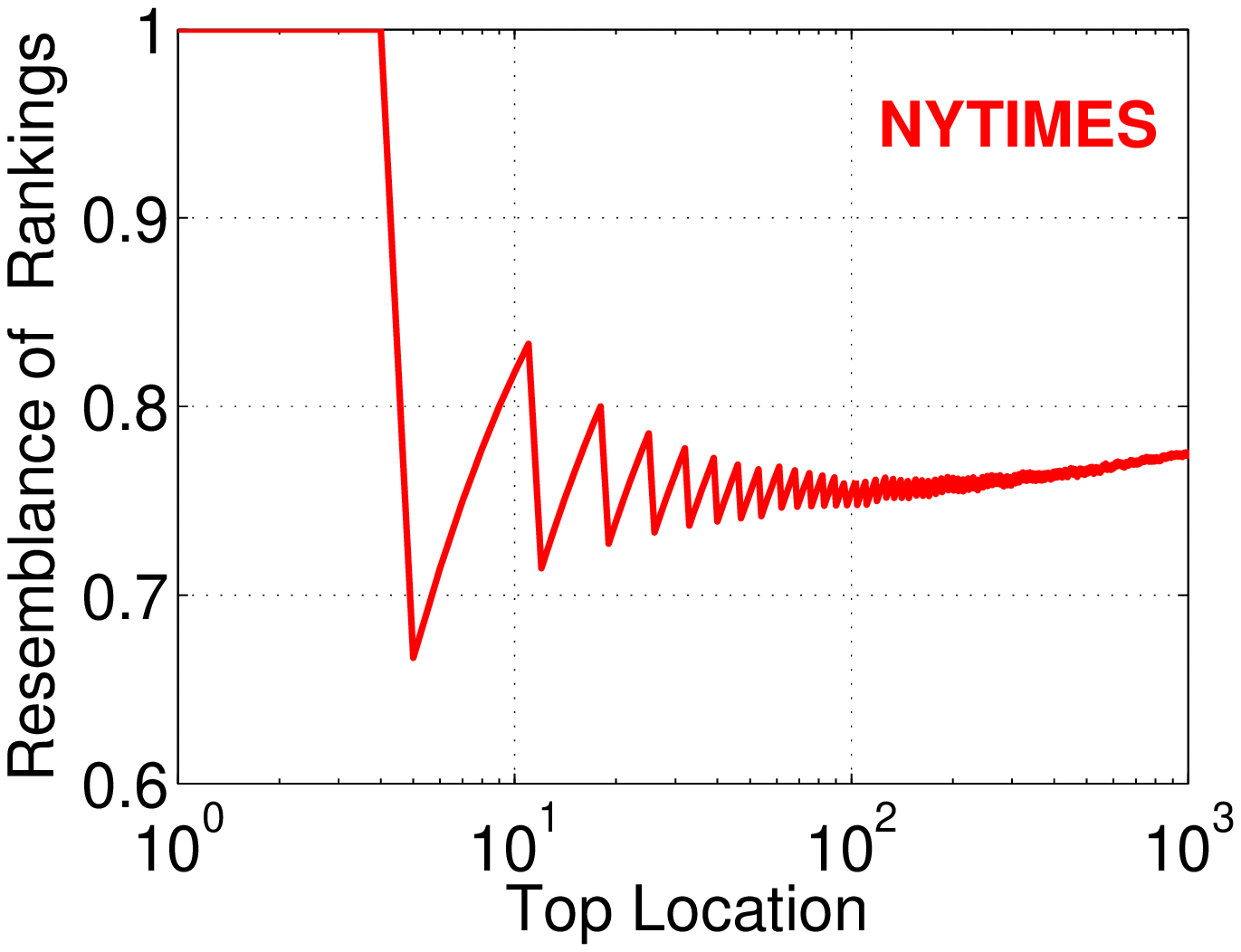}
}

\vspace{-0.03in}

\mbox{
\includegraphics[width=1.63in]{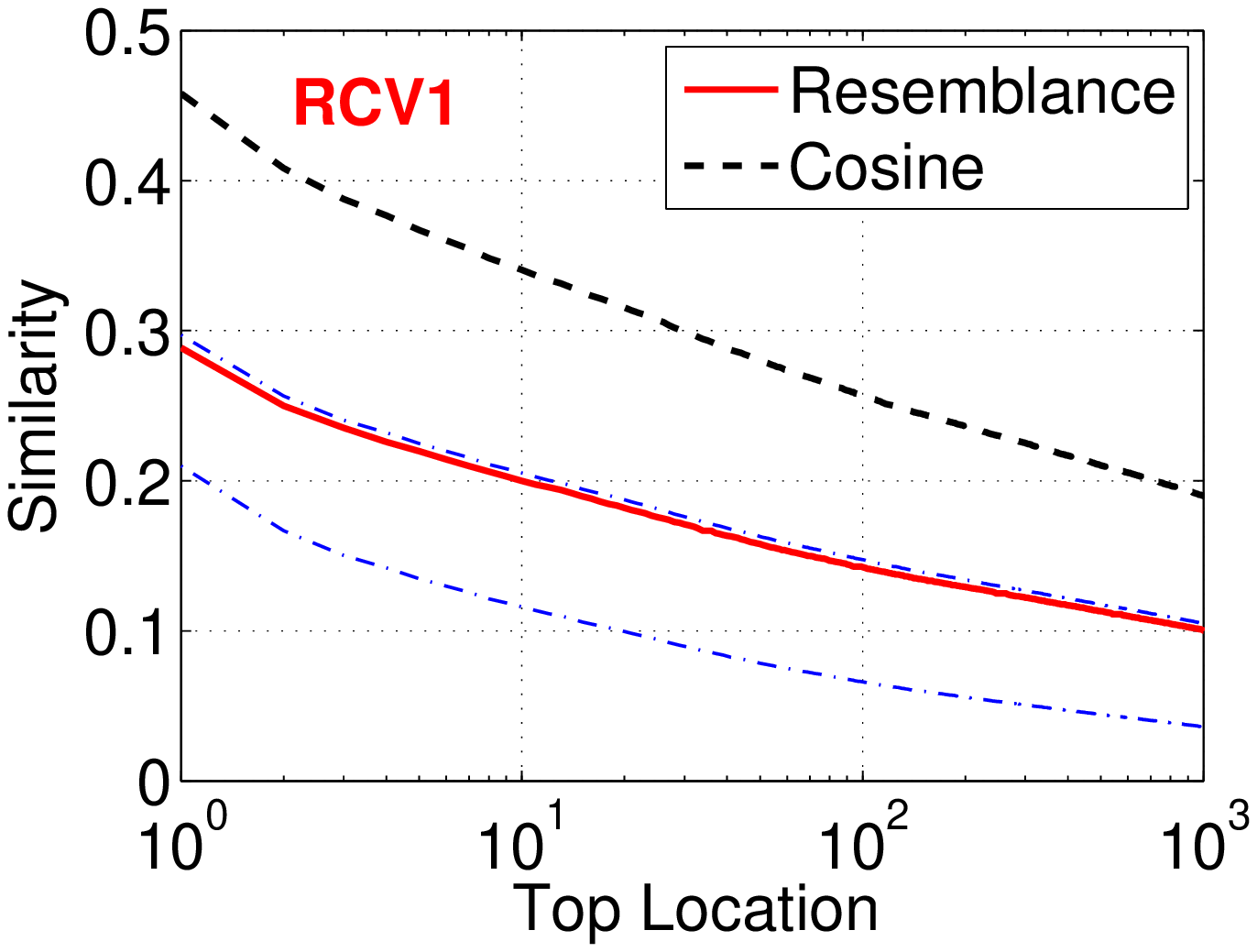}\hspace{-0.13in}
\includegraphics[width=1.63in]{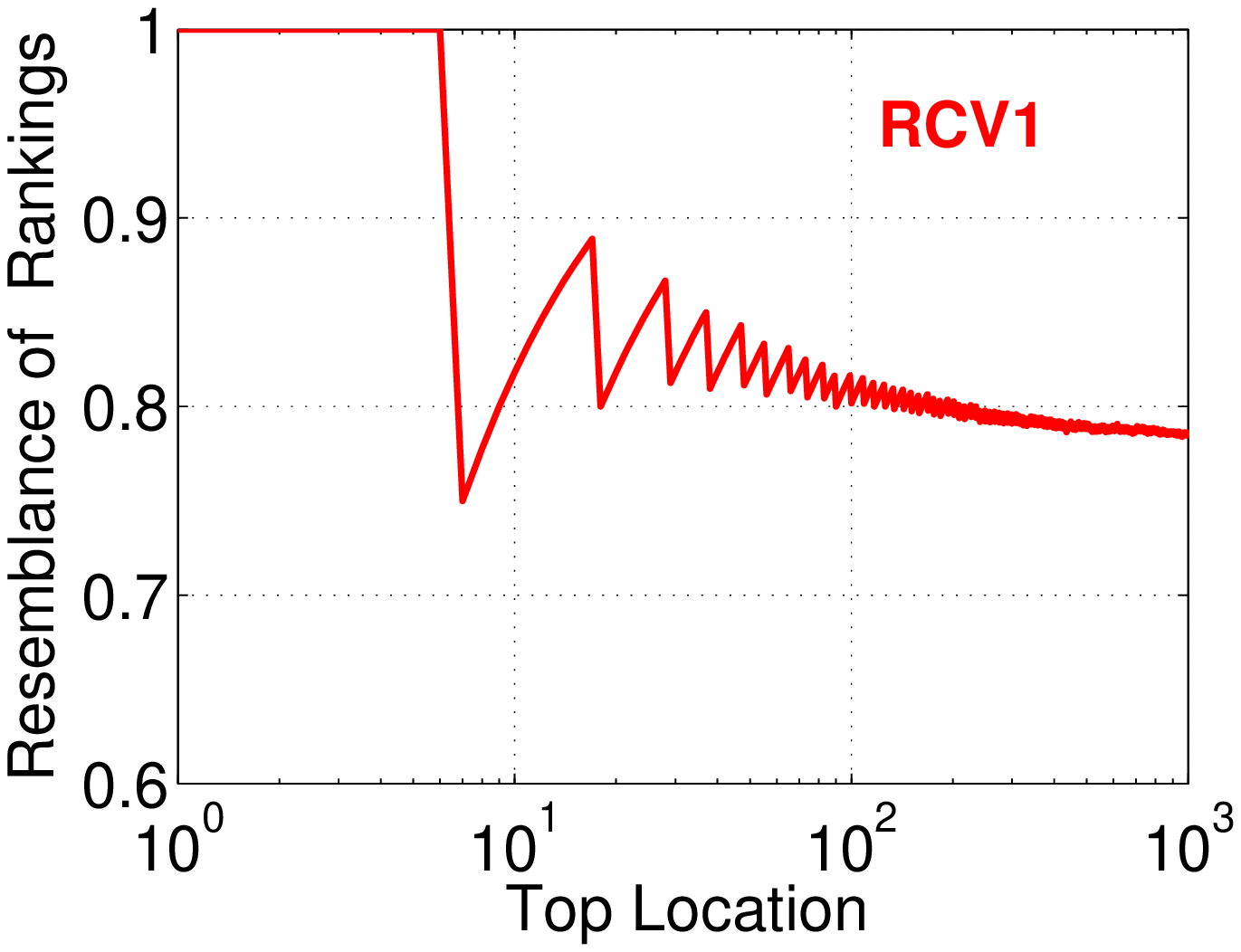}
}


\mbox{
\includegraphics[width=1.63in]{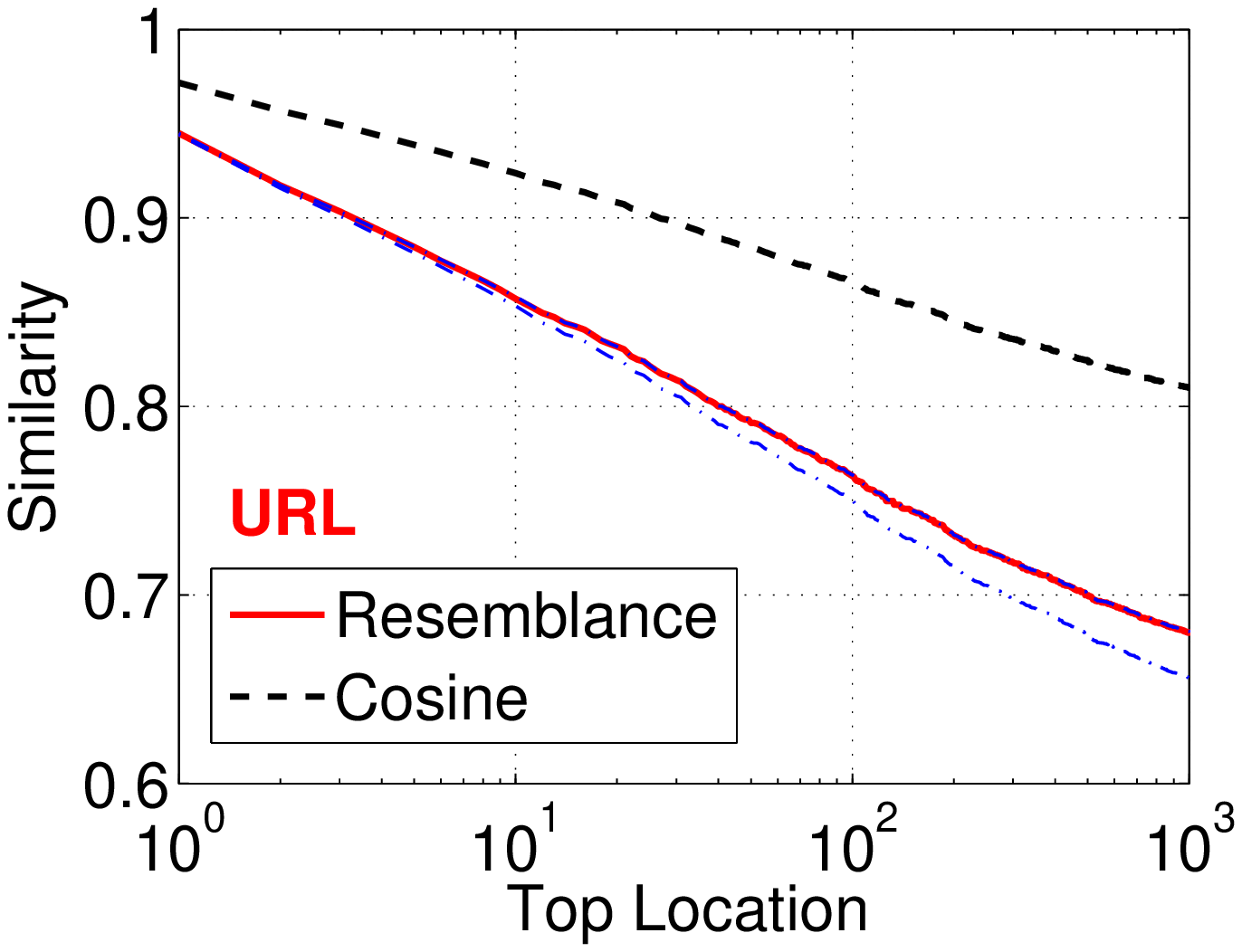}\hspace{-0.13in}
\includegraphics[width=1.63in]{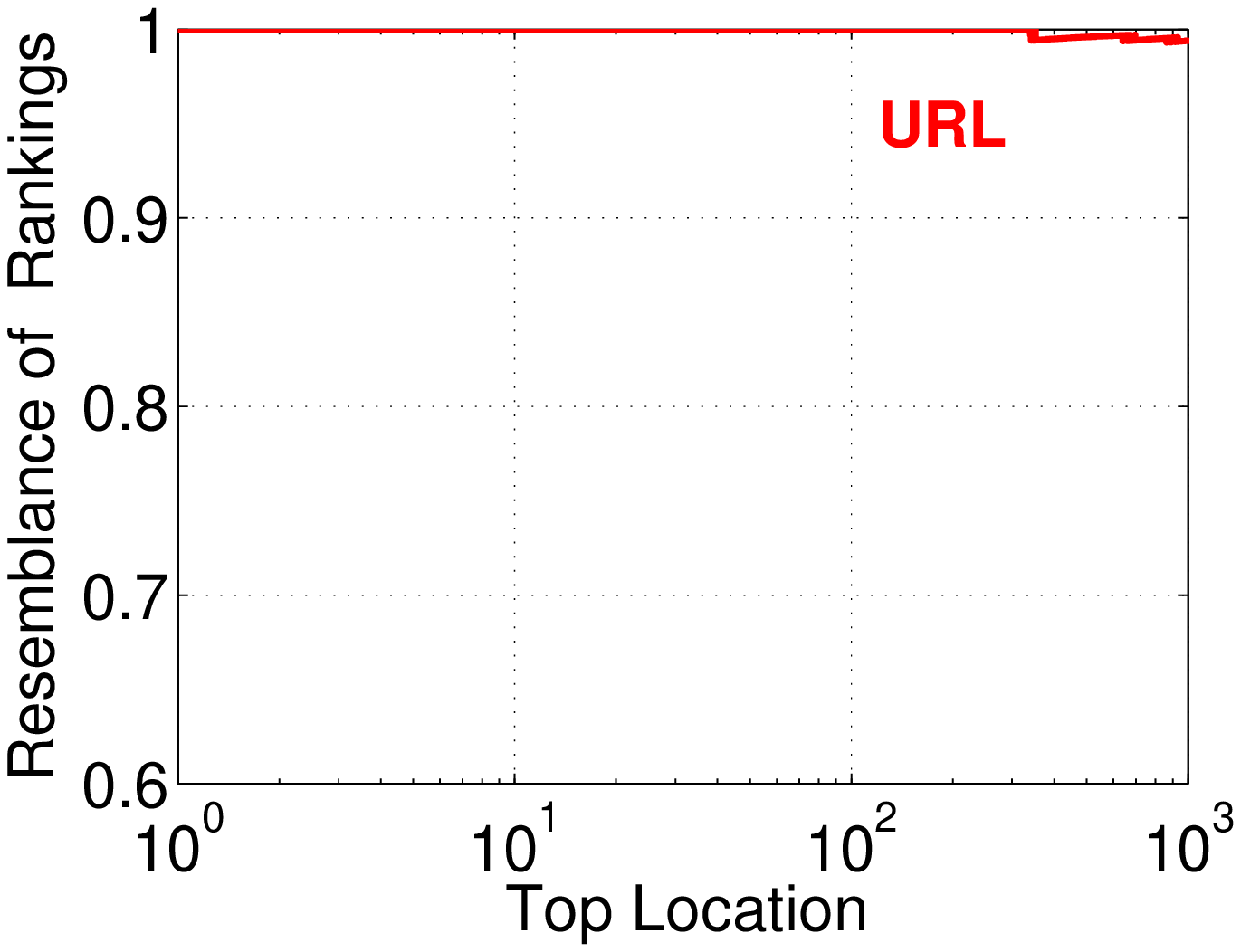}
}


\mbox{
\includegraphics[width=1.63in]{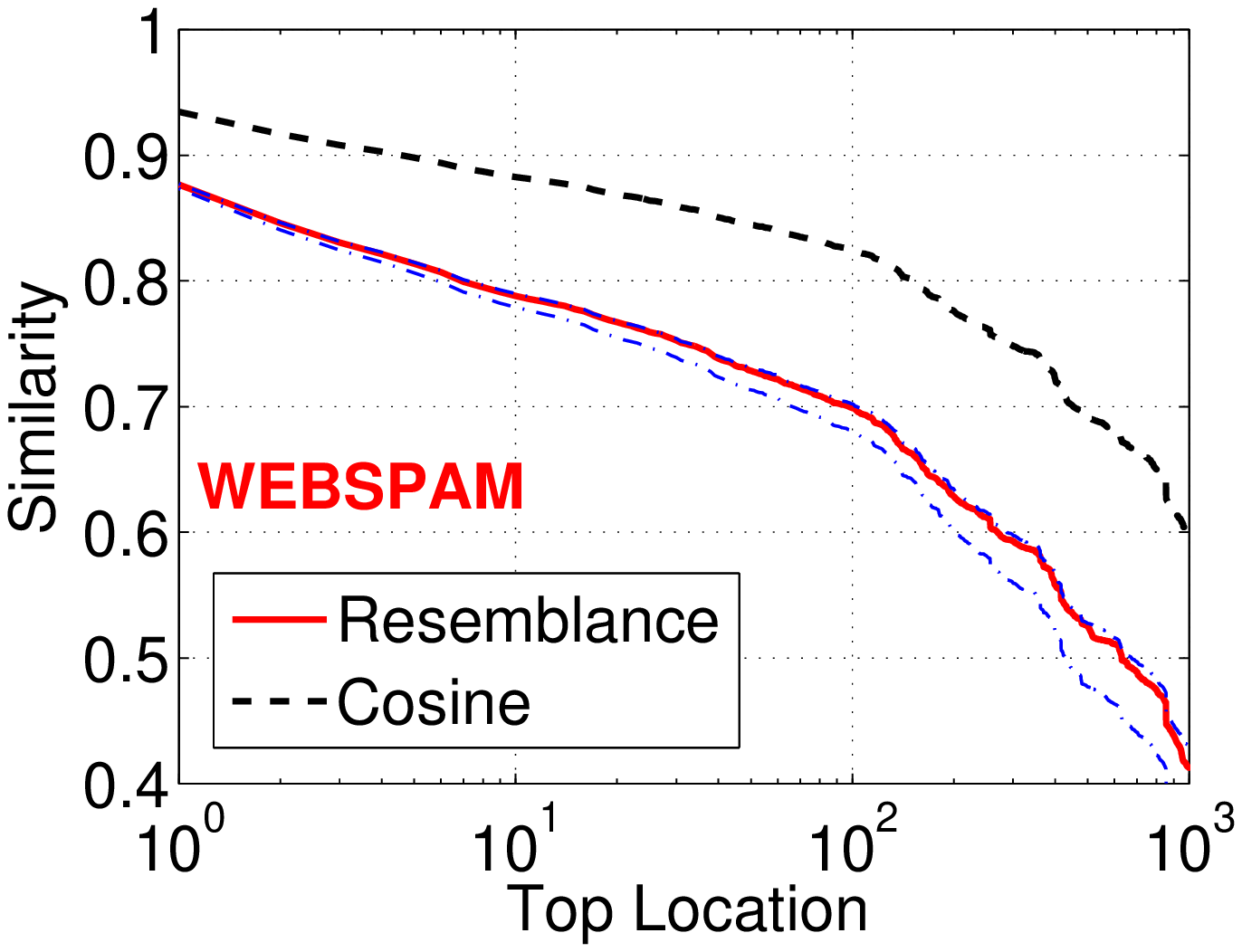}\hspace{-0.13in}
\includegraphics[width=1.63in]{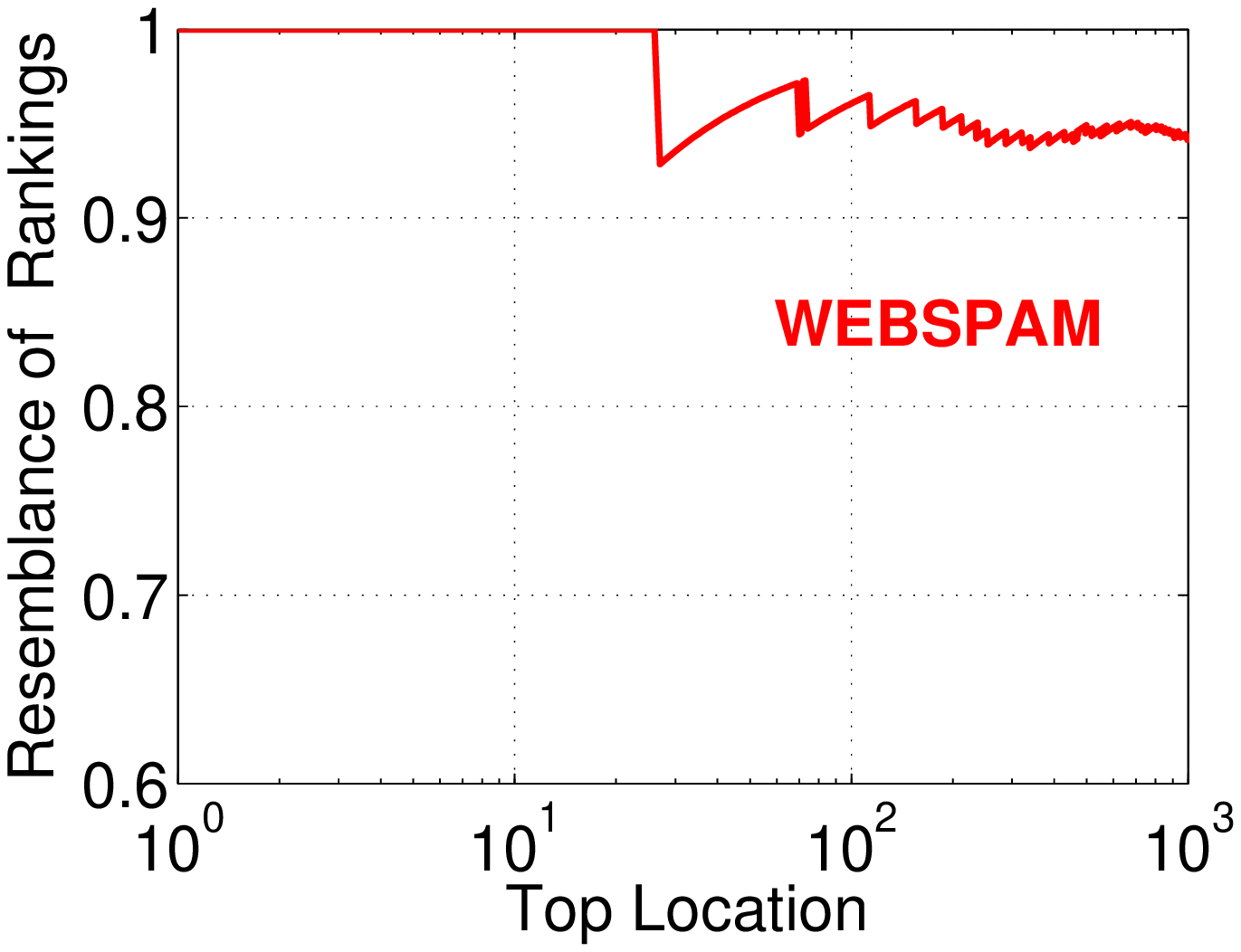}
}

\end{center}
\vspace{-0.2in}
\caption{\textbf{Left panels}:  For each query point, we rank its similarities to all  training points in descending order. For every top location, we plot the median (among all query points) of the similarities, separately for cosine (dashed) and resemblance (solid), together with the lower and upper bounds of $\mathcal{R}$ (dot-dashed). \textbf{Right Panels}:  For every query point, we rank the training points in descending order of similarities, separately for cosine and resemblance.  We plot the resemblance of two ranked lists at top-$T$ ($T=1$ to 1000).  }\label{fig_Top1000}\vspace{-0.1in}
\end{figure}

For each dataset, we compute both cosine and resemblance for every query-train pair (e.g.,  $10000 \times 60000$ pairs for MNIST dataset). For each query point, we rank its similarities to all  training points in descending order. We examine the top-1000 locations as in Figure~\ref{fig_Top1000}. In the left panels, for every top location, we plot the median (among all query points) of the similarities, separately for cosine (dashed) and resemblance (solid), together with the lower and upper bounds of $\mathcal{R}$ (dot-dashed). We can see  for NEWS20, NYTIMES, and RCV1, the data are not too similar. Interestingly, for all six datasets, $\mathcal{R}$ matches fairly well with the upper bound $\frac{\mathcal{S}}{2-\mathcal{S}}$. In other words, the lower bound $\mathcal{S}^2$ can be very conservative  even in low similarity region.

The right panels of Figure~\ref{fig_Top1000} present the comparisons of the orderings of similarities in an interesting way. For every query point, we rank the training points in descending order of similarities, separately for cosine and resemblance. This way, for every query point we have two lists of numbers (of the data points). We truncate the lists at top-$T$ and compute the resemblance between the two lists. By varying $T$ from 1 to 1000, we obtain a curve which roughly measures the ``similarity'' of  cosine and resemblance. We present the averaged curve over all query points. Clearly Figure~\ref{fig_Top1000} shows there is a strong correlation between the two measures in all datasets, as one would expect.

\vspace{-0.1in}

\section{ Locality Sensitive Hashing (LSH)}
\vspace{-0.1in}

A common formalism for approximate near neighbor problem is the $c$-approximate near neighbor or $c$-NN.

\textbf{Definition}: ($c$-Approximate Near Neighbor or $c$-NN). Given a set of points in a $d$-dimensional space $\mathbb{R}^d$, and parameters $S_0 > 0$, $\delta > 0$, construct a data structure which, given any query point $q$, does the following with probability $1- \delta$: if there exist an $S_0$-near neighbor of $q$ in $P$, it reports some $cS_0$-near neighbor of $q$ in $P$.

The usual notion of $S_0$-near neighbor is in terms of the distance function. Since we are dealing with similarities, we can equivalently define $S_0$-near neighbor of point $q$ as a point $p$ with $Sim(q,p) \ge S_0$, where $Sim$ is the similarity function of interest.

 A popular technique  for $c$-NN, uses the underlying theory of \emph{Locality Sensitive Hashing} (LSH)~\cite{Proc:Indyk_STOC98}. LSH is a family of functions, with the property that similar input objects in the domain of these functions have a higher probability of colliding in the range space than non-similar ones. In formal terms, consider $\mathcal{H}$ a family of hash functions mapping $\mathbb{R}^D$ to some set $\mathcal{S}$.

\textbf{Definition: Locality Sensitive Hashing}\ A family $\mathcal{H}$ is called $(S_0,cS_0,p_1,p_2)$-sensitive if for any two points $x,y \in \mathbb{R}^d$  and $h$ chosen uniformly from $\mathcal{H}$ satisfies the following:
\begin{itemize}
\item if $Sim(x,y)\ge S_0$ then ${Pr}_\mathcal{H}(h(x) = h(y)) \ge p_1$
\item if $ Sim(x,y)\le cS_0$ then ${Pr}_\mathcal{H}(h(x) = h(y)) \le p_2$
\end{itemize}
For approximate nearest neighbor search typically, $p_1 > p_2$ and $c < 1$ is needed. Since we are defining neighbors in terms of similarity we have $c < 1$. To get distance analogy we can use the transformation $D(x,y) = 1- Sim(x,y)$ with a requirement of  $c > 1$.

The definition of LSH family $\mathcal{H}$ is tightly linked with the similarity function of interest $Sim$. An LSH allows us to construct data structures that give provably efficient query time algorithms for $c$-NN problem.

\textbf{Fact}:   Given a family of $(S_0,cS_0,p_1,p_2)$ -sensitive hash functions, one can construct a data structure for $c$-NN with $O(n^\rho \log_{1/p_2}{n})$ query time, where $\rho = \frac{\log{p_1}}{\log{p_2}}$.

The quantity $\rho < 1$ measures the efficiency of a given LSH, the smaller the better.  In theory, in the worst case, the number of points scanned by a given LSH to find a $c$-approximate near neighbor is $O(n^\rho)$~\cite{Proc:Indyk_STOC98}, which is dependent on $\rho$.  Thus given two LSHs, for the same $c$-NN problem, the LSH with smaller value of $\rho$ will achieve the same approximation guarantee and at the same time will have faster query time. LSH with lower value of $\rho$ will report fewer points from the database as the potential near neighbors. These reported points need additional re-ranking to find the true $c$-approximate near neighbor, which is a costly step.  It should be noted that the efficiency of an LSH scheme, the $\rho$ value,  is dependent on many things. It depends on the similarity threshold $S_0$ and the value of $c$ which is the approximation parameter.

\vspace{-0.1in}
\subsection{Resemblance Similarity and MinHash}
\vspace{-0.1in}
\label{sec:minhash}

Minwise hashing~\cite{Proc:Broder_STOC98} is the LSH for resemblance similarity.
The minwise hashing family applies a random permutation $\pi:\Omega \rightarrow \Omega$, on the given set $W$, and stores only the minimum value after the permutation mapping.  Formally MinHash is defined as:
 \begin{equation}h_{\pi}^{min}(W) = \min(\pi(W)).\end{equation}
Given sets $W_1$ and $W_2$, it can be shown by elementary probability argument that
\begin{equation}
\label{eq:minhash}
Pr({h_{\pi}^{min}(W_1) = h_{\pi}^{min}(W_2)) =  \frac{|W_1 \cap W_2|}{| W_1 \cup W_2|}} = \mathcal{R}.
\end{equation}

It follows from (\ref{eq:minhash}) that minwise hashing is  $(\mathcal{R}_0,c\mathcal{R}_0,\mathcal{R}_0,c\mathcal{R}_0)$ sensitive family of hash function when the similarity function of interest is resemblance i.e $\mathcal{R}$. It has efficiency $\rho = \frac{\log{\mathcal{R}_0}}{\log{c\mathcal{R}_0}}$ for approximate resemblance based search.

\vspace{-0.1in}
\subsection{SimHash and Cosine Similarity}
\vspace{-0.1in}
SimHash is another popular LSH for the cosine similarity measure, which originates from the concept of {\em sign random projections (SRP)}~\cite{Proc:Charikar}. Given a vector $x$, SRP utilizes a random vector $w$  with each component generated from i.i.d. normal, i.e., $w_i \sim N(0,1)$, and only stores the sign of the projected data. Formally, SimHash is given by
 \begin{align}
 h^{sim}_w(x) = sign(w^Tx)
 \end{align}
It was shown in~\cite{Article:Goemans} that the collision under SRP satisfies the following equation:
\begin{equation}\label{eq:srp}
Pr(h^{sim}_w(x) = h^{sim}_w(y)) = 1 - \frac{\theta}{\pi},
\end{equation}
where  $\theta = cos^{-1}\left( \frac{x^Ty}{||x||_2 ||y||_2}\right)$. The term $\frac{x^Ty}{||x||_2 ||y||_2}$, is the cosine similarity for data vectors $x$ and $y$, which becomes $\mathcal{S}=\frac{a}{\sqrt{f_1f_2}}$ when the data  are binary.

Since $1 - \frac{\theta}{\pi}$ is monotonic with respect to cosine similarity $\mathcal{S}$.  Eq. (\ref{eq:srp}) implies that SimHash is  a $\bigg(\mathcal{S}_0,c\mathcal{S}_0,\left(1 - \frac{cos^{-1}(\mathcal{S}_0)}{\pi}\right), \left(1 - \frac{cos^{-1}(c\mathcal{S}_0)}{\pi}\right) \bigg)$ sensitive hash function with efficiency $\rho = \frac{\log{\left(1 - \frac{cos^{-1}(\mathcal{S}_0)}{\pi}\right)}}{\log{\left(1 - \frac{cos^{-1}(c\mathcal{S}_0)}{\pi}\right)}}$.

\vspace{-0.1in}
\section{Theoretical Comparisons}
\vspace{-0.1in}

We would like to highlight here that the $\rho$ values for MinHash and SimHash, shown in the previous section, are not directly comparable because they are in the context of different similarity measures. Consequently, it was not clear, before our work, if there is any theoretical way of finding conditions under which MinHash is preferable over SimHash and vice versa. It turns out that the two sided bounds  in Theorem~\ref{thm_ineq} allow us to prove  MinHash is also an LSH for cosine similarity.

\vspace{-0.1in}
\subsection{MinHash as an LSH for Cosine Similarity}
\vspace{-0.1in}
We fix our gold standard similarity measure to be the cosine similarity $Sim = \mathcal{S}$. Theorem~\ref{thm_ineq} leads to two simple corollaries:
\begin{corollary}
\label{cor:p1}
 If $\mathcal{S}(x,y) \ge S_0$, then we have $Pr(h^{min}_\pi (x) = h^{min}_\pi (y))= \mathcal{R}(x,y) \ge S_0^2$
\end{corollary}

\begin{corollary}
\label{cor:p2}
 If $\mathcal{S}(x,y) \le cS_0$, then we have $Pr(h^{min}_\pi (x) = h^{min}_\pi (y))= \mathcal{R}(x,y) \le \frac{cS_0}{2 - cS_0}$
\end{corollary}

Immediate consequence of these two corollaries combined with the definition of LSH is the following:
\begin{theorem}
\label{theo:minrho}
For binary data, MinHash is $(S_0,cS_0,S_0^2,\frac{cS_0}{2 - cS_0})$ sensitive family of hash function for cosine similarity with $\rho = \frac{\log{S_0^2}}{\log{\frac{cS_0}{2 - cS_0}}}$.
\end{theorem}

\vspace{-0.1in}
\subsection{1-bit Minwise Hashing}
\vspace{-0.1in}
SimHash  generates a single bit output (only the signs) whereas MinHash generates an integer value. Recently proposed $b$-bit minwise hashing~\cite{Proc:Li_Internetware13}
provides simple strategy  to generate an informative single bit output from MinHash, by using the parity of MinHash values:
\begin{equation}h_{\pi}^{min,1bit}(W_1) = \begin{cases}1 & \mbox{if $h_{\pi}^{min}(W_1)$ is odd} \\ 0& otherwise \end{cases}\end{equation}

For 1-bit  MinHash and very sparse data (i.e., $\frac{f_1}{D}\rightarrow0$, $\frac{f_2}{D}\rightarrow0$), we have the following collision probability
\begin{align}
{Pr}(h_{\pi}^{min,1bit}(W_1) = h_{\pi}^{min,1bit}(W_2))= \frac{\mathcal{R}+1}{2}
\end{align}
The analysis presented in previous sections allows us to theoretically analyze this new scheme. The inequality in Theorem~\ref{thm_ineq} can be modified for $\frac{\mathcal{R}+1}{2}$ and using similar arguments as for MinHash we obtain
\begin{theorem}
\label{theo:1bitmin}
For binary data, 1-bit MH (minwise hashing) is $(S_0,cS_0,\frac{S_0^2+1}{2},\frac{1}{2 - cS_0})$ sensitive family of hash function for cosine similarity with $\rho = \frac{\log{\frac{2}{S_0^2+1}}}{\log{(2 - cS_0)}}$.
\end{theorem}

\vspace{-0.1in}
\subsection{Worst Case Gap Analysis}
\vspace{-0.1in}

\begin{figure}[h!]
\begin{center}
\mbox{
\includegraphics[width=1.7in]{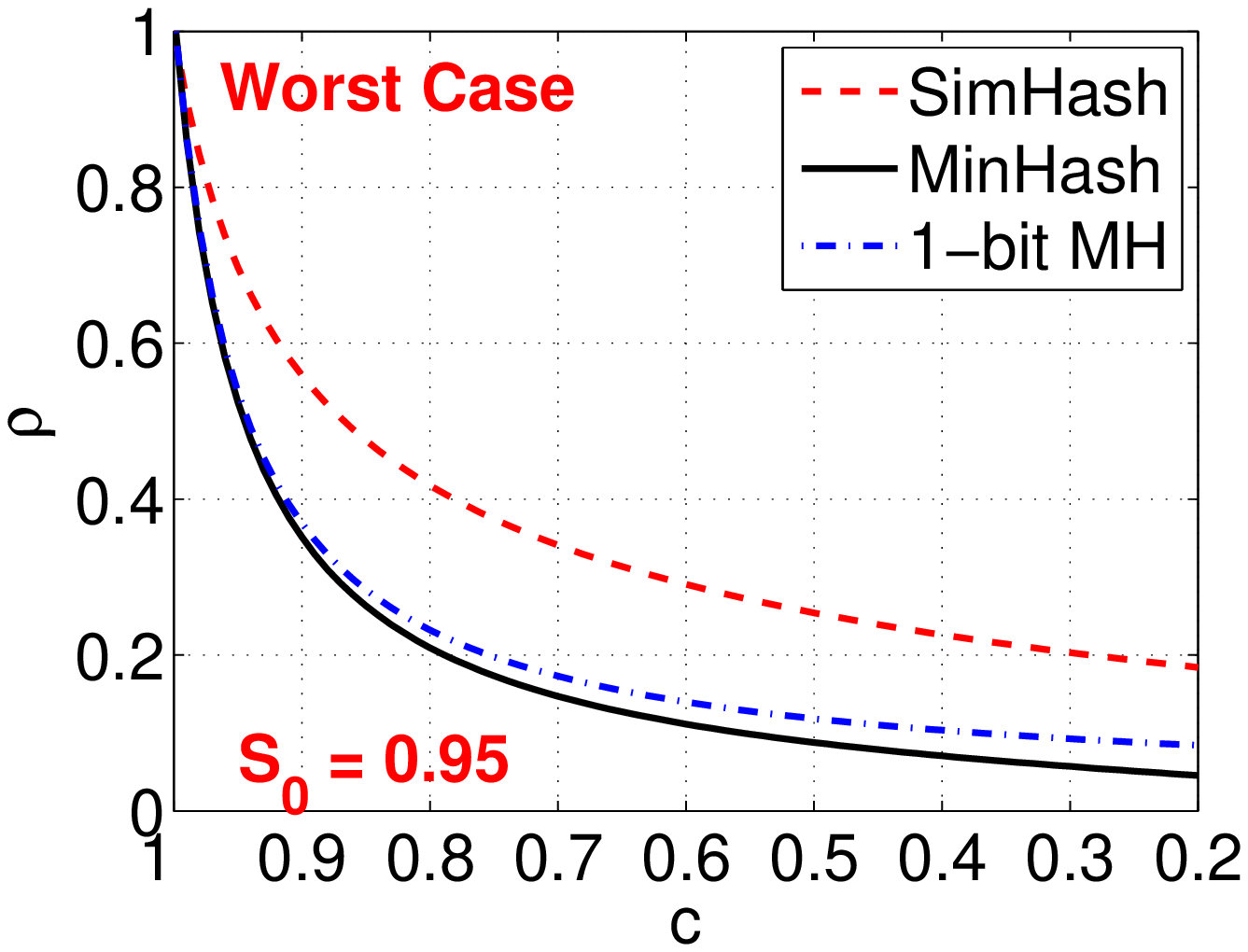}\hspace{-0.15in}
\includegraphics[width=1.7in]{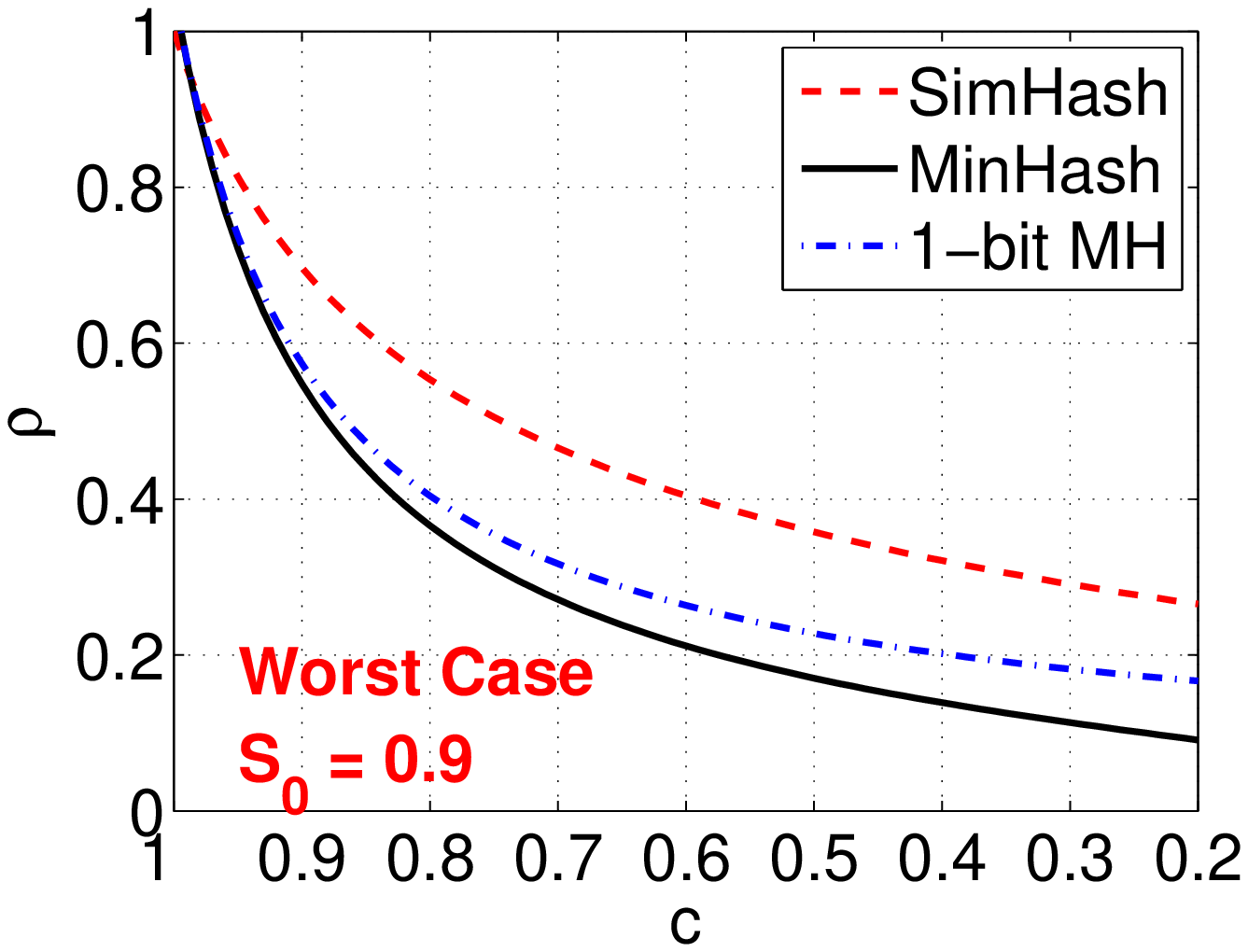}
}

\mbox{
\includegraphics[width=1.7in]{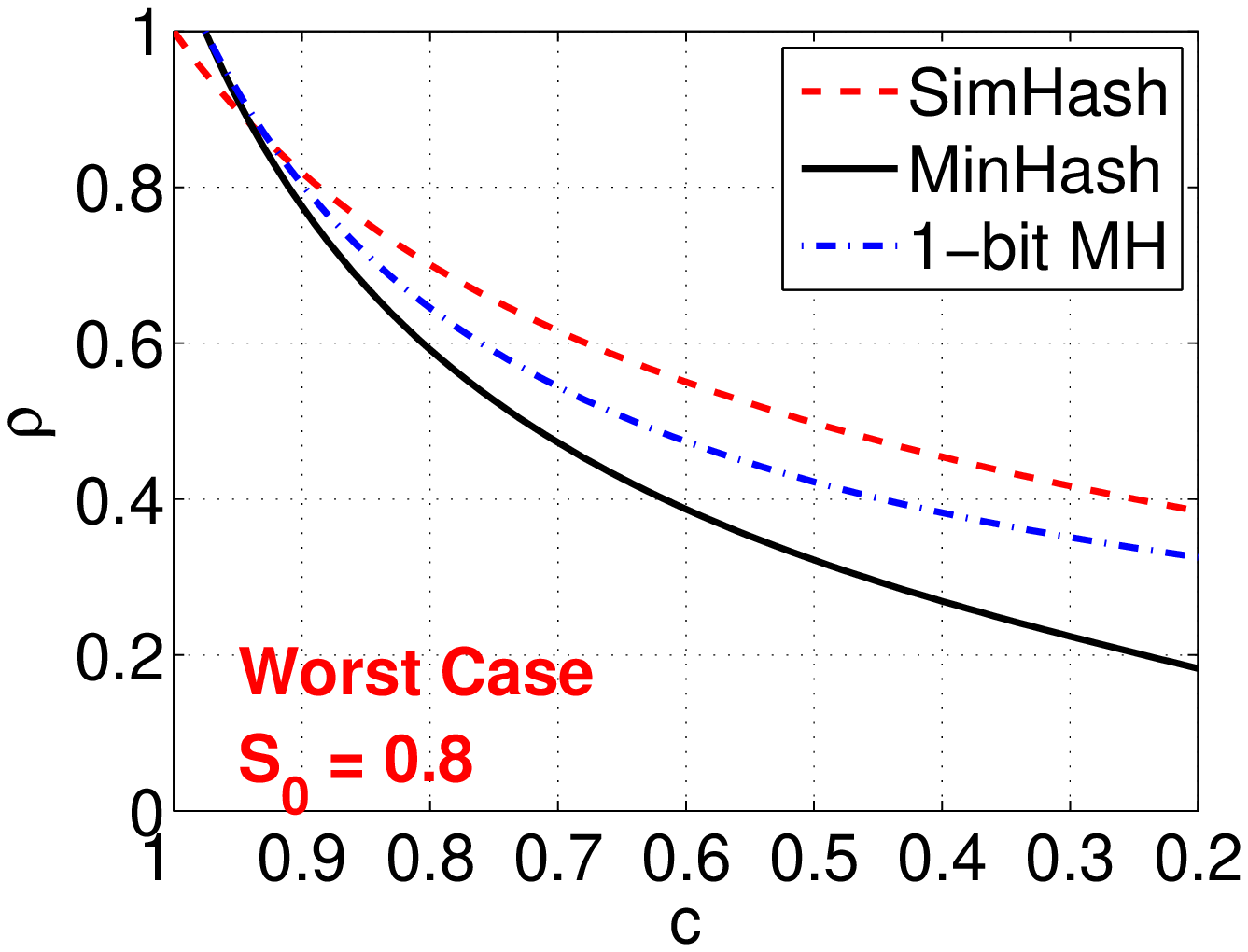}\hspace{-0.15in}
\includegraphics[width=1.7in]{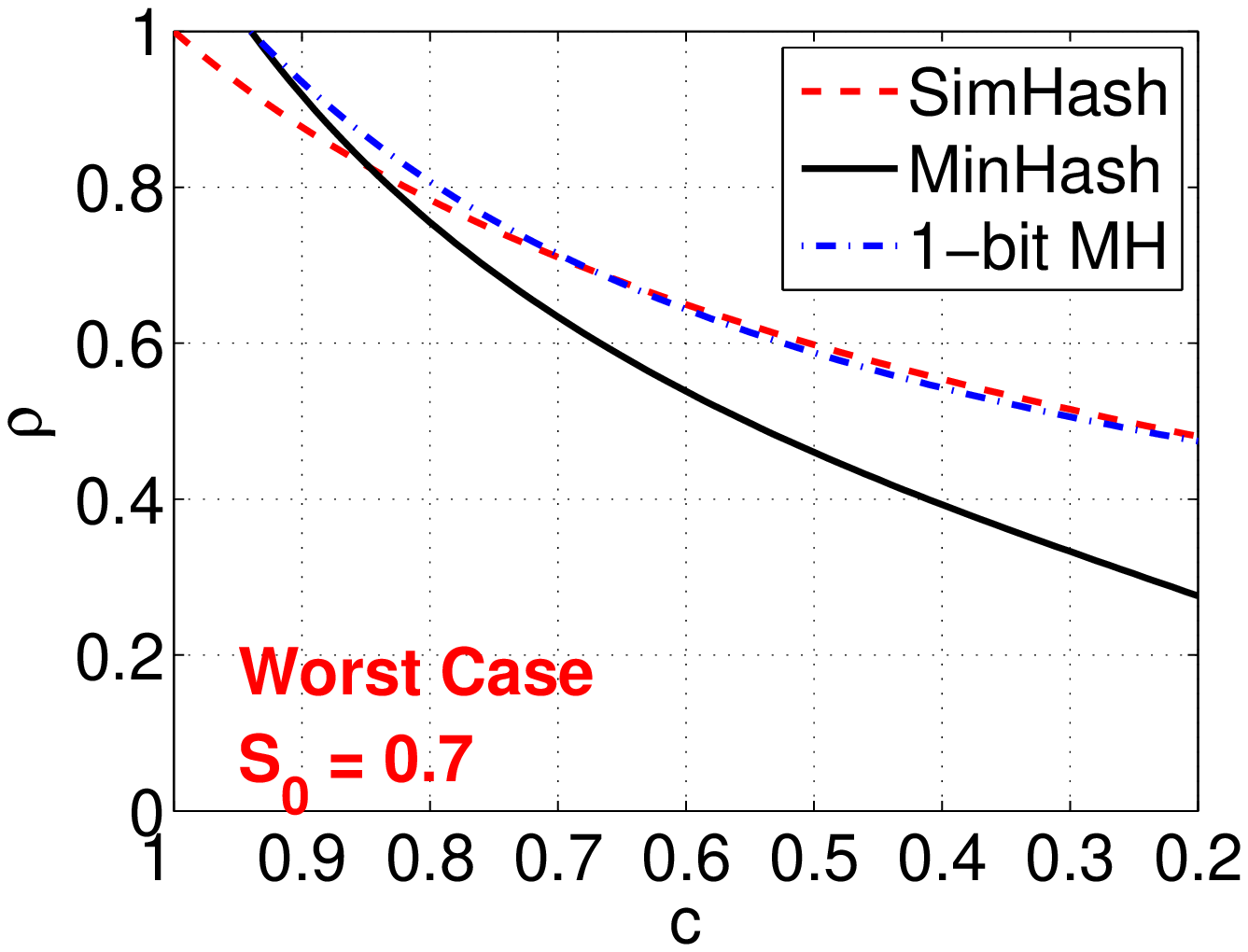}
}
\end{center}
\vspace{-0.2in}
\caption{Worst case gap ($\rho$) analysis, i.e., (\ref{eqn_RhoSH}) (\ref{eqn_RhoMH}) (\ref{eqn_Rho1bitMH}), for high similarity region; lower is better. }\label{fig_WorstRho}
\end{figure}

We will compare the gap ($\rho$) values of the three hashing methods we have studied:
\begin{align}\label{eqn_RhoSH}
&\text{SimHash:} \ \ \rho = \frac{\log{\left(1 - \frac{cos^{-1}(\mathcal{S}_0)}{\pi}\right)}}{\log{\left(1 - \frac{cos^{-1}(c\mathcal{S}_0)}{\pi}\right)}} \\\label{eqn_RhoMH}
&\text{MinHash:} \ \ \rho = \frac{\log{S_0^2}}{\log{\frac{cS_0}{2 - cS_0}}} \\\label{eqn_Rho1bitMH}
&\text{1-bit MH:} \ \rho = \frac{\log{\frac{2}{S_0^2+1}}}{\log{(2 - cS_0)}}
\end{align}
This is a worst case analysis. We know the lower bound $\mathcal{S}^2\leq \mathcal{R}$ is usually very conservative in real data when the similarity level is low.  Nevertheless, for high similarity region, the comparisons of the $\rho$ values indicate that MinHash significantly outperforms SimHash as shown in Figure~\ref{fig_WorstRho}, at least for $\mathcal{S}_0\geq 0.8$.

\vspace{-0.1in}
\subsection{Restricted Worst Case Gap Analysis}
\vspace{-0.1in}

\begin{figure}[h!]
\begin{center}
\mbox{
\includegraphics[width=1.7in]{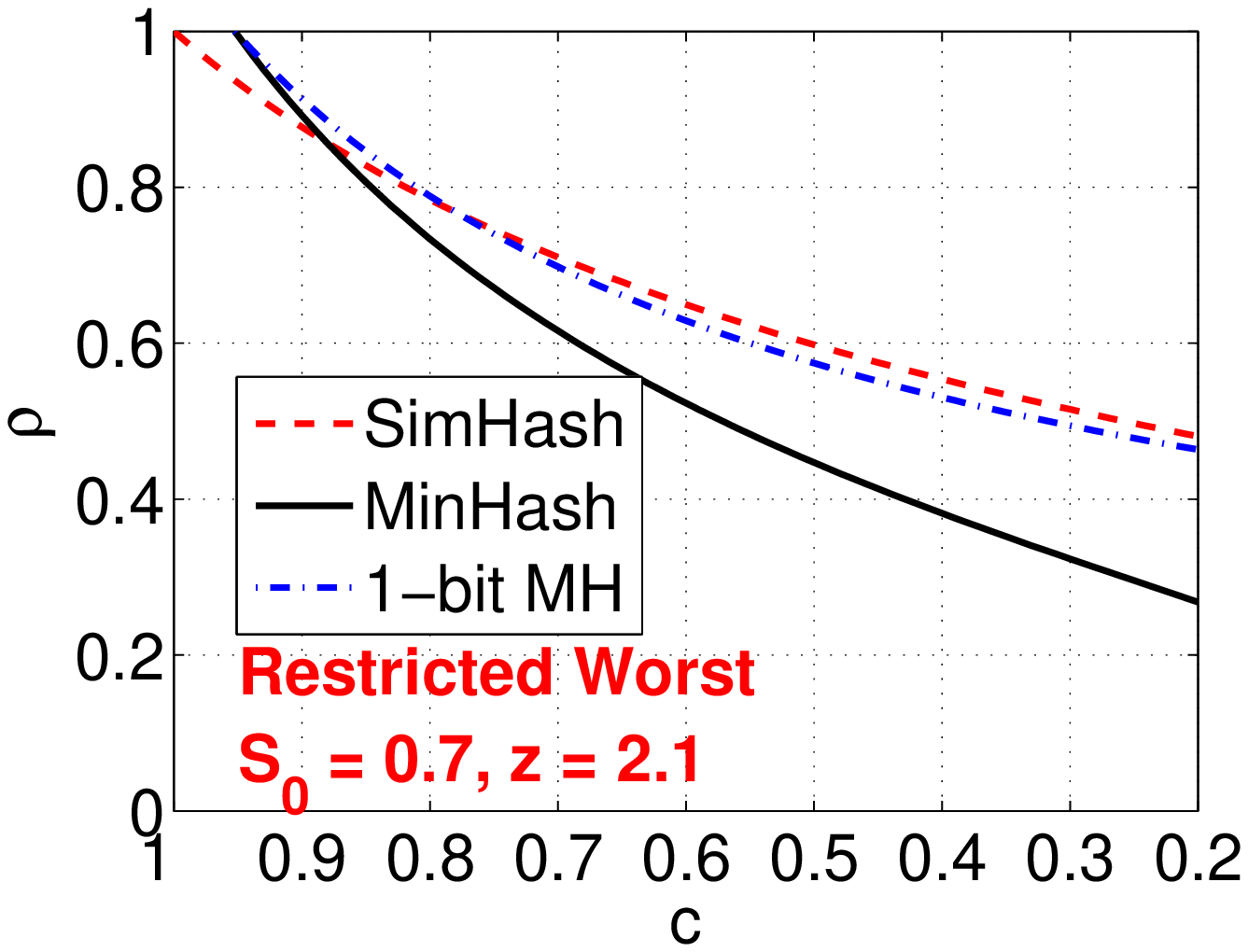}\hspace{-0.15in}
\includegraphics[width=1.7in]{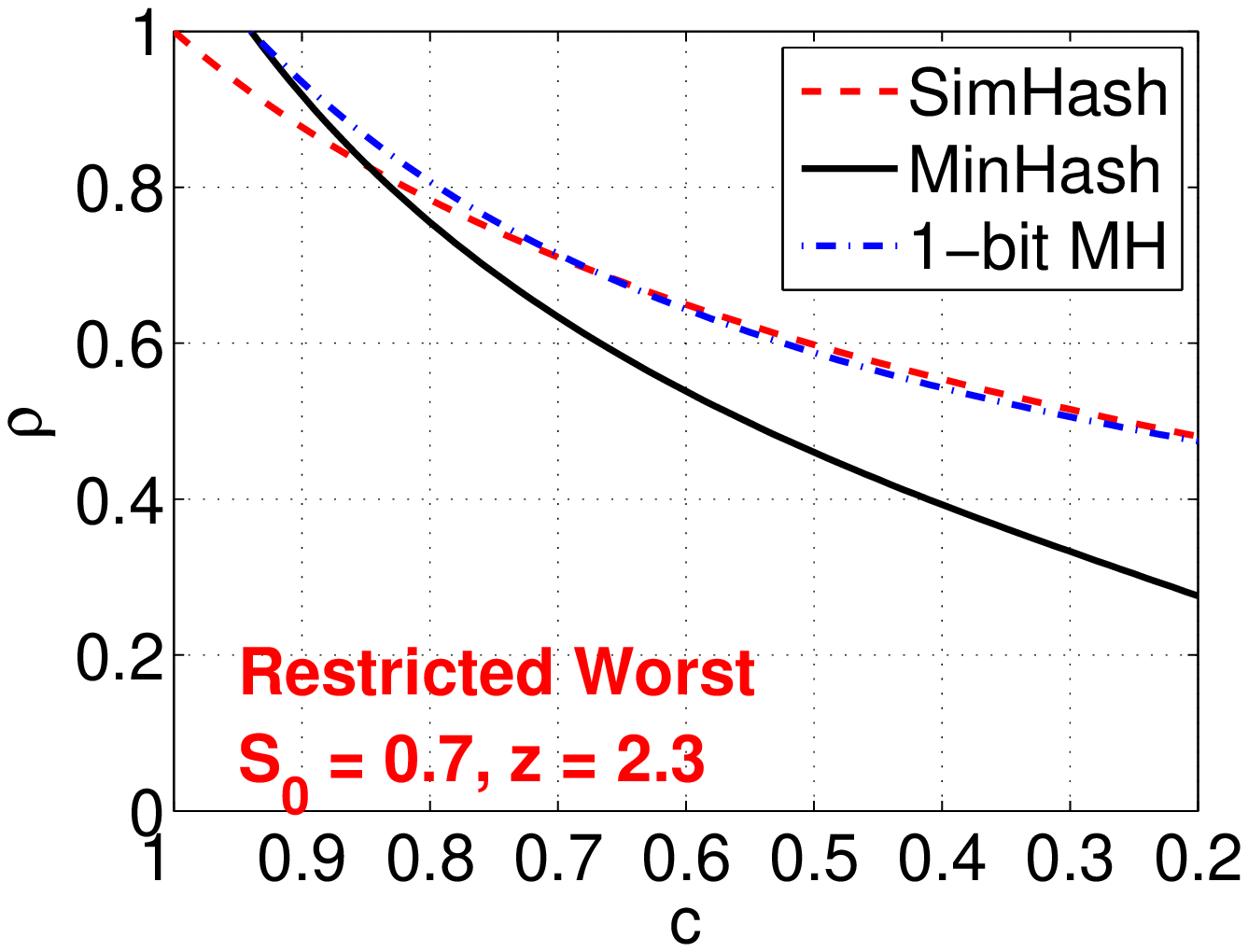}
}

\mbox{
\includegraphics[width=1.7in]{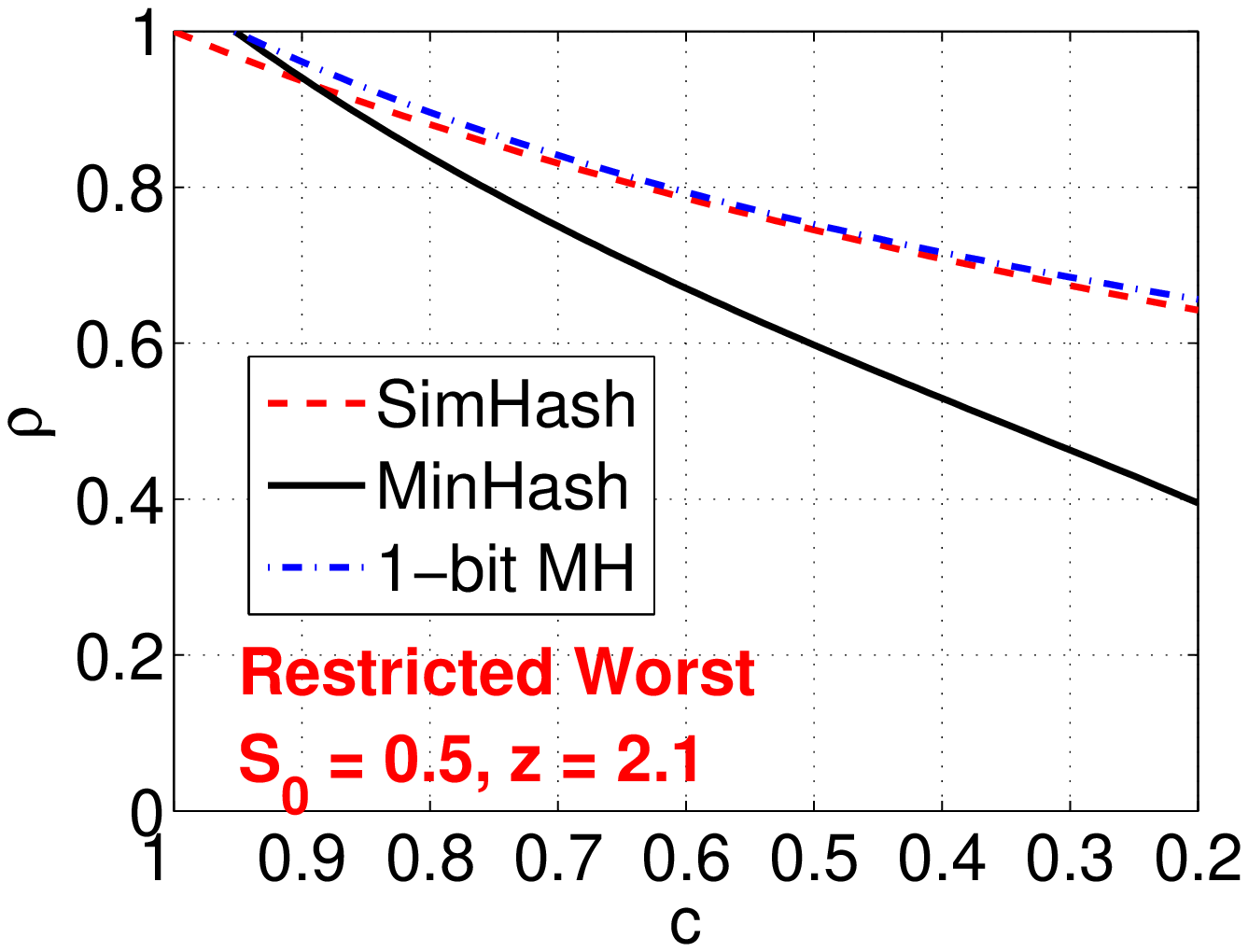}\hspace{-0.15in}
\includegraphics[width=1.7in]{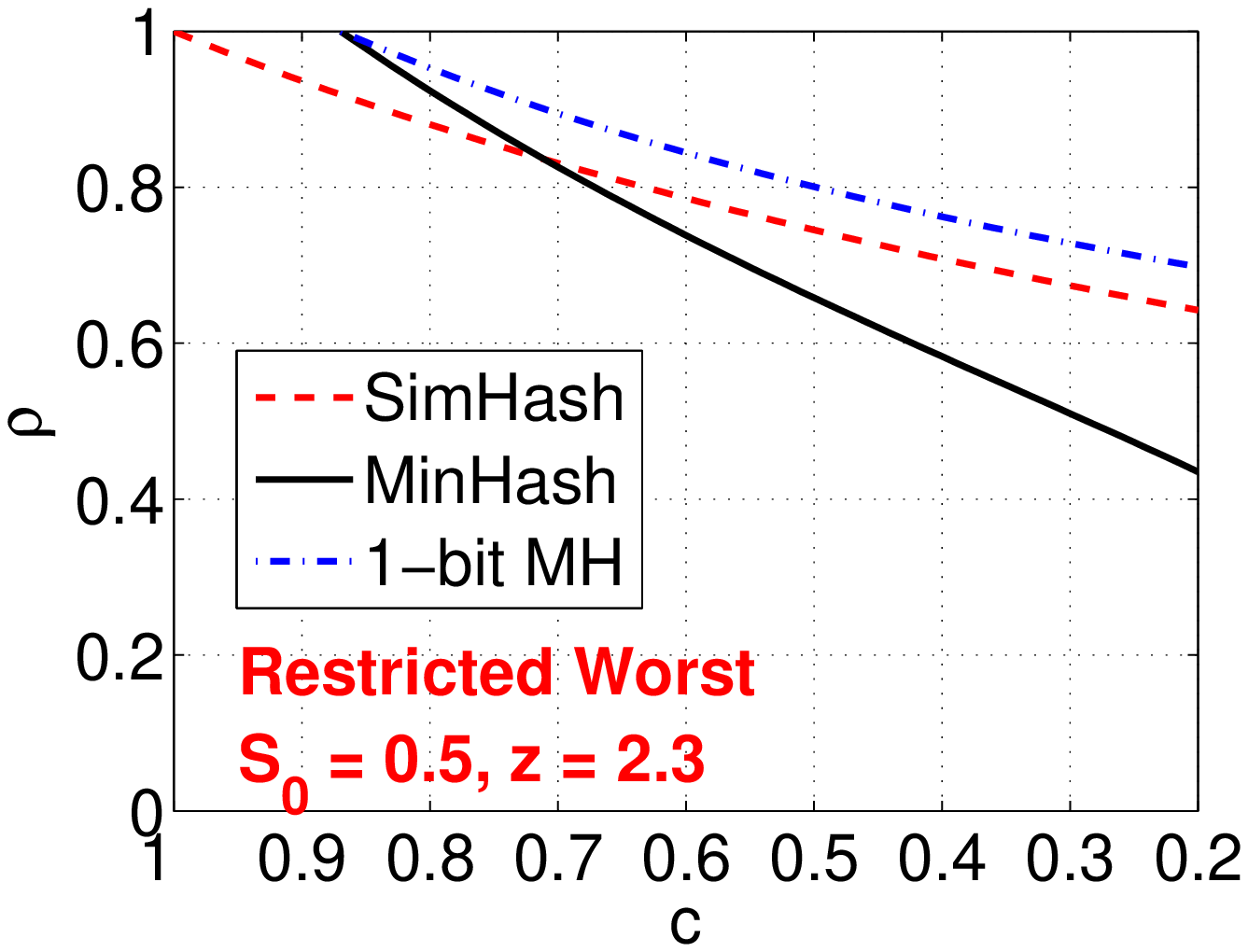}
}

\mbox{
\includegraphics[width=1.7in]{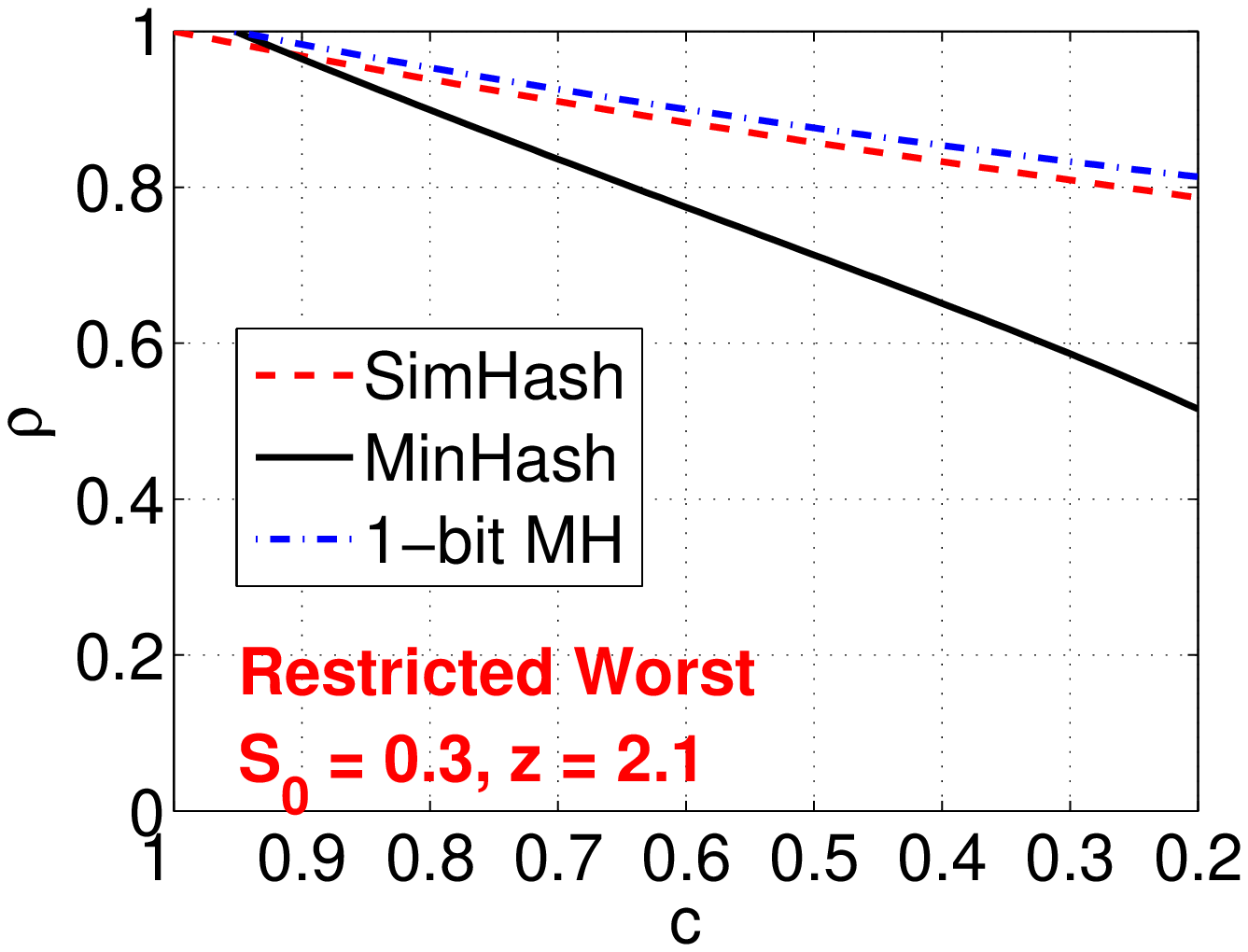}\hspace{-0.15in}
\includegraphics[width=1.7in]{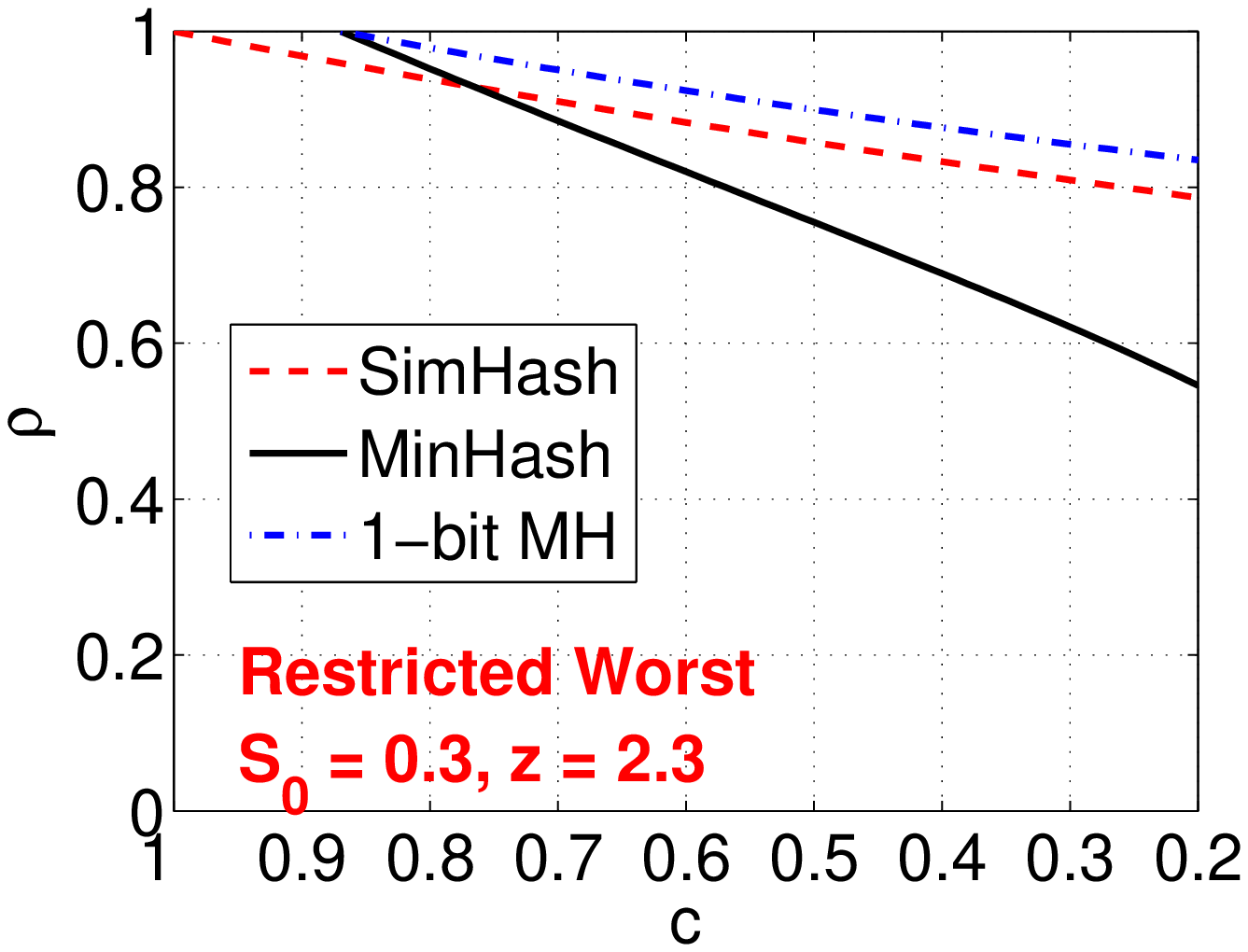}
}
\end{center}
\vspace{-0.2in}
\caption{Restricted worst case gap ($\rho$) analysis by assuming the data satisfy $\frac{\mathcal{S}}{z-\mathbf{S}}\leq R\leq \frac{\mathcal{S}}{2-\mathbf{S}}$, where $z$ is defined in (\ref{eqn_z}). The $\rho$ values for MinHash and 1-bit MinHash are expressed in (\ref{eqn_RhoMHRes}) and (\ref{eqn_Rho1bitMHRes}), respectively.}\label{fig_ResWorstRho}
\end{figure}

The worst case analysis does not make any assumption on the data. It is obviously too conservative when the data are not too similar. Figure~\ref{fig_z} has demonstrated that in real data, we can fairly safely replace the lower bound $\mathcal{S}^2$ with $\frac{\mathcal{S}}{z-\mathcal{S}}$ for some $z$ which, defined in (\ref{eqn_z}), is very close to 2 (for example, 2.1). If we are willing to make this assumption, then we can go through the same analysis for MinHash as an LSH for cosine and compute the corresponding $\rho$ values:
\begin{align}\label{eqn_RhoMHRes}
&\text{MinHash:} \ \ \rho = \frac{\log{\frac{\mathcal{S}_0}{z - \mathcal{S}_0}}}{\log{\frac{c\mathcal{S}_0}{2 - c\mathcal{S}_0}}} \\\label{eqn_Rho1bitMHRes}
&\text{1-bit MH:} \ \rho = \frac{\log{\frac{2(z-\mathcal{S}_0)}{z}}}{\log{(2 - c\mathcal{S}_0)}}
\end{align}
Note that this is still a worst case analysis (and hence can still be very conservative). Figure~\ref{fig_ResWorstRho} presents the $\rho$ values for this restricted worst case gap analysis, for two values of $z$  (2.1 and 2.3) and $\mathcal{S}_0$ as small as 0.2. The results confirms that MinHash still significantly outperforms SimHash even in low similarity region.

\begin{figure}[h!]
\begin{center}
\mbox{
\includegraphics[width=1.7in]{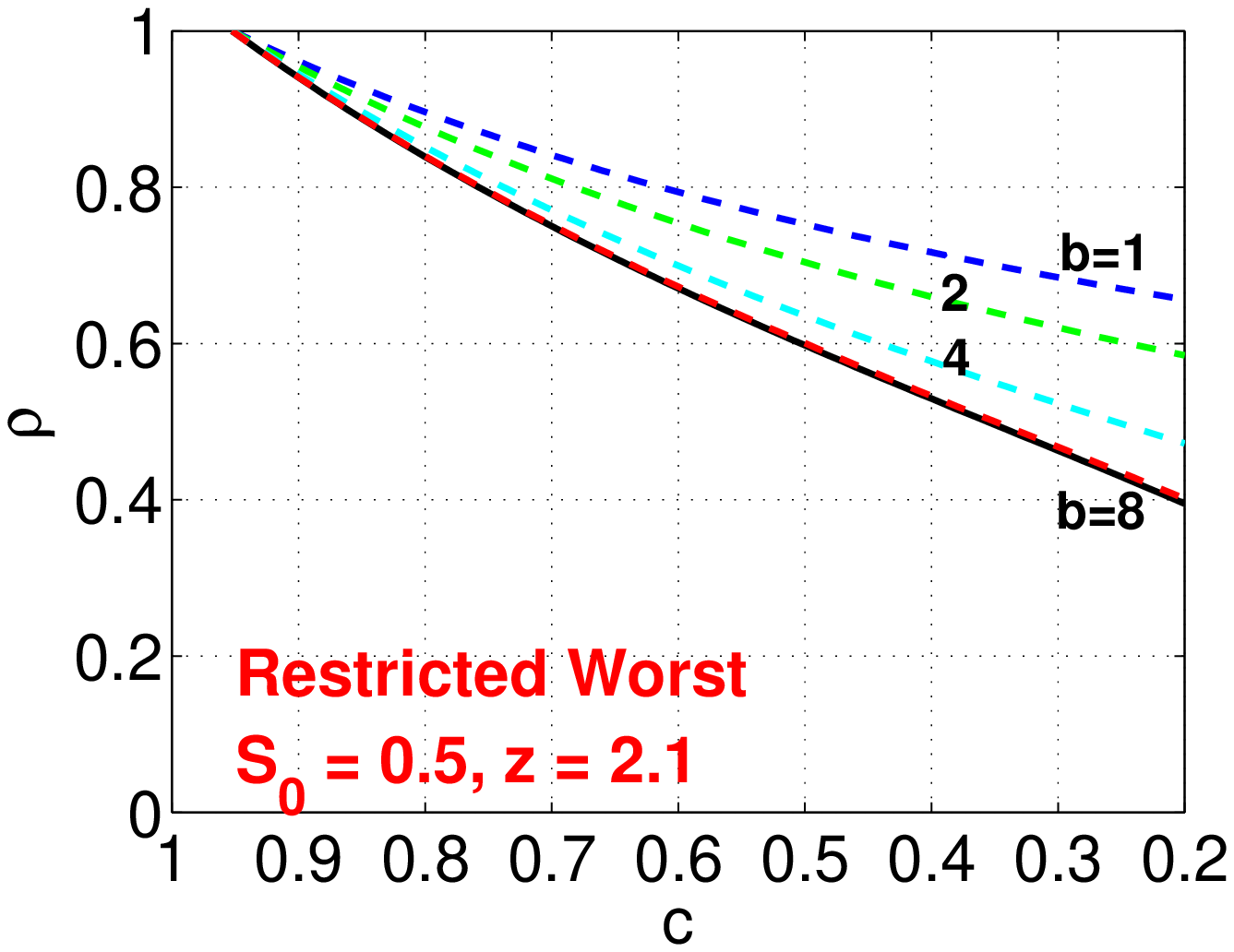}\hspace{-0.15in}
\includegraphics[width=1.7in]{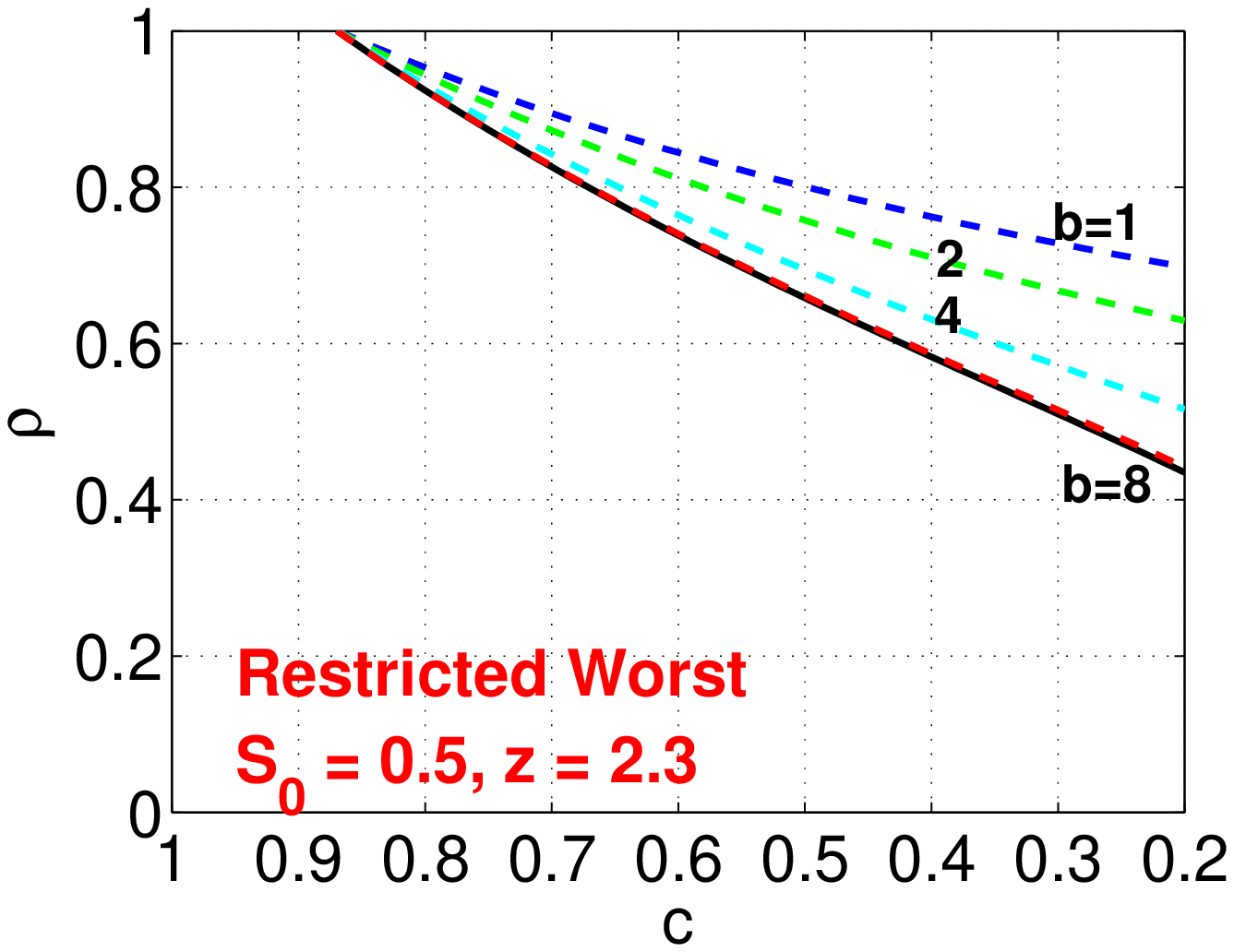}
}
\mbox{
\includegraphics[width=1.7in]{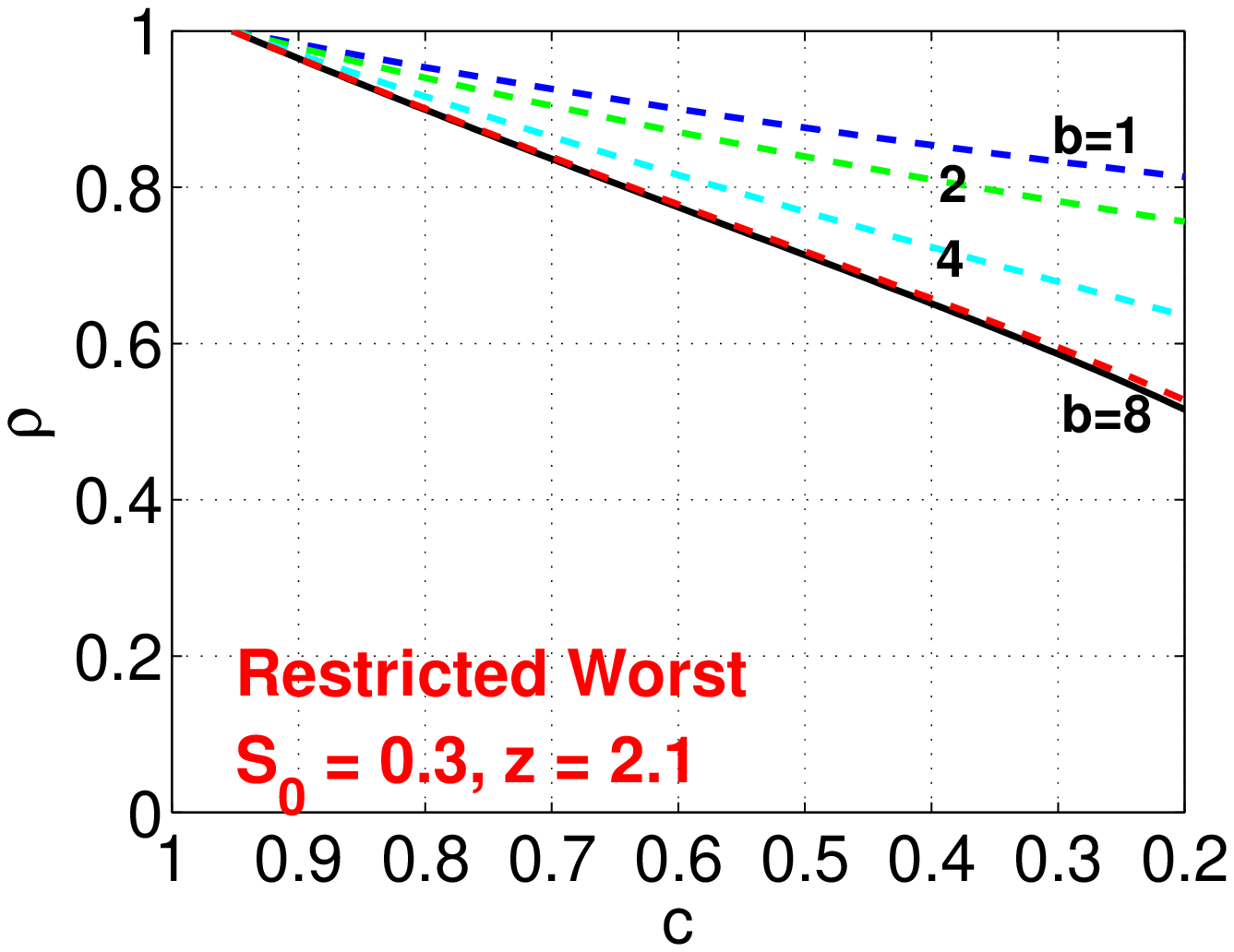}\hspace{-0.15in}
\includegraphics[width=1.7in]{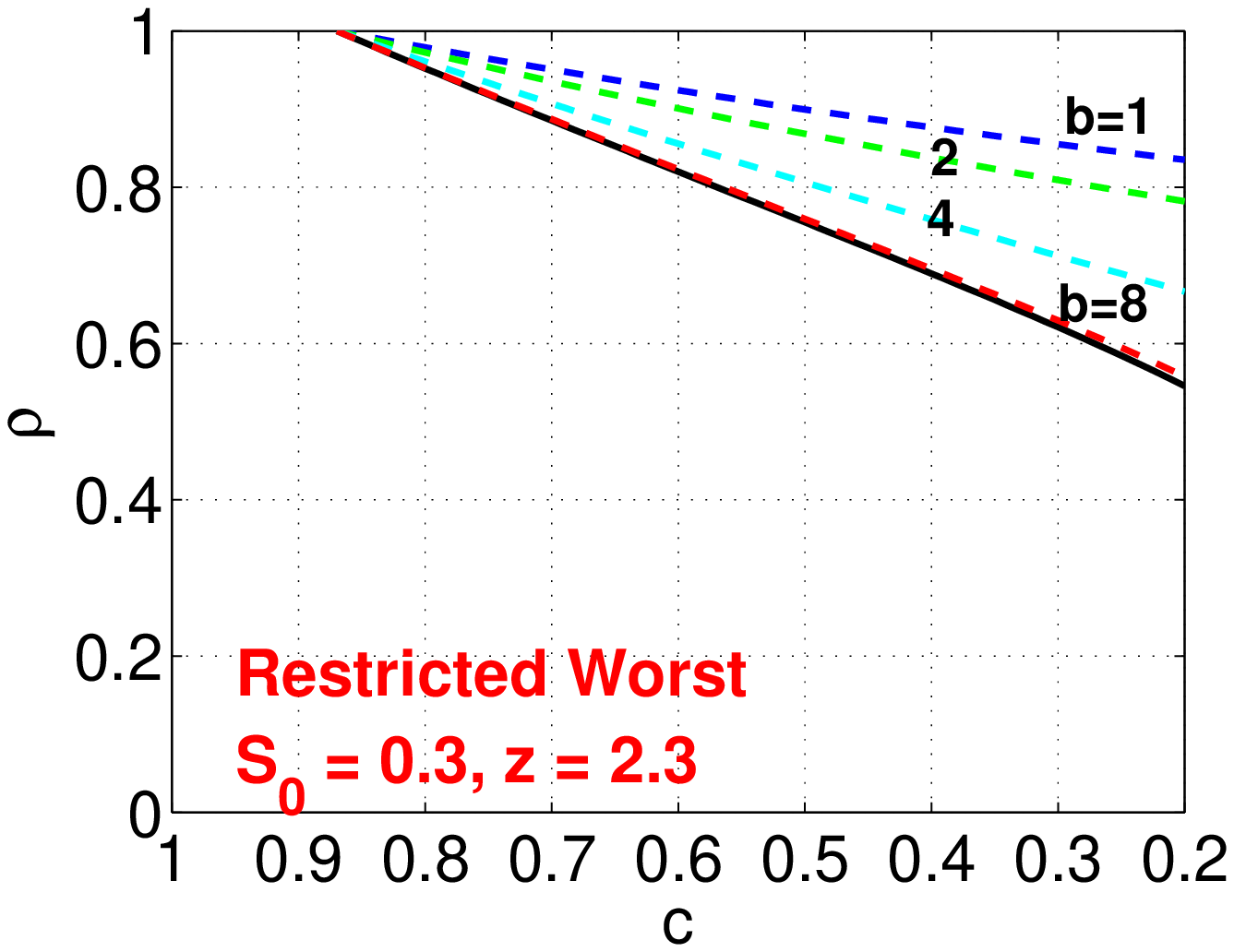}
}
\end{center}
\vspace{-0.2in}
\caption{Restricted worst case gap ($\rho$) analysis for $b$-bit minwise hashing for $b=1$, 2, 4, 8.}\label{fig_bbitMH}\vspace{-0.1in}
\end{figure}

Both Figure~\ref{fig_WorstRho} and Figure~\ref{fig_ResWorstRho} show that 1-bit MinHash can be less competitive when the similarity is not high. This is expected as analyzed in the original paper of $b$-bit minwise hashing~\cite{Proc:Li_Konig_WWW10}. The remedy is to use more bits. As shown in Figure~\ref{fig_bbitMH}, once we use $b=8$ (or even $b=4$) bits, the performance of $b$-bit minwise hashing is not much different from MinHash.

\vspace{-0.1in}

\vspace{-0.1in}
\subsection{Idealized Case Gap Analysis}
\vspace{-0.1in}

\begin{figure}[h!]
\begin{center}
\mbox{
\includegraphics[width=1.7in]{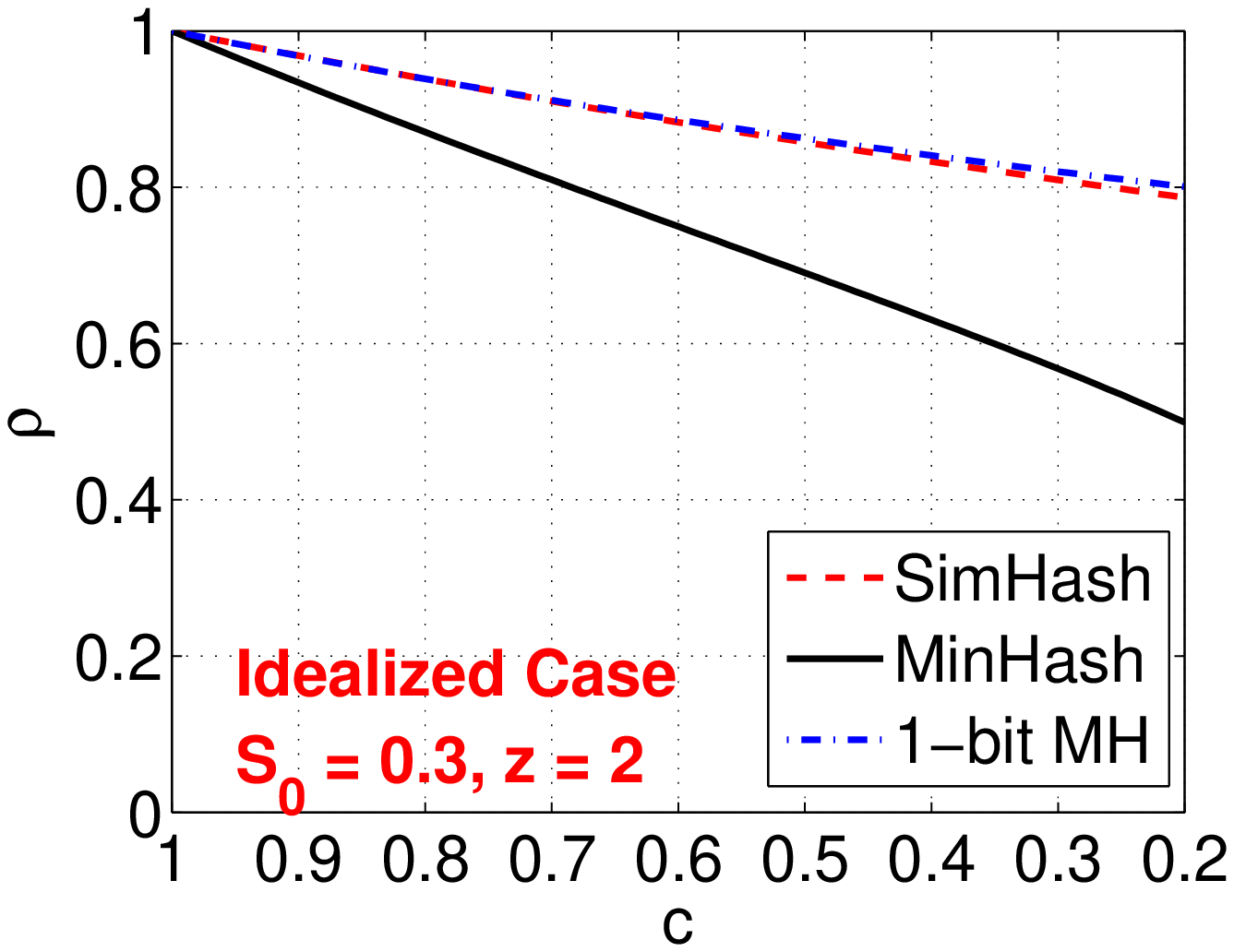}\hspace{-0.15in}
\includegraphics[width=1.7in]{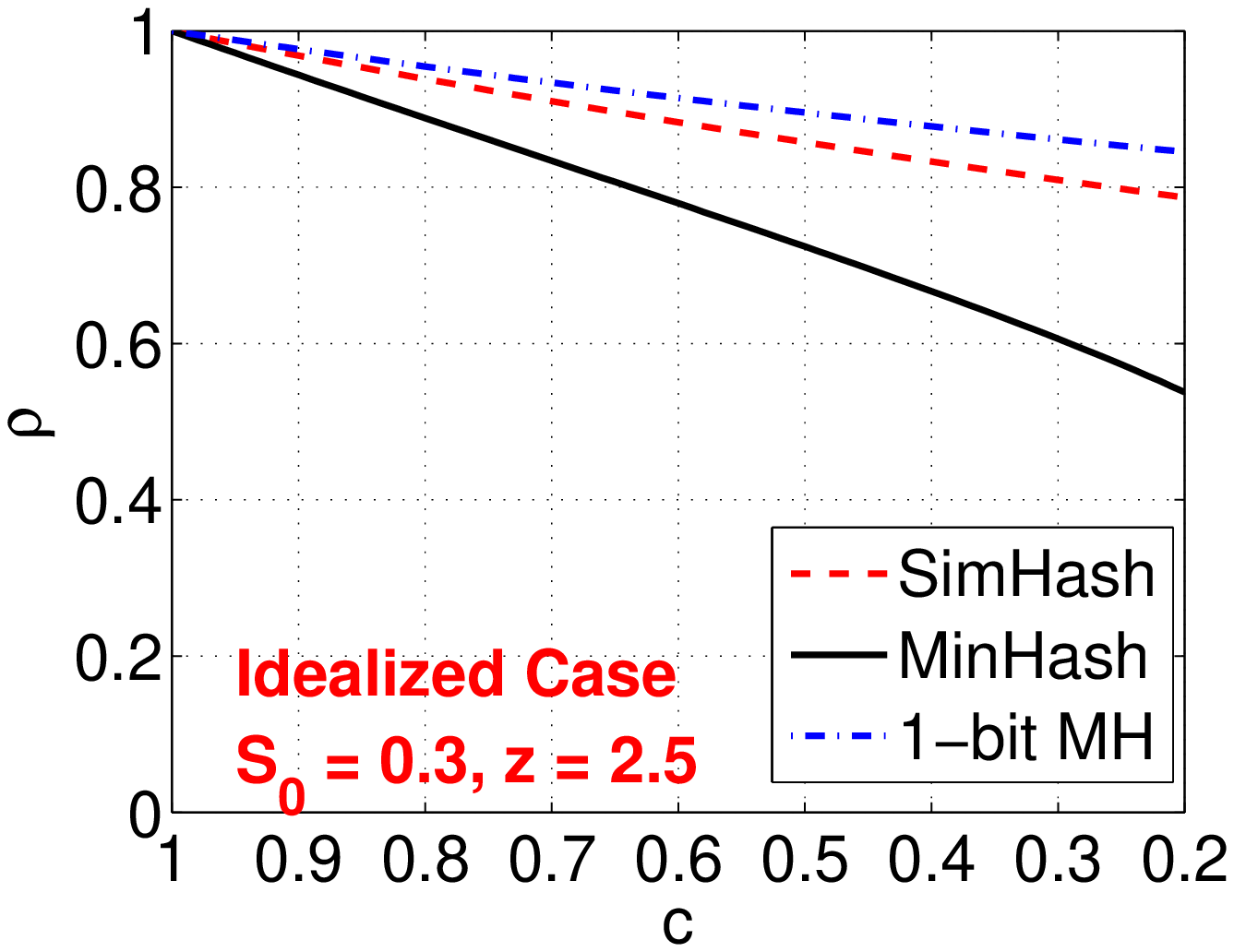}
}
\mbox{
\includegraphics[width=1.7in]{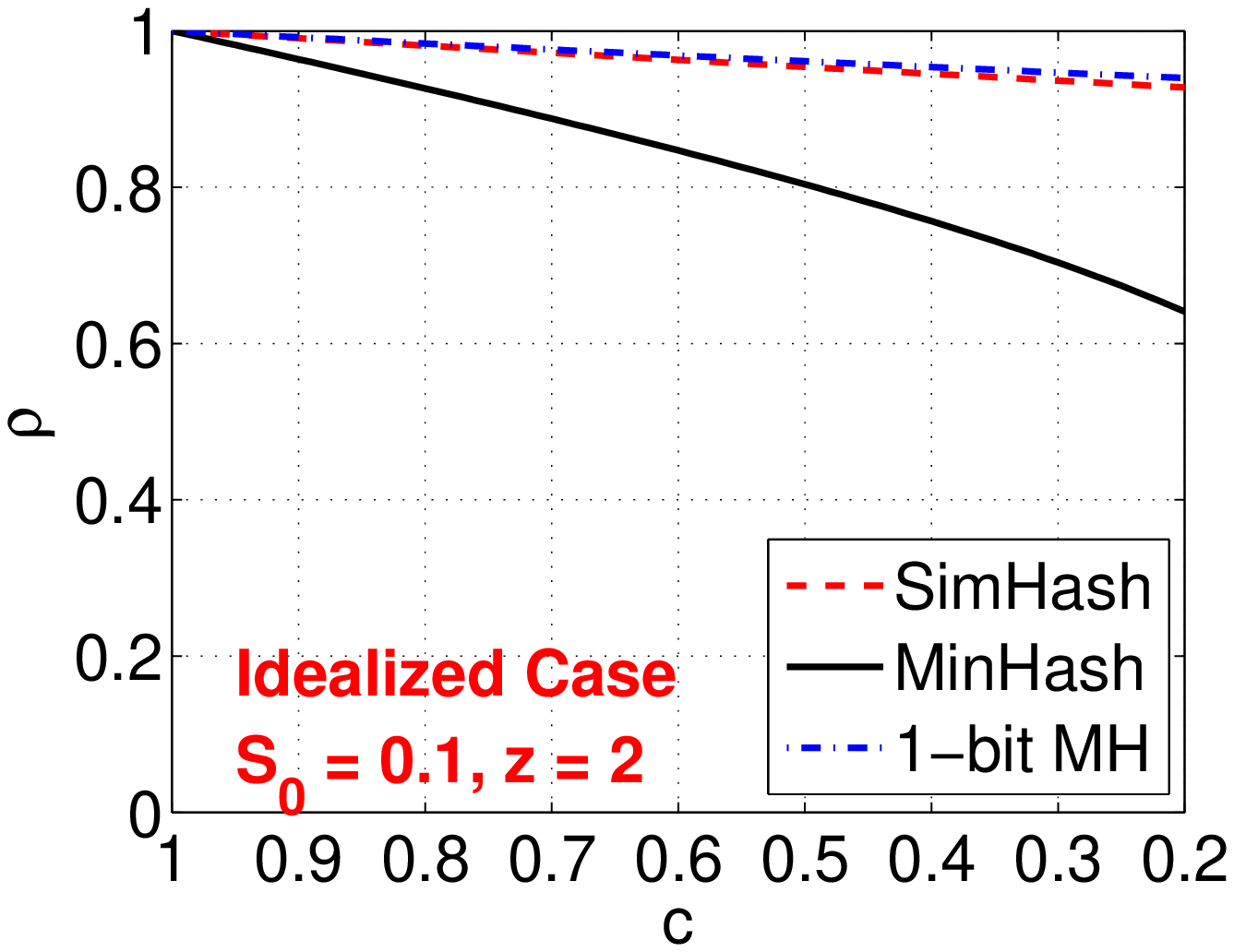}\hspace{-0.15in}
\includegraphics[width=1.7in]{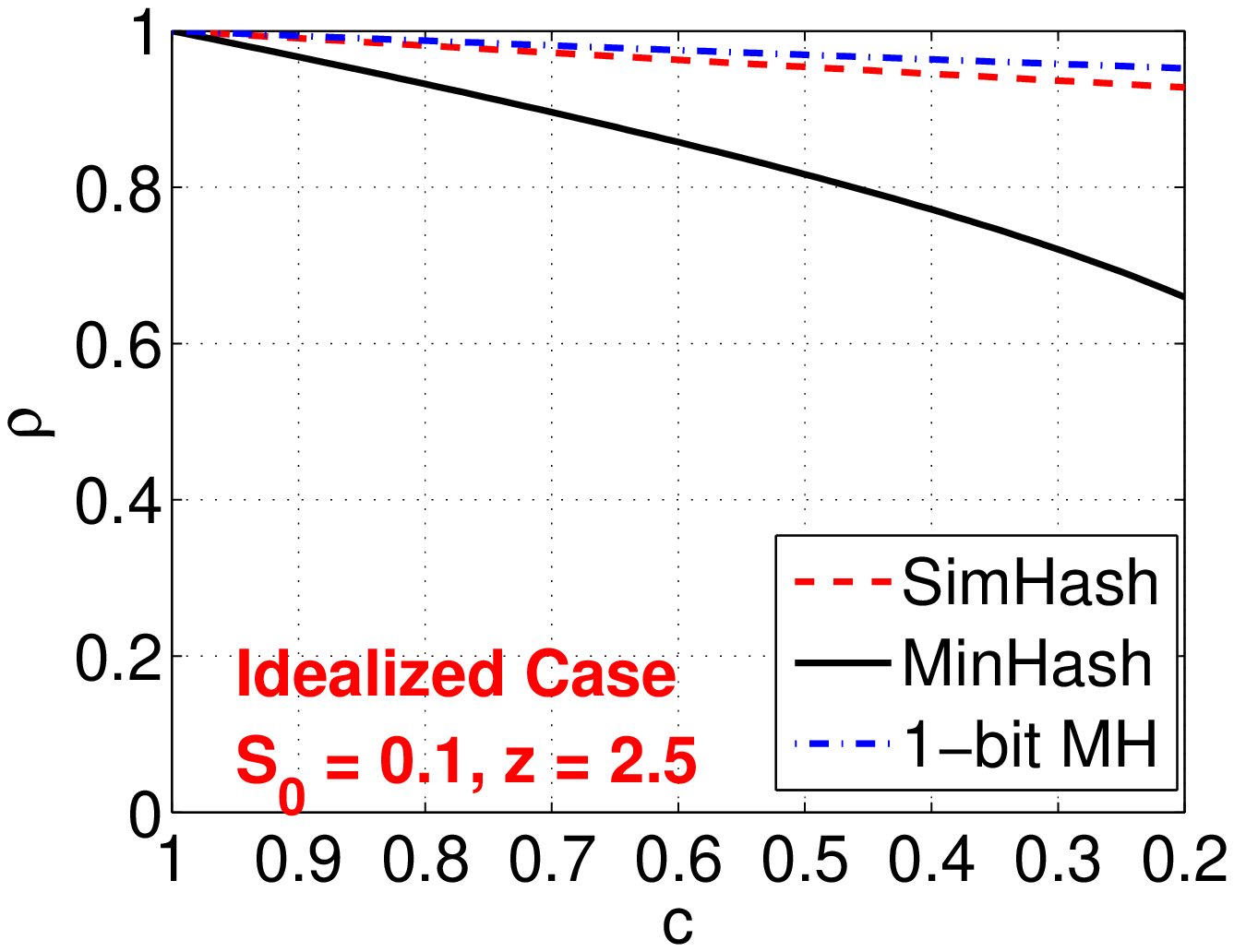}
}
\end{center}
\vspace{-0.2in}
\caption{Idealized case gap ($\rho$)  analysis by assuming $\mathcal{R} = \frac{\mathcal{S}}{z-\mathcal{S}}$ for a fixed $z$  ($z=2$ and $z=2.5$ in the plots). }
\label{fig_IdealRho}\vspace{-0.1in}
\end{figure}

The restricted worst case analysis can still be very conservative and may not fully explain the stunning performance of MinHash in our experiments on datasets of low similarities. Here, we also provide an analysis based on fixed $z$ value. That is, we only analyze the gap $\rho$ by assuming $\mathcal{R} = \frac{\mathcal{S}}{z-\mathcal{S}}$ for a fixed $z$. We call this idealized gap analysis. Not surprisingly, Figure~\ref{fig_IdealRho} confirms that, with this assumption, MinHash significantly outperform SimHash even for extremely low similarity. We should keep in mind that this idealized gap analysis can be somewhat optimistic and should only be used as some side information.

\vspace{-0.1in}
\section{Experiments}
\vspace{-0.1in}

We evaluate both MinHash and SimHash in the actual task of retrieving top-$k$ near neighbors. We implemented the standard $(K,L)$ parameterized LSH~\cite{Proc:Indyk_STOC98} algorithms with both MinHash and SimHash. That is, we concatenate $K$ hash functions to form a new hash function for each table, and we generate $L$ such tables (see~\cite{Report:E2LSH} for more details about the implementation).   We used all the six binarized datasets with the query and training partitions as shown in Table~\ref{tab_data}. For each dataset, elements from training partition were used for constructing hash tables, while the elements of the query partition were used as query for top-$k$ neighbor search. For every query, we compute the gold standard top-$k$ near neighbors using the cosine similarity as the underlying similarity measure.

In standard $(K,L)$ parameterized bucketing scheme the choice of $K$ and $L$ is dependent on the similarity thresholds and the hash function under consideration. In the task of top-$k$ near neighbor retrieval, the similarity thresholds vary with the datasets. Hence, the actual choice of ideal $K$ and $L$ is difficult to determine. To ensure that this choice does not affect our evaluations, we implemented all the combinations of $K\in\{1, 2, ..., 30\}$ and $L\in\{1,2,...,200\}$. These combinations include  the reasonable choices for both the hash function and different threshold levels.

\begin{figure*}[h!]
\begin{center}

\mbox{
\includegraphics[width=1.7in]{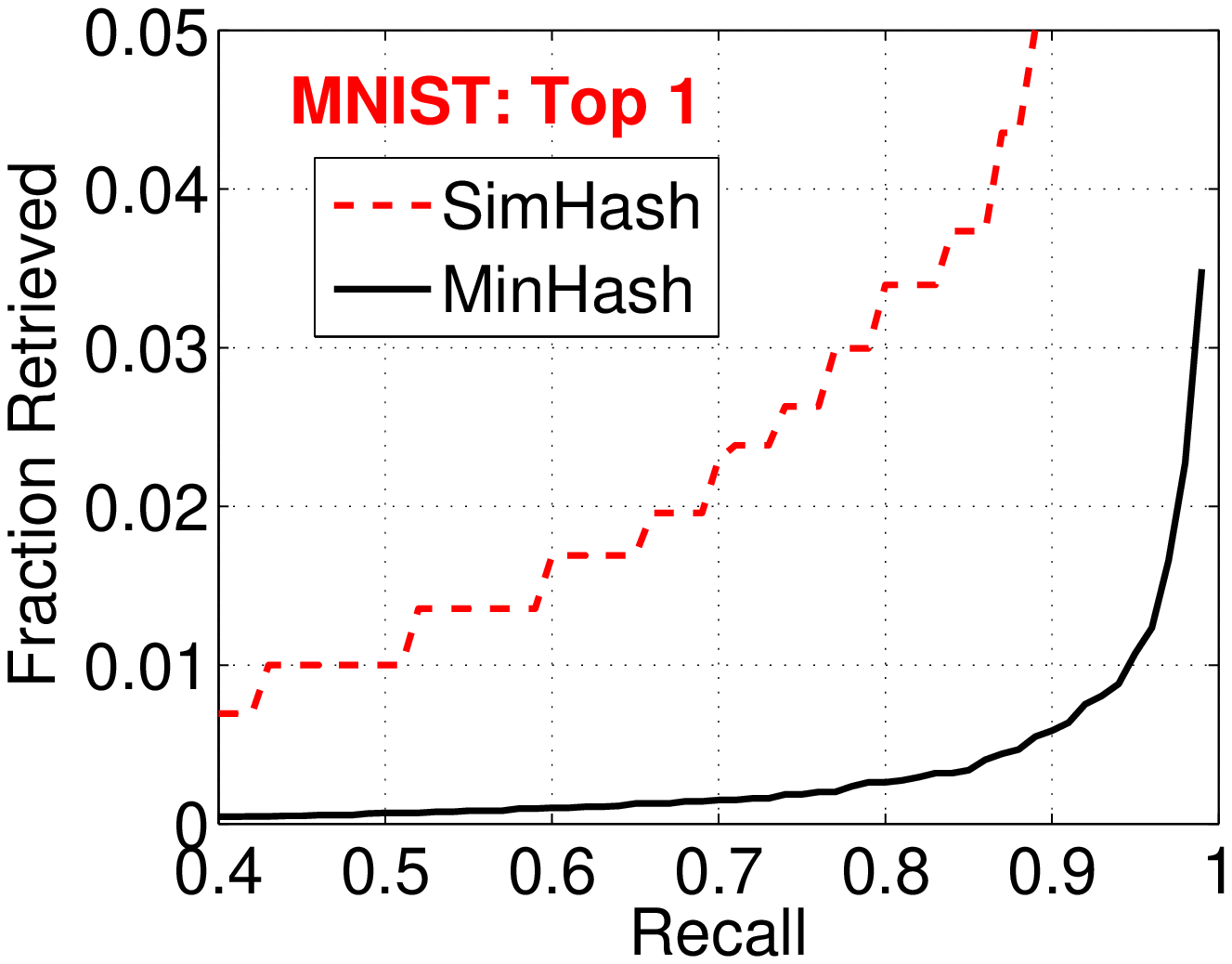}\hspace{-0.13in}
\includegraphics[width=1.7in]{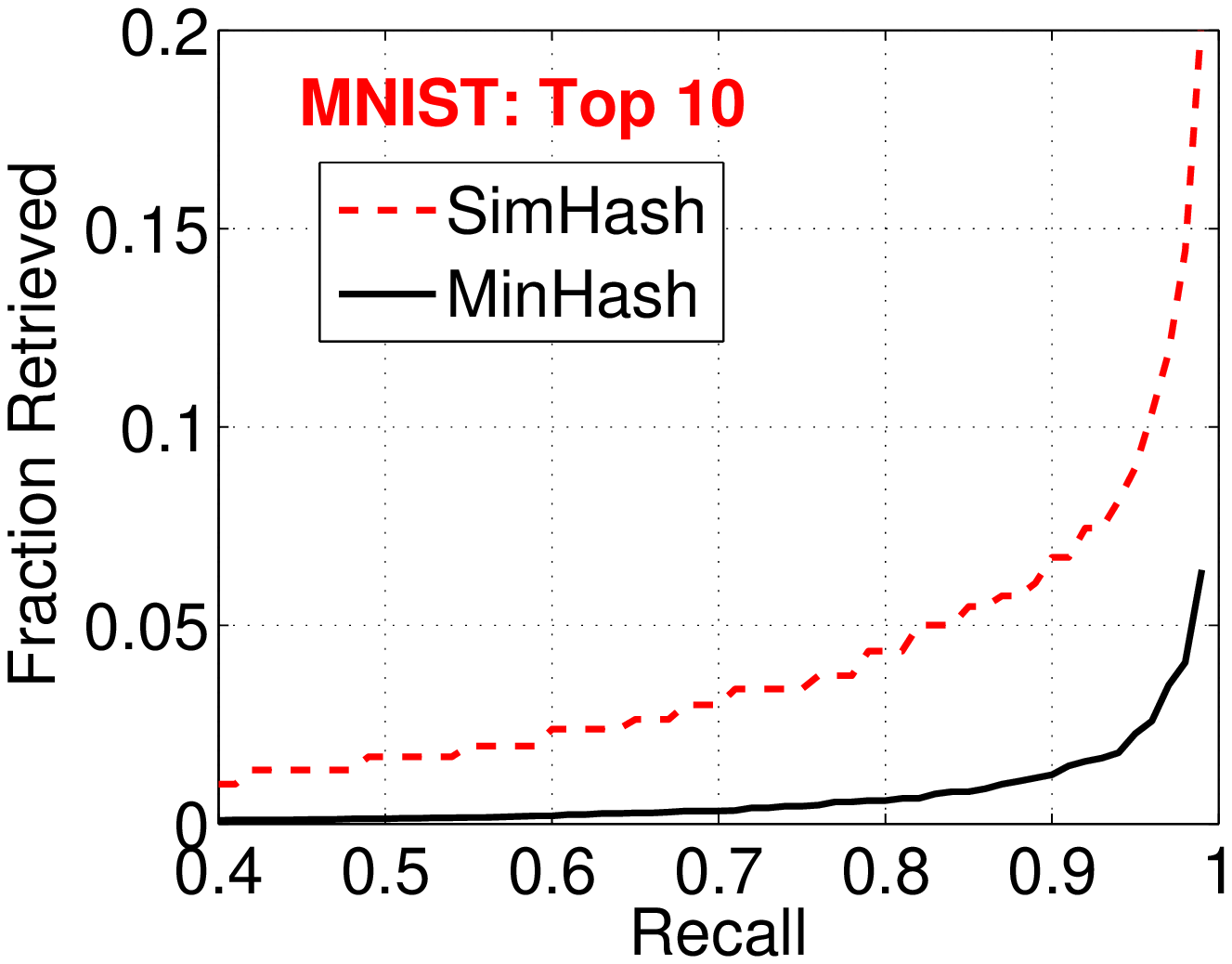}\hspace{-0.13in}
\includegraphics[width=1.7in]{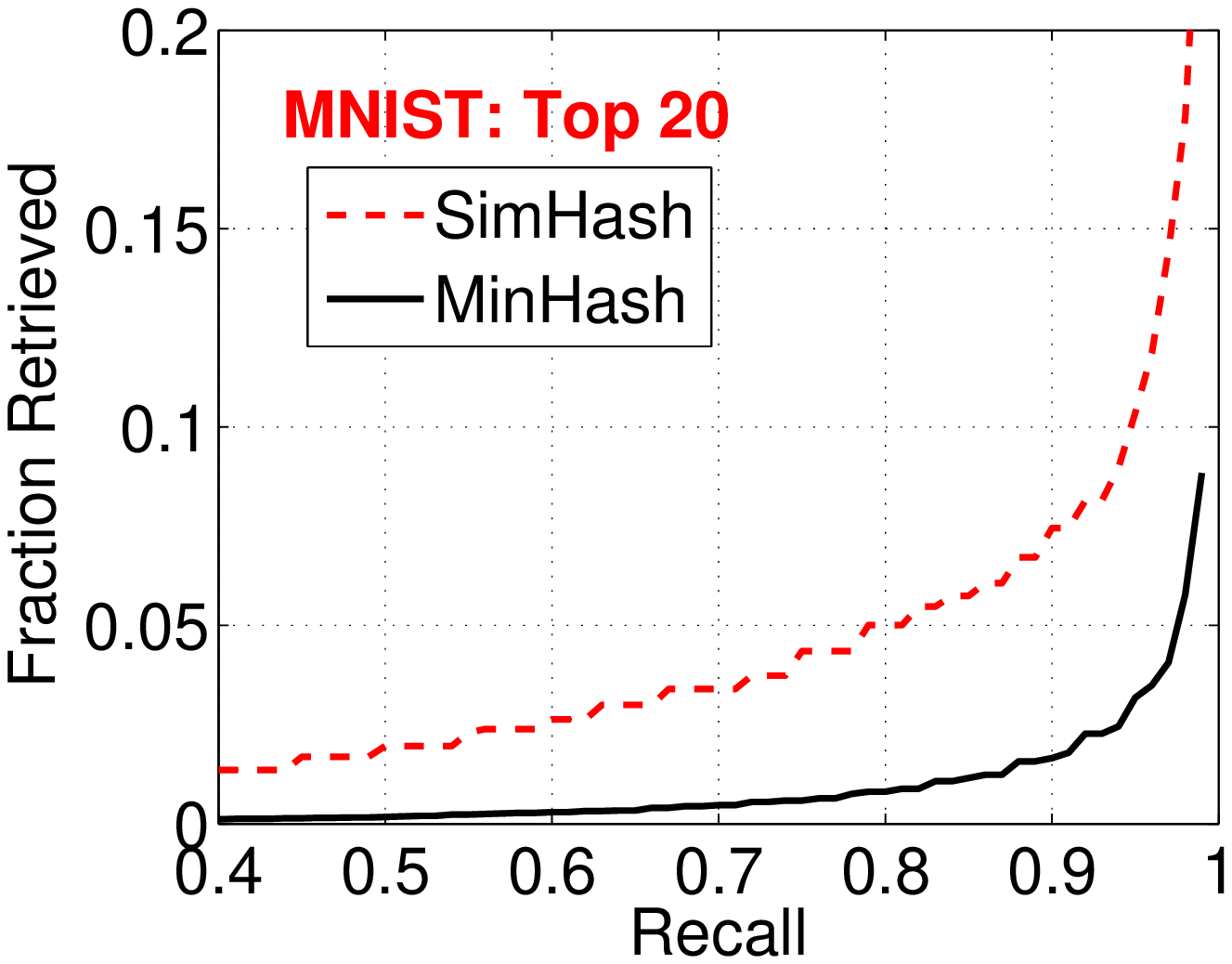}\hspace{-0.13in}
\includegraphics[width=1.7in]{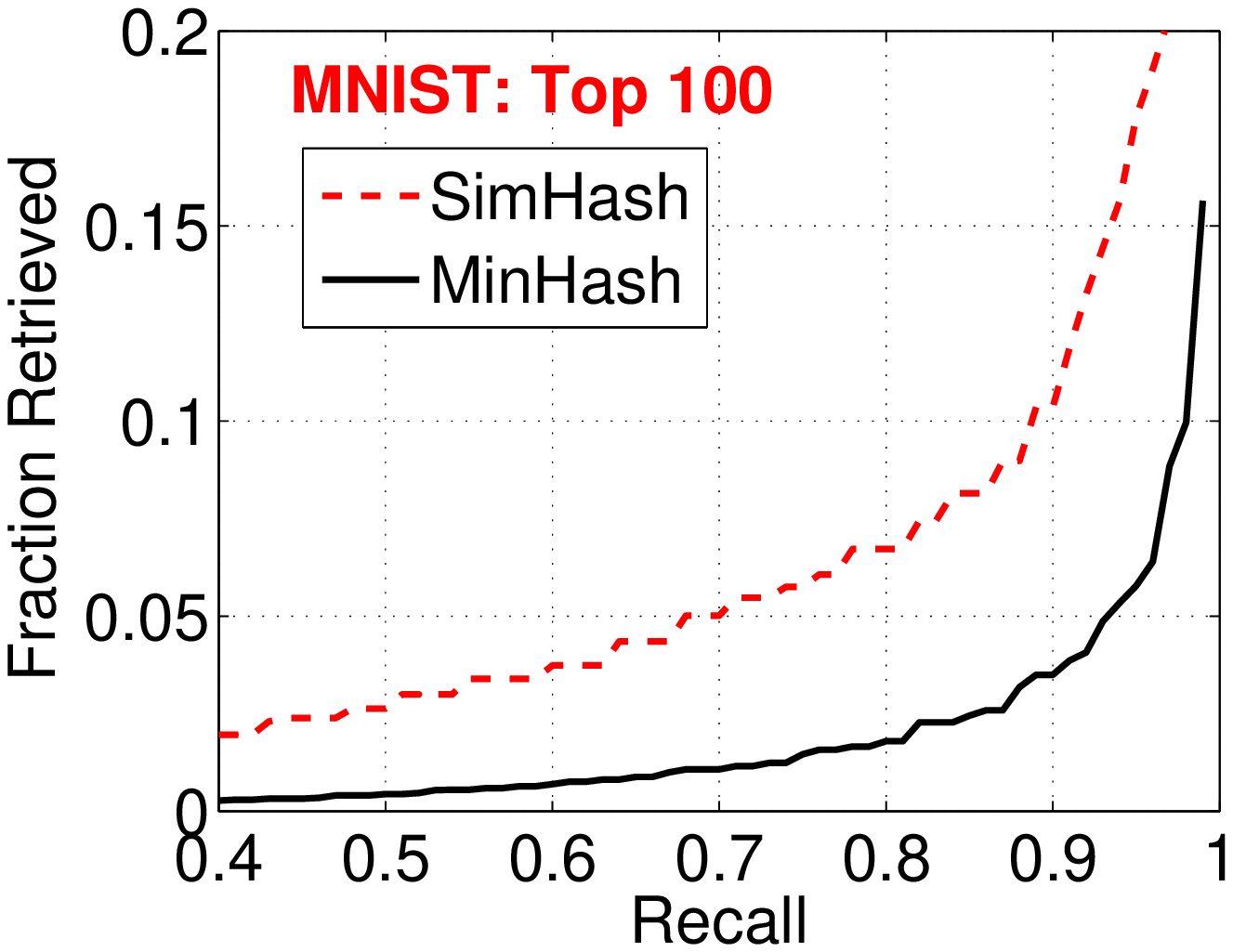}
}

\mbox{
\includegraphics[width=1.7in]{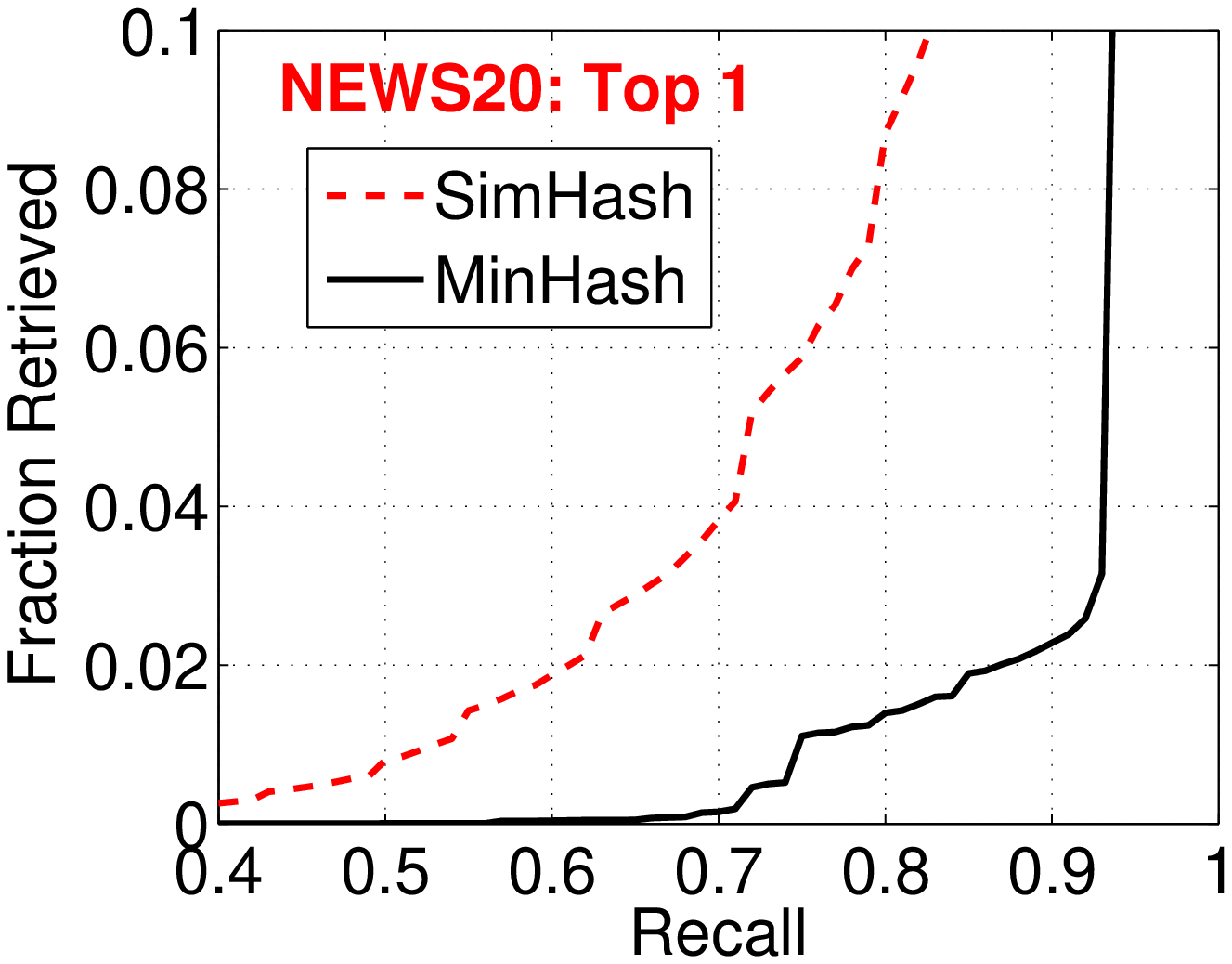}\hspace{-0.13in}
\includegraphics[width=1.7in]{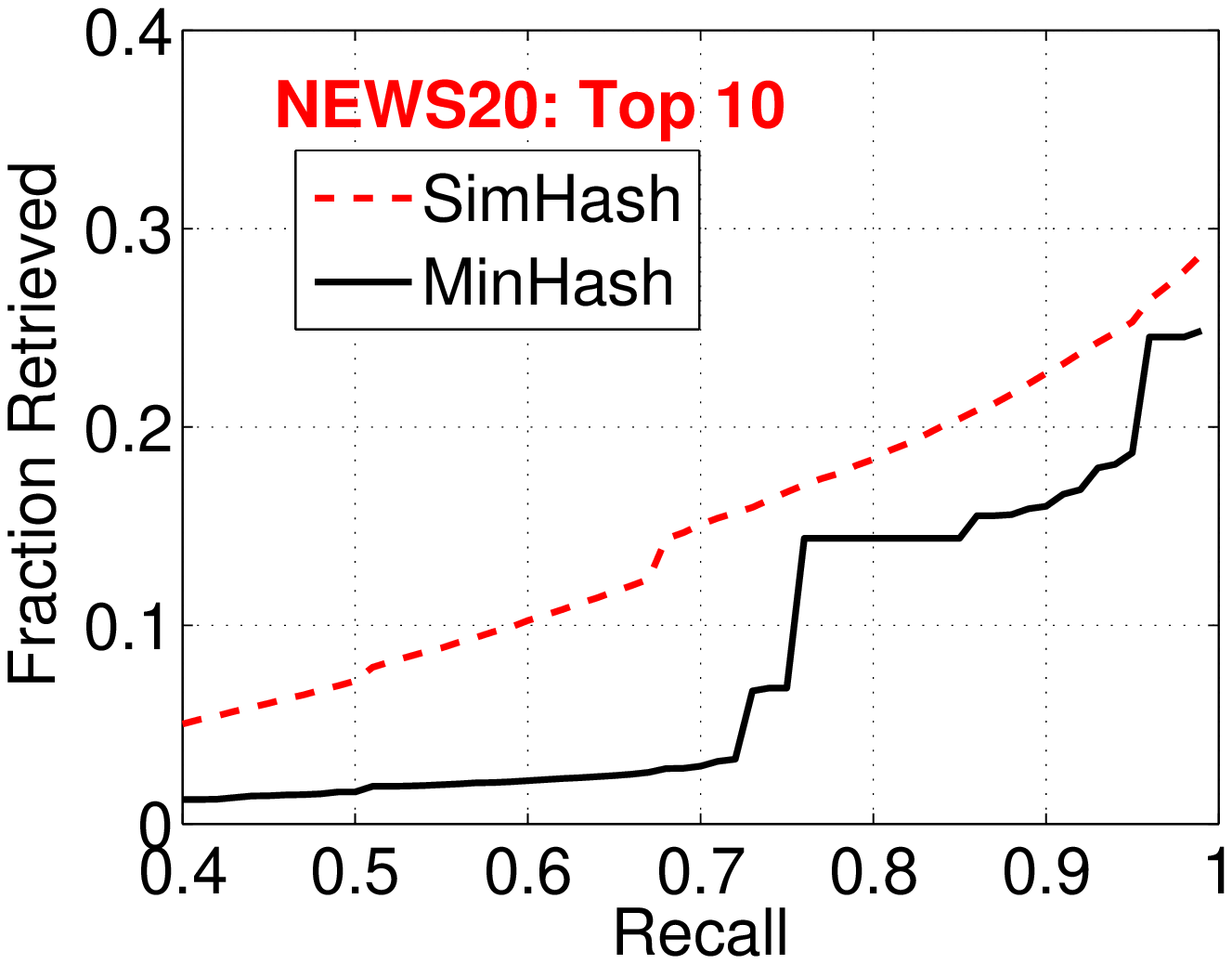}\hspace{-0.13in}
\includegraphics[width=1.7in]{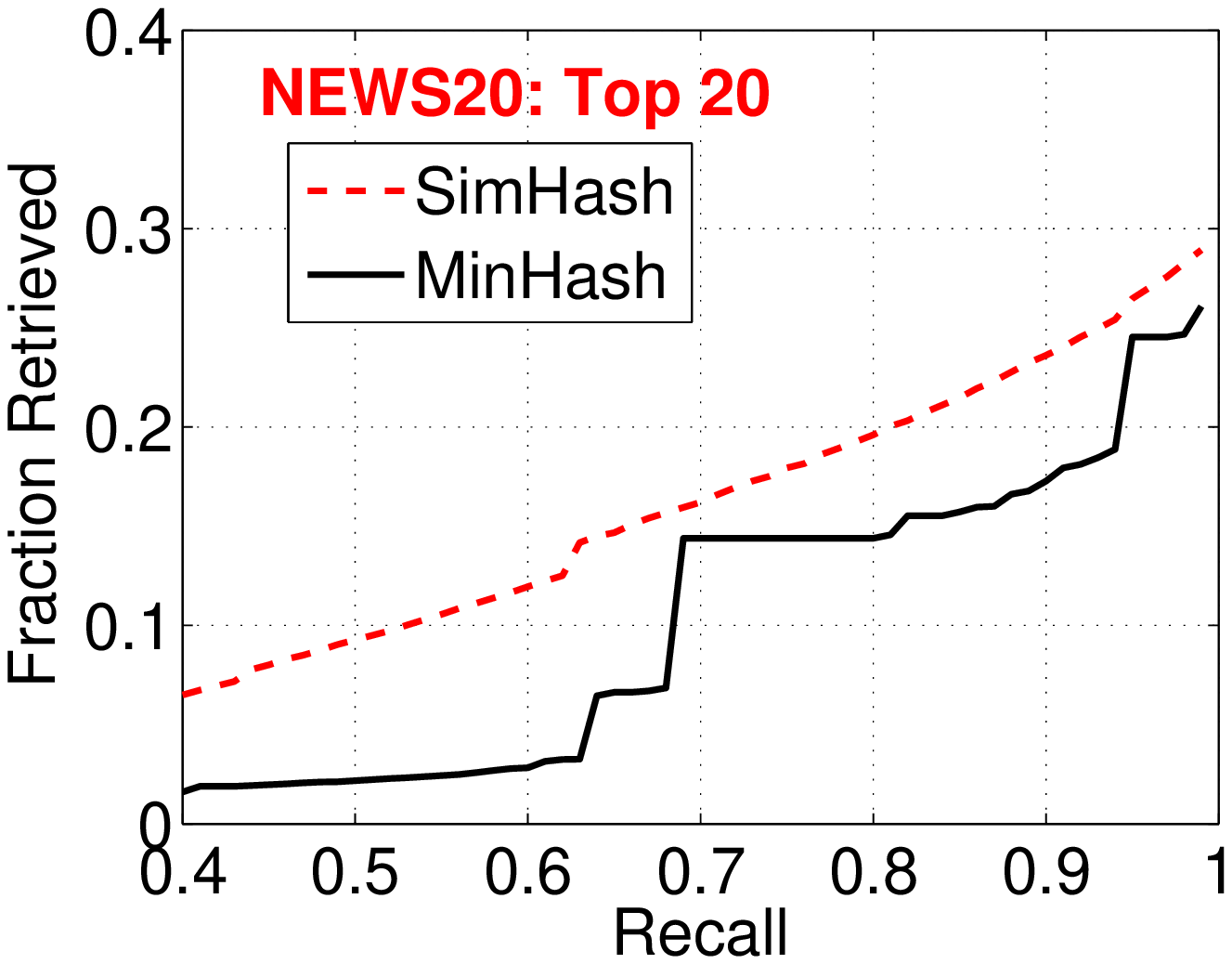}\hspace{-0.13in}
\includegraphics[width=1.7in]{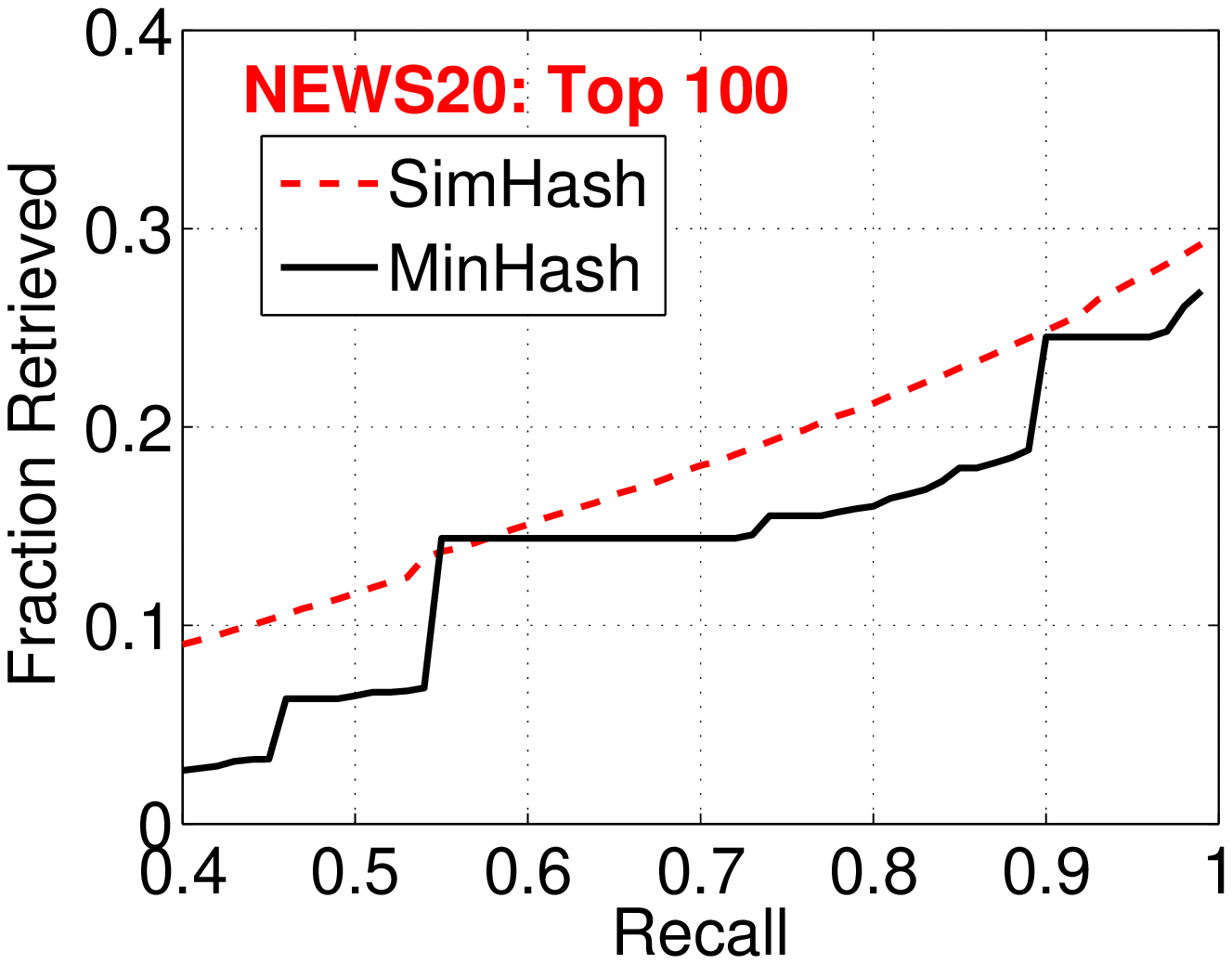}
}

\mbox{
\includegraphics[width=1.7in]{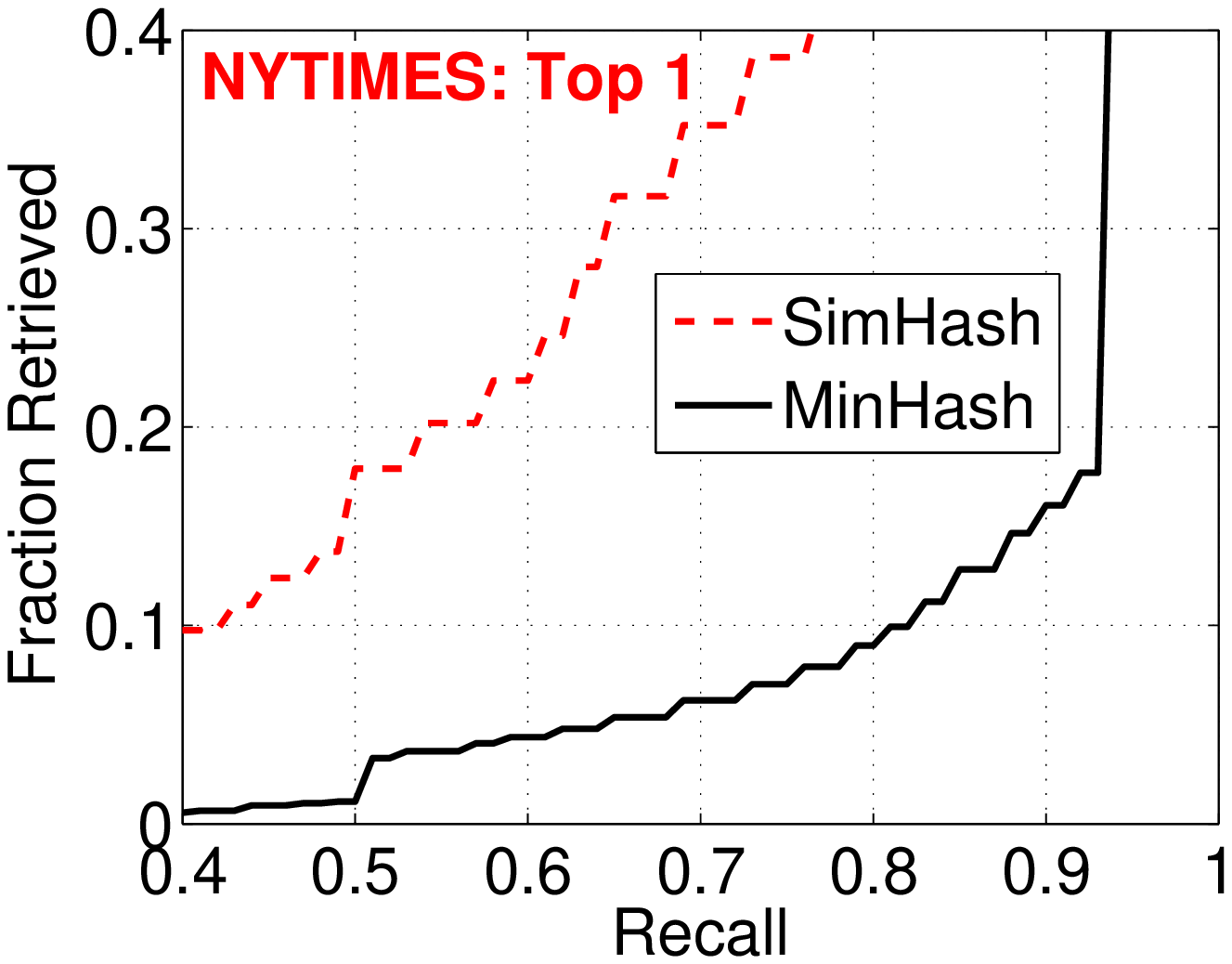}\hspace{-0.13in}
\includegraphics[width=1.7in]{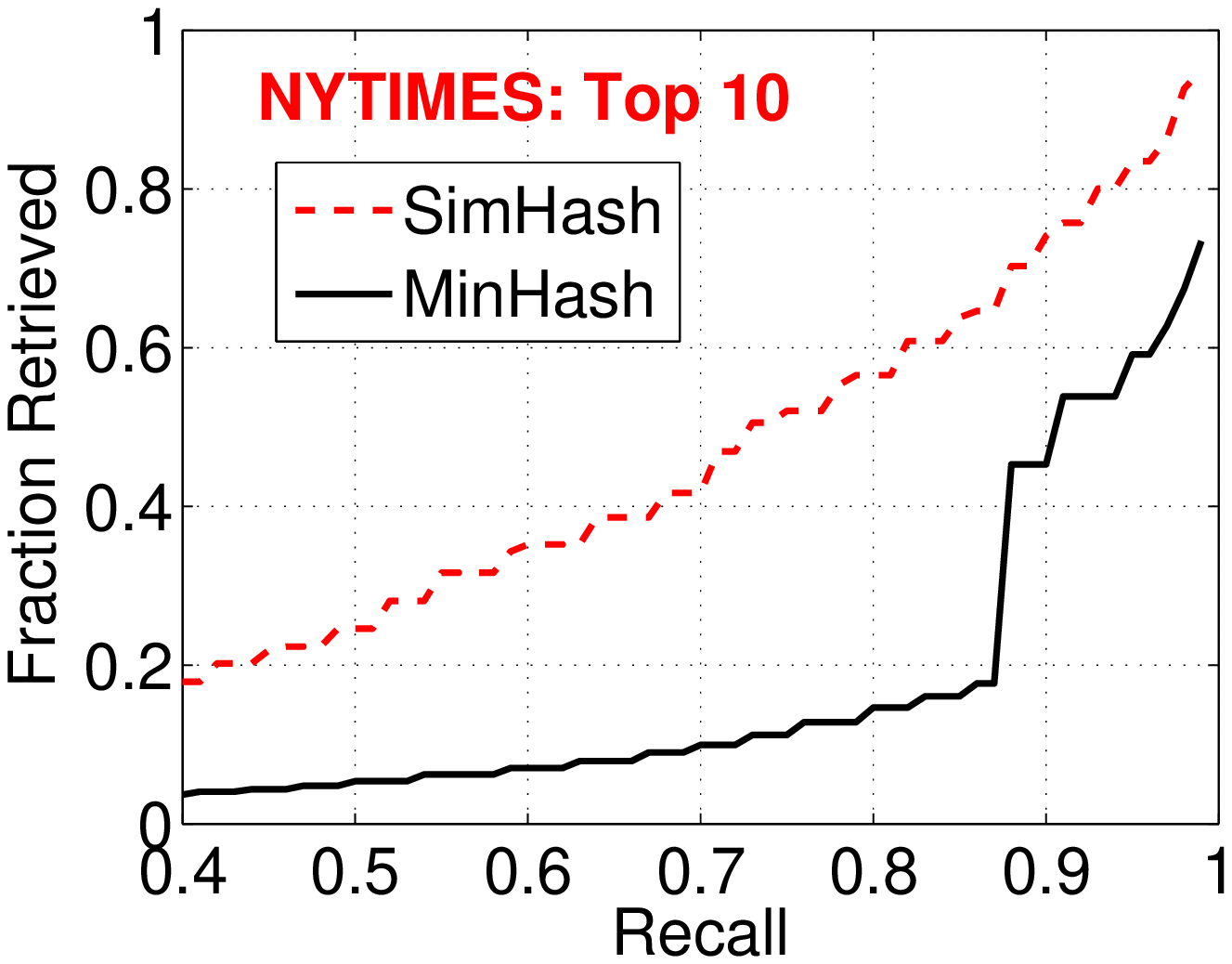}\hspace{-0.13in}
\includegraphics[width=1.7in]{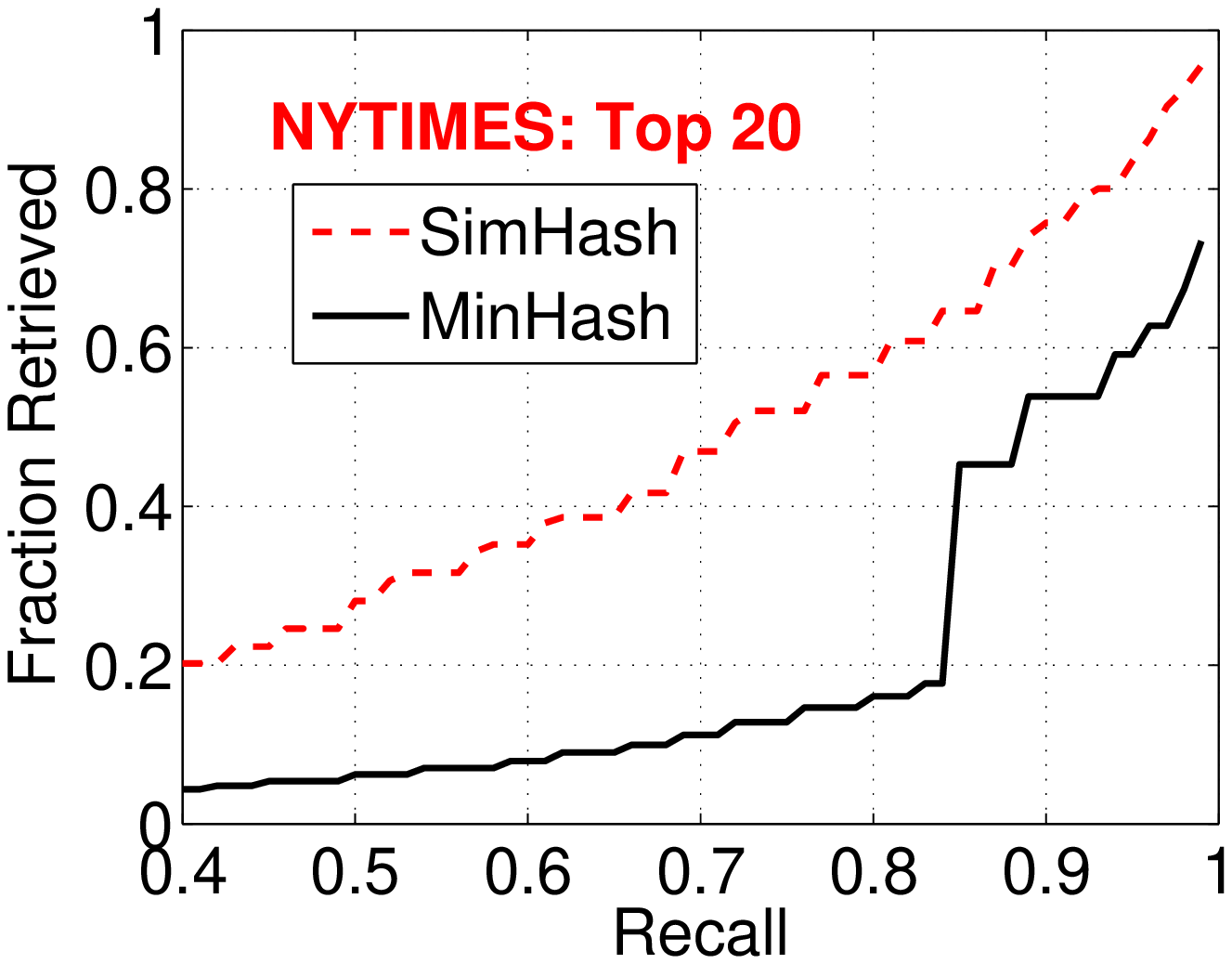}\hspace{-0.13in}
\includegraphics[width=1.7in]{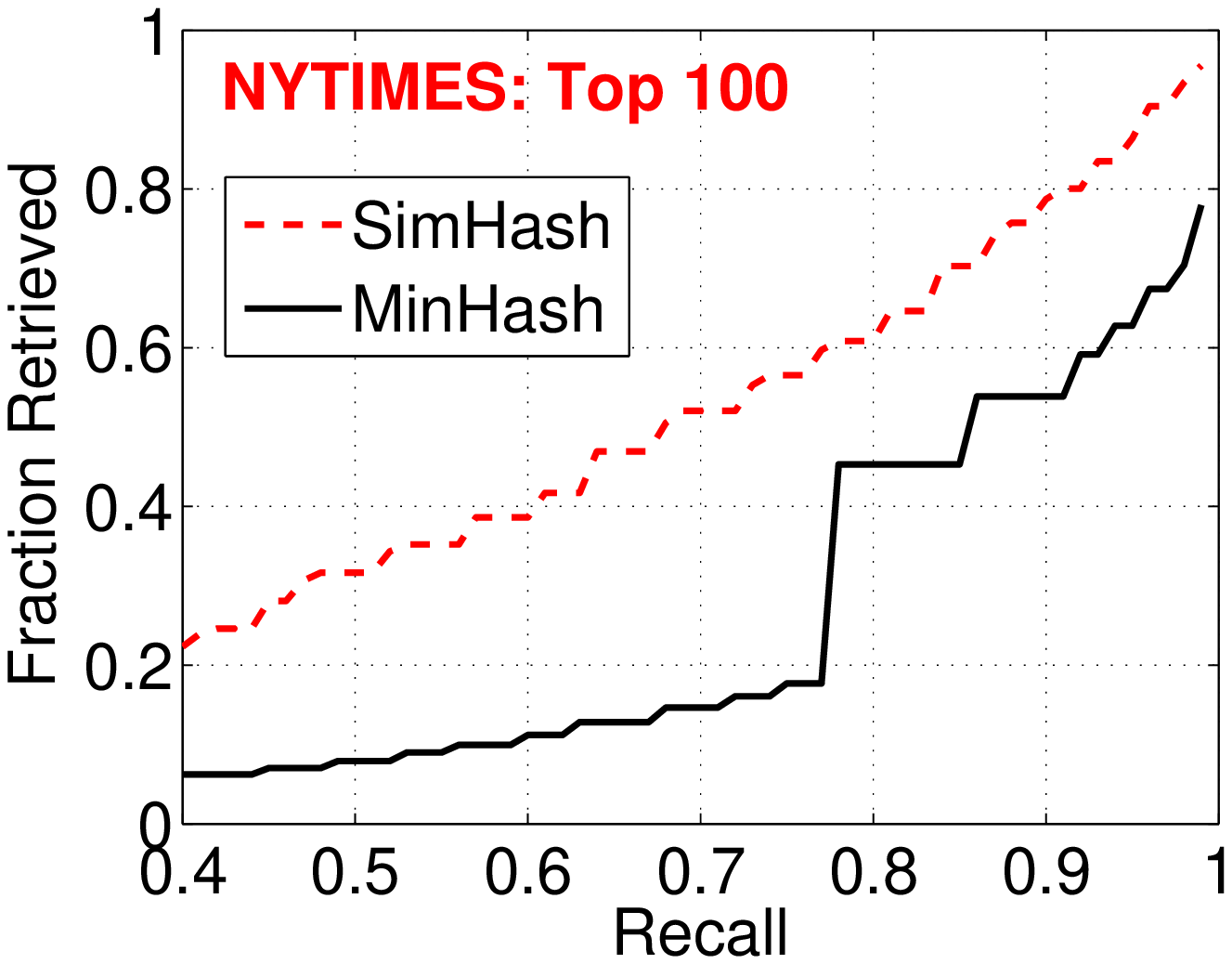}
}
\mbox{
\includegraphics[width=1.7in]{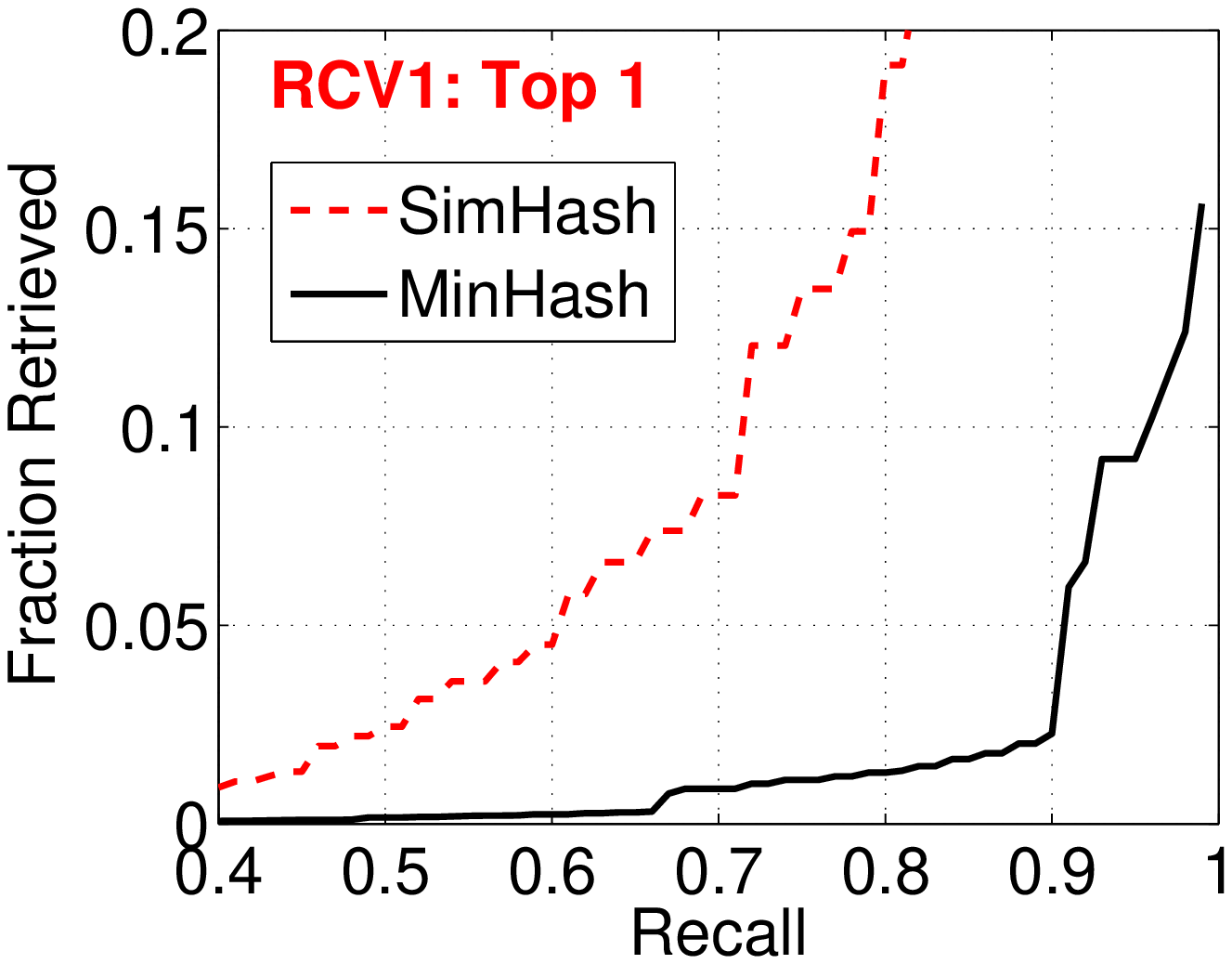}\hspace{-0.13in}
\includegraphics[width=1.7in]{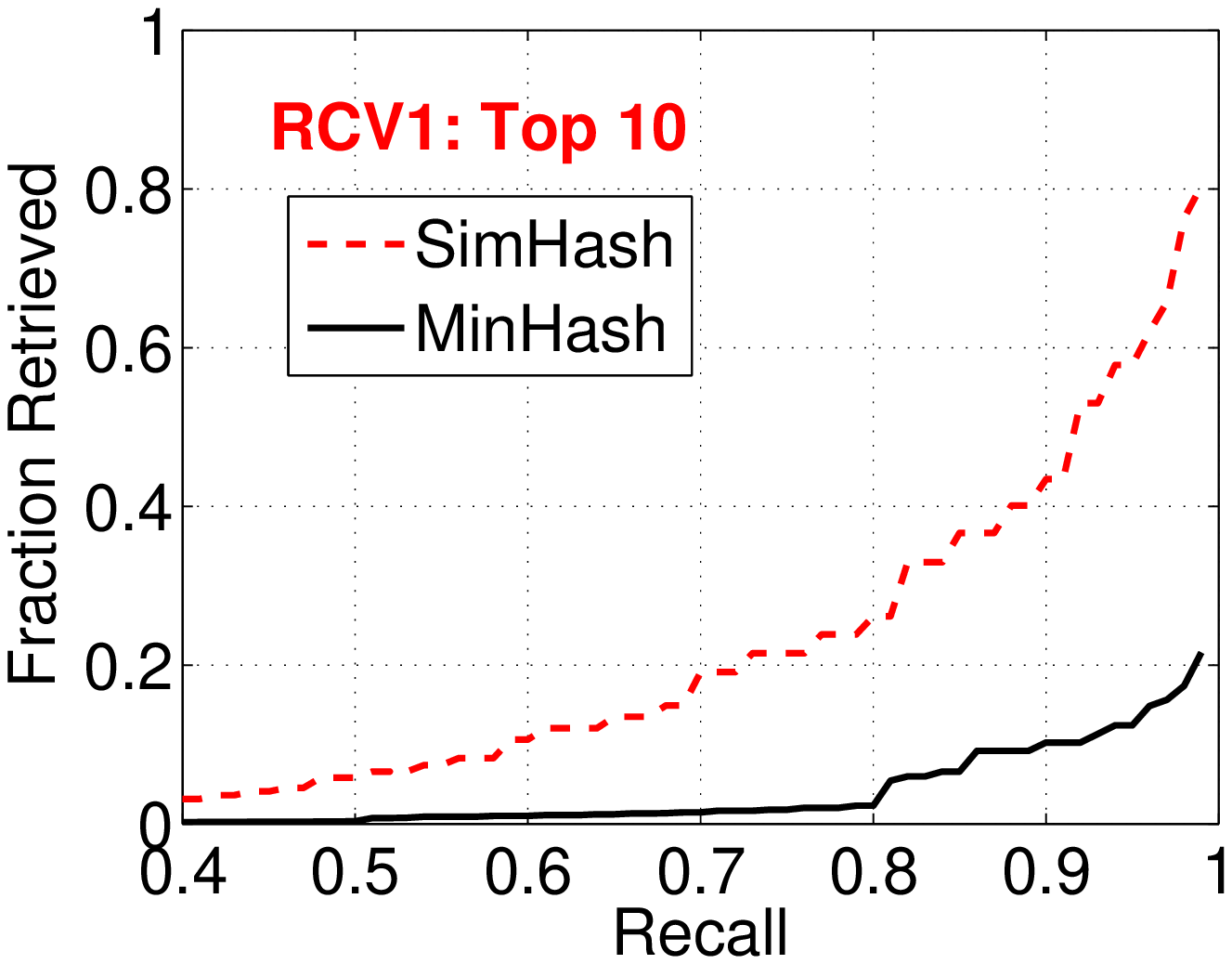}\hspace{-0.13in}
\includegraphics[width=1.7in]{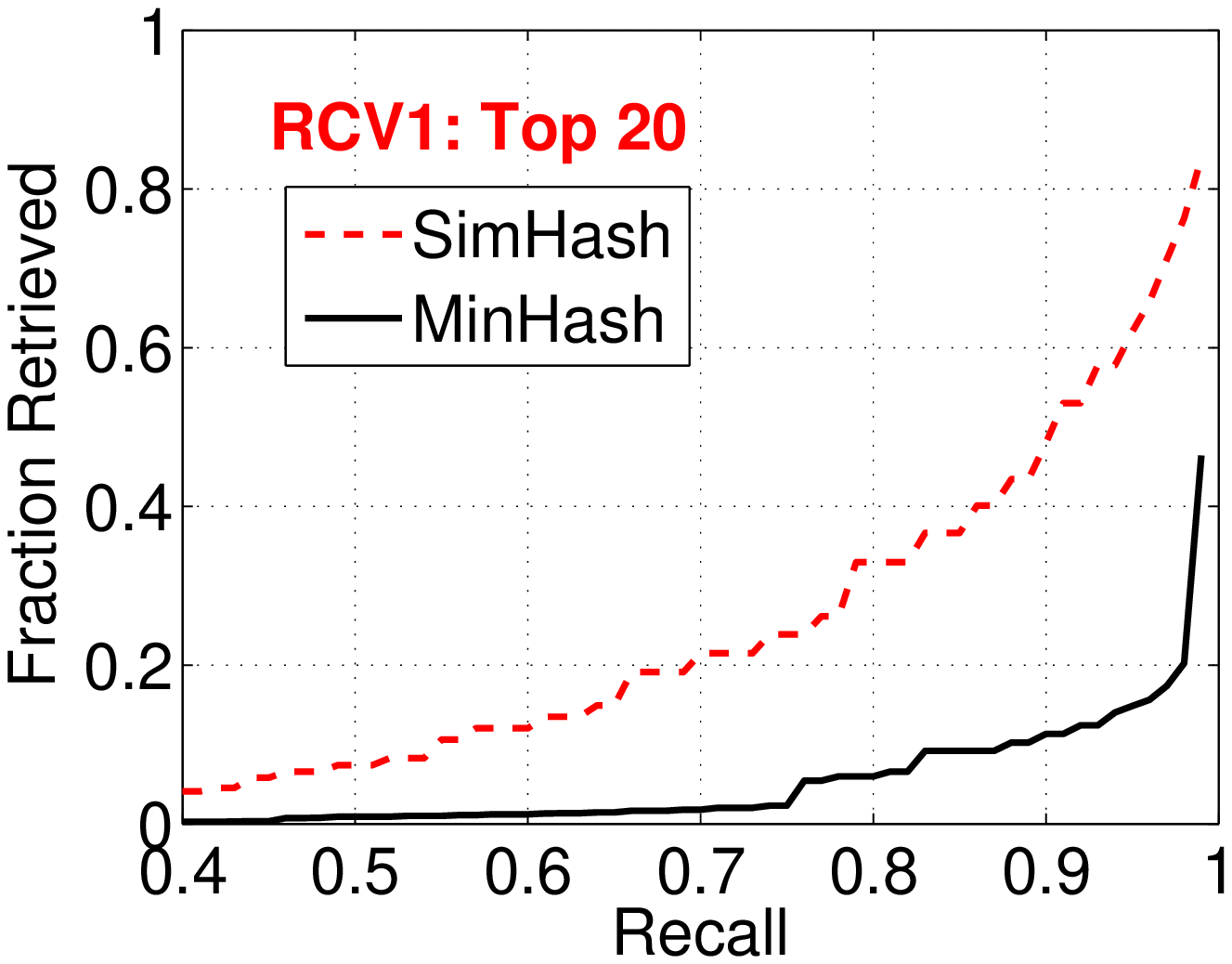}\hspace{-0.13in}
\includegraphics[width=1.7in]{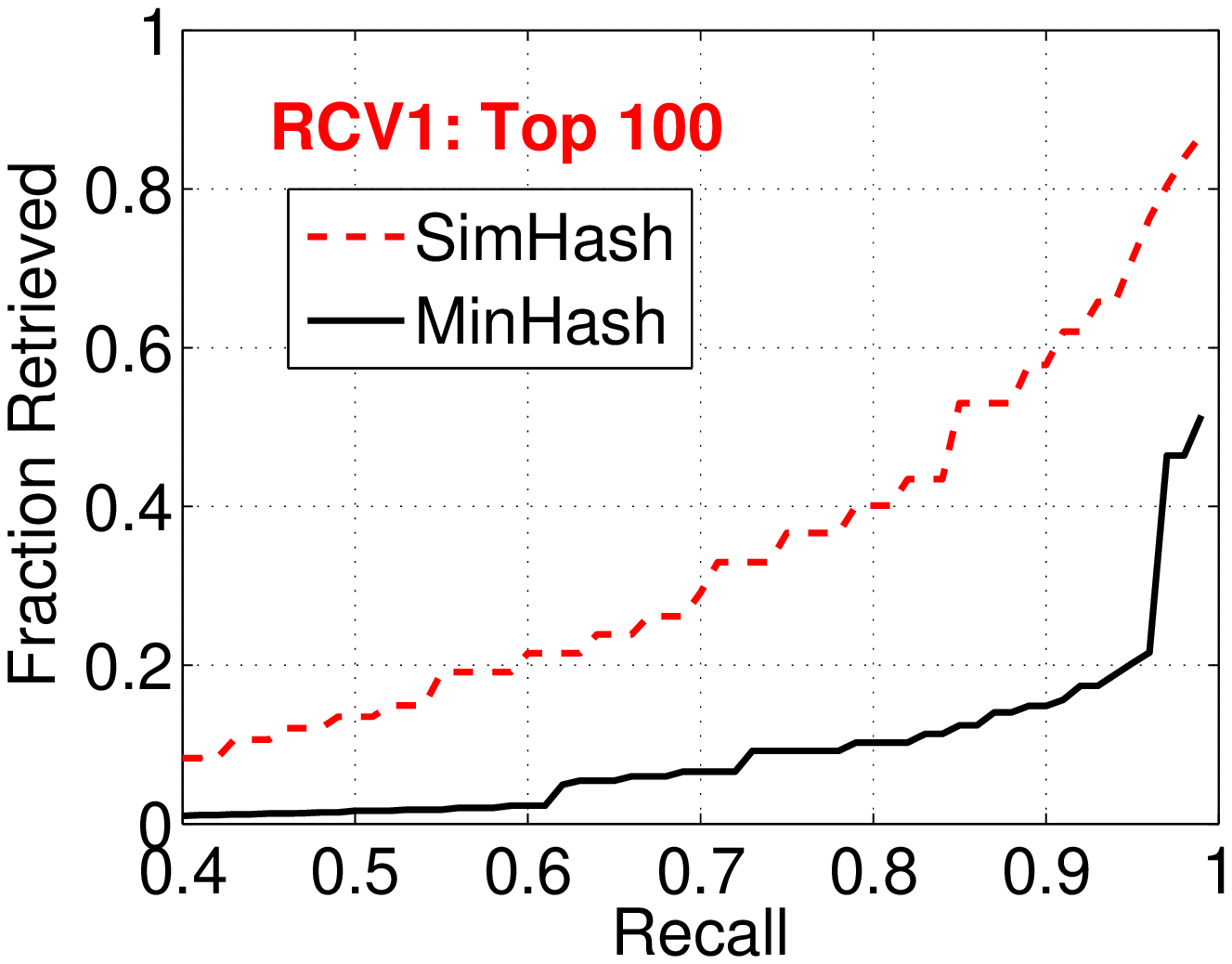}

}
\mbox{
\includegraphics[width=1.7in]{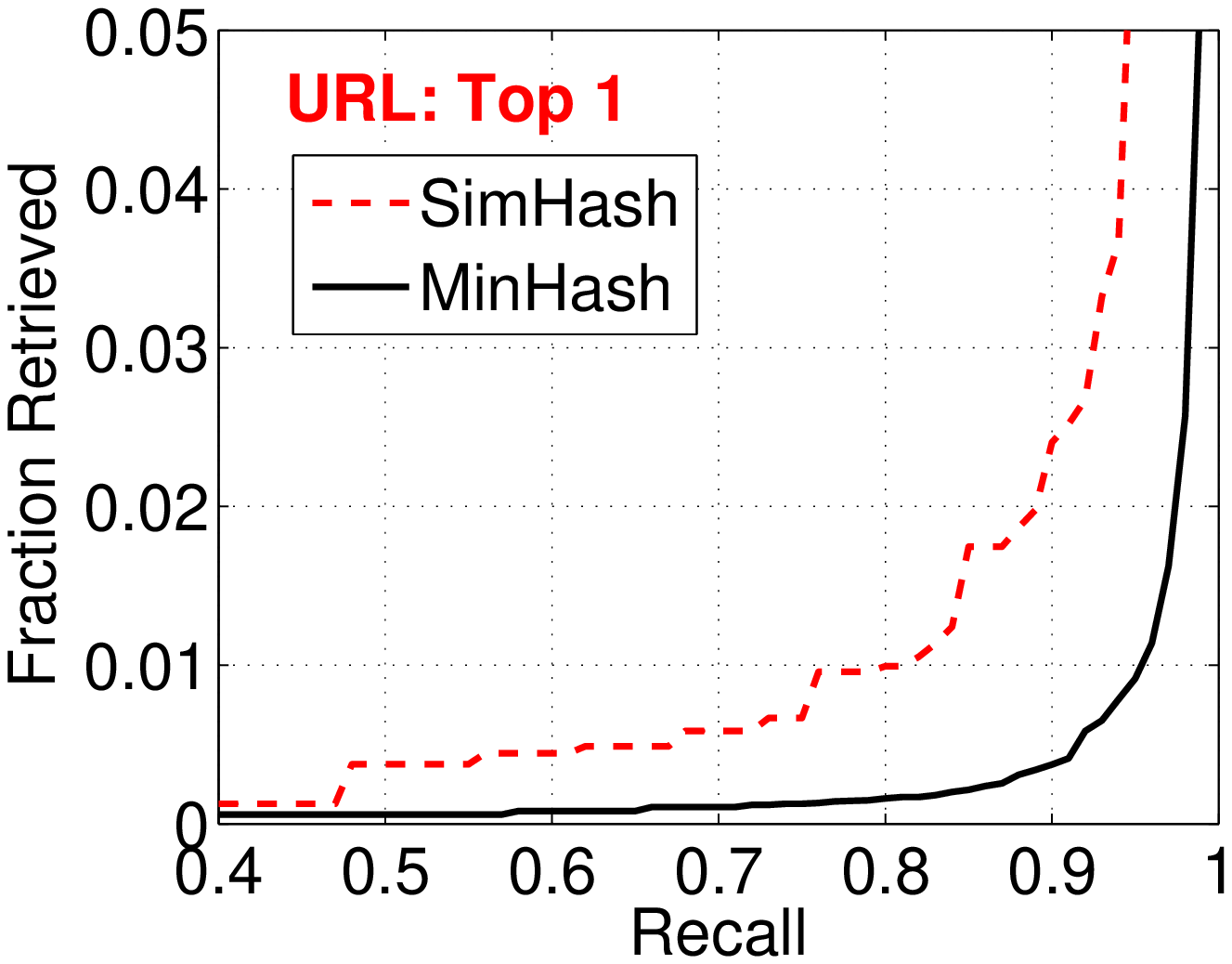}\hspace{-0.13in}
\includegraphics[width=1.7in]{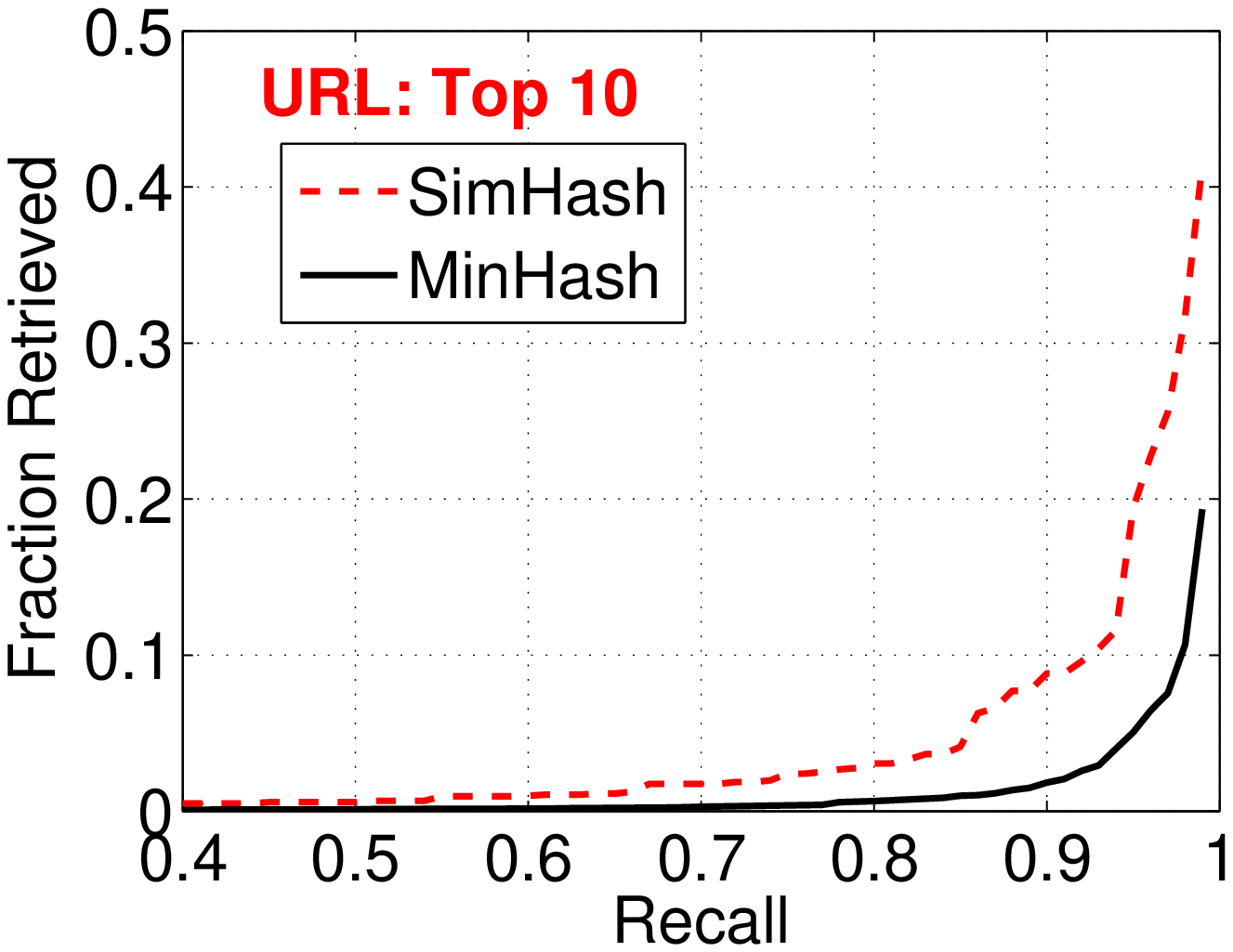}\hspace{-0.13in}
\includegraphics[width=1.7in]{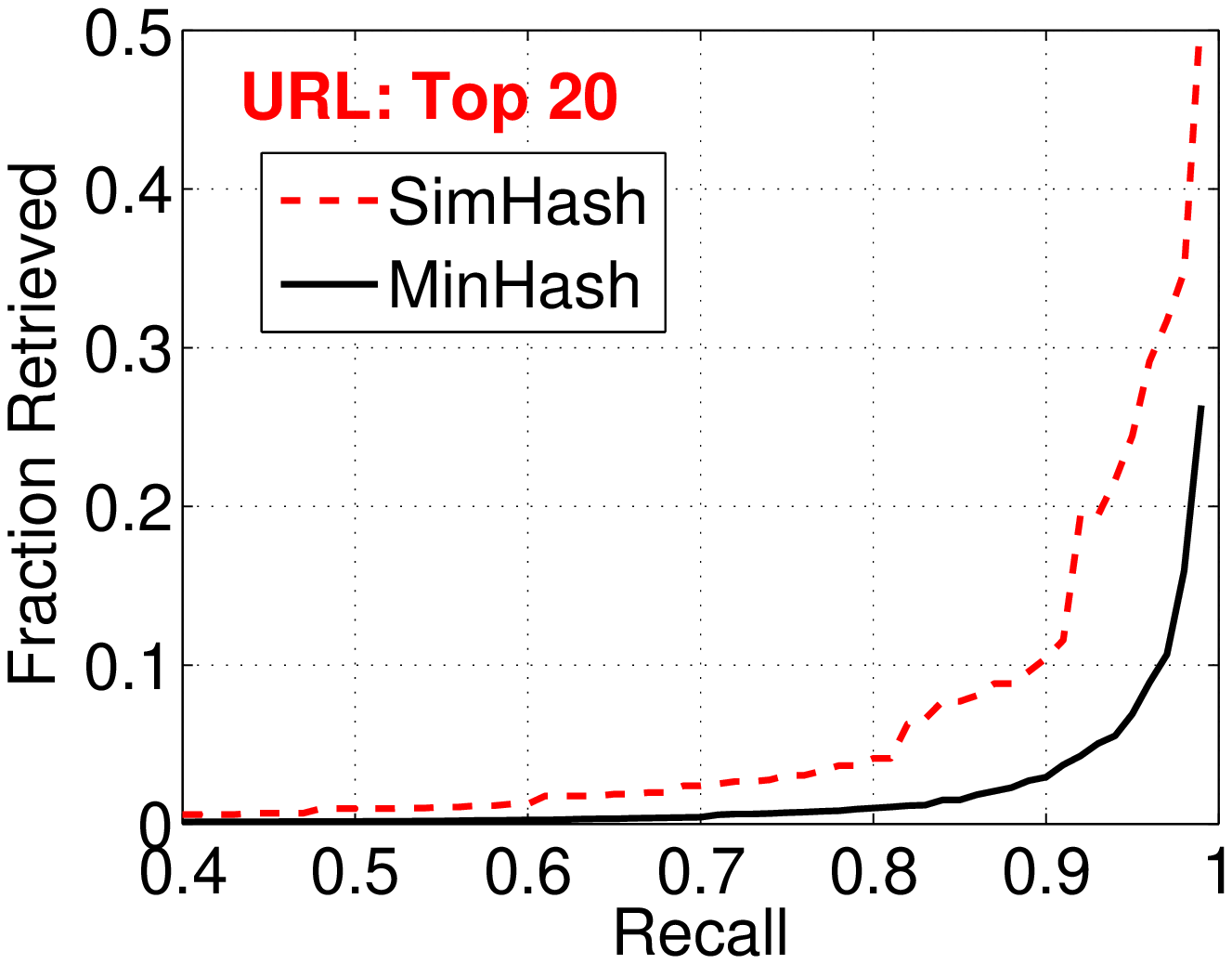}\hspace{-0.13in}
\includegraphics[width=1.7in]{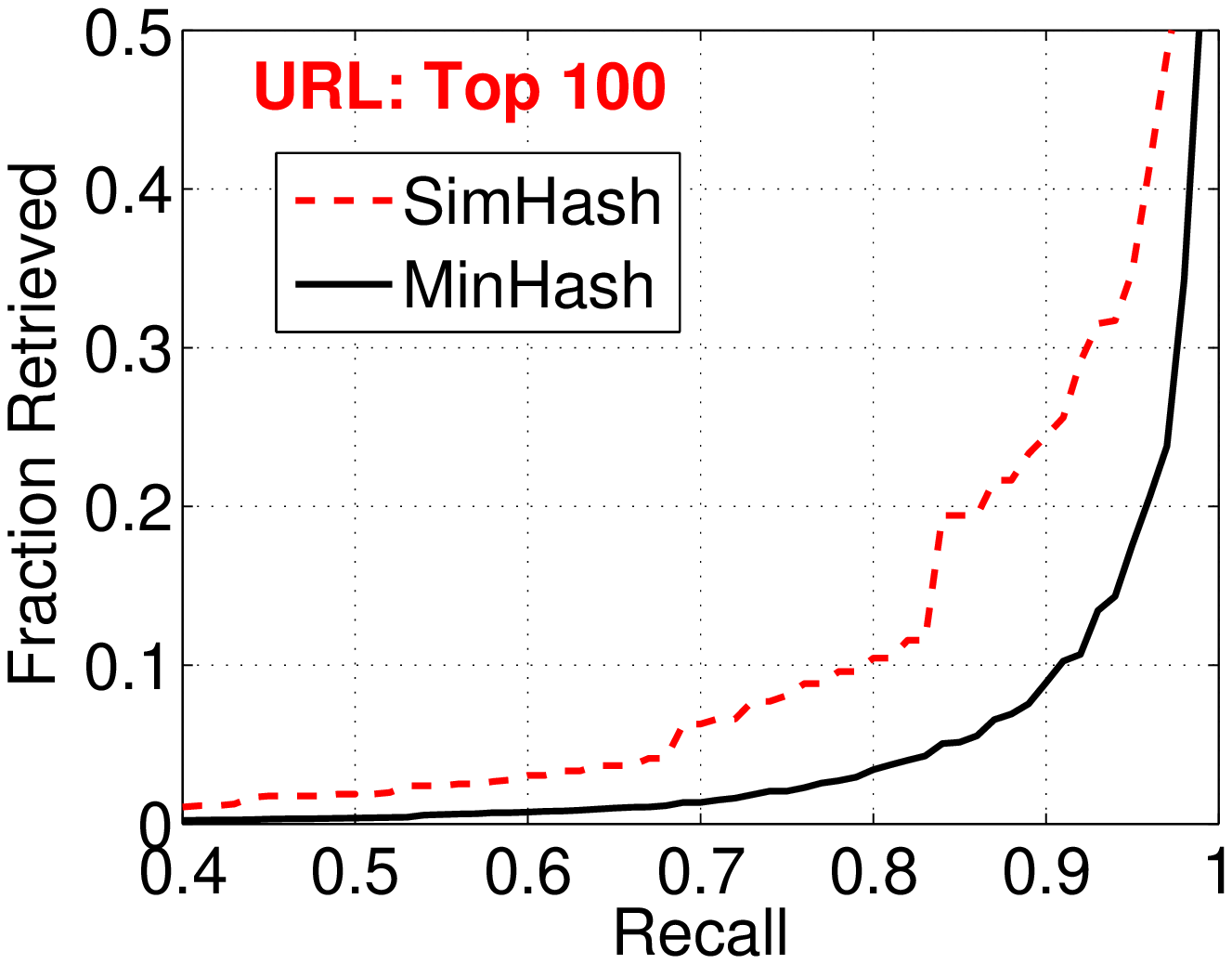}
}

\mbox{
\includegraphics[width=1.7in]{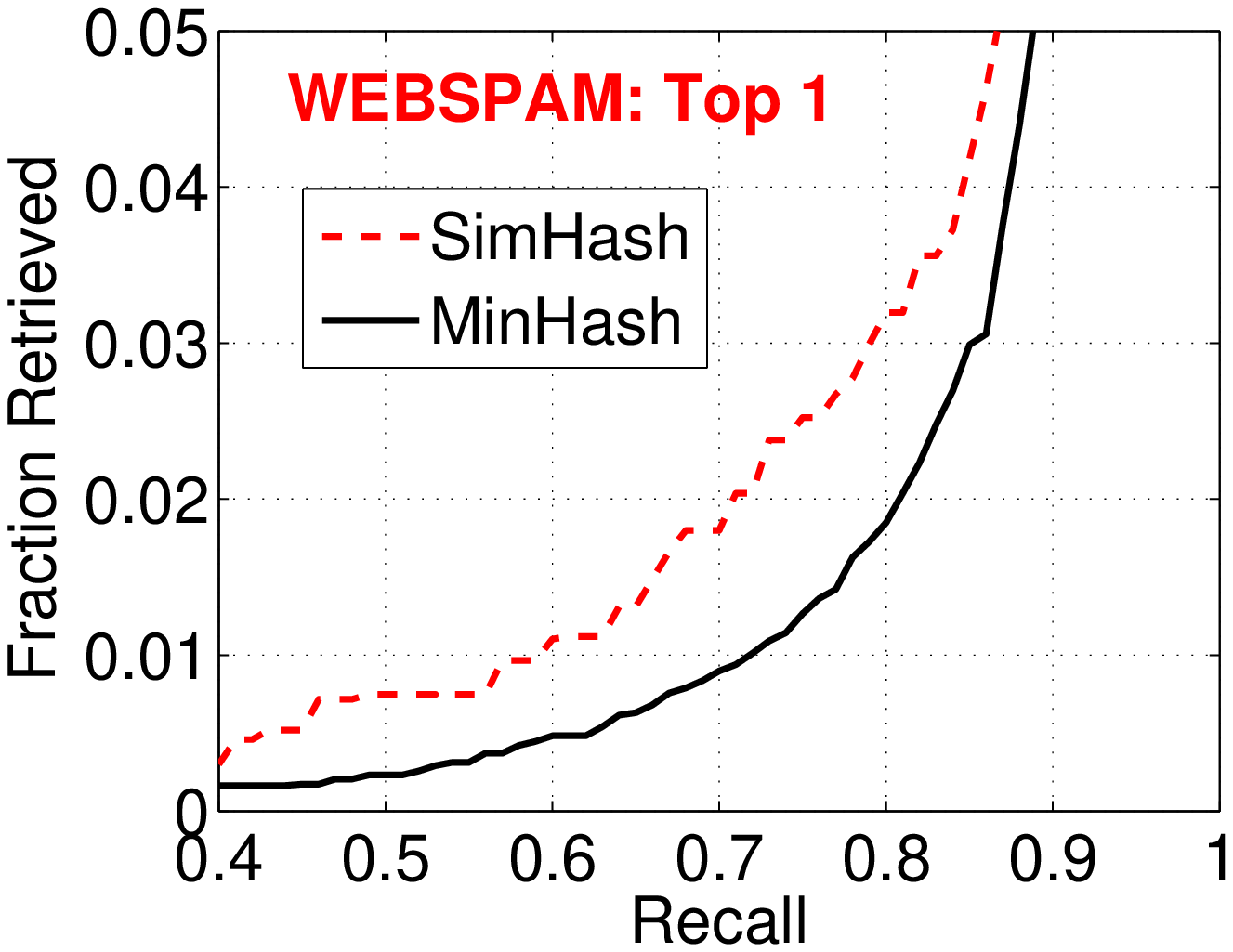}\hspace{-0.13in}
\includegraphics[width=1.7in]{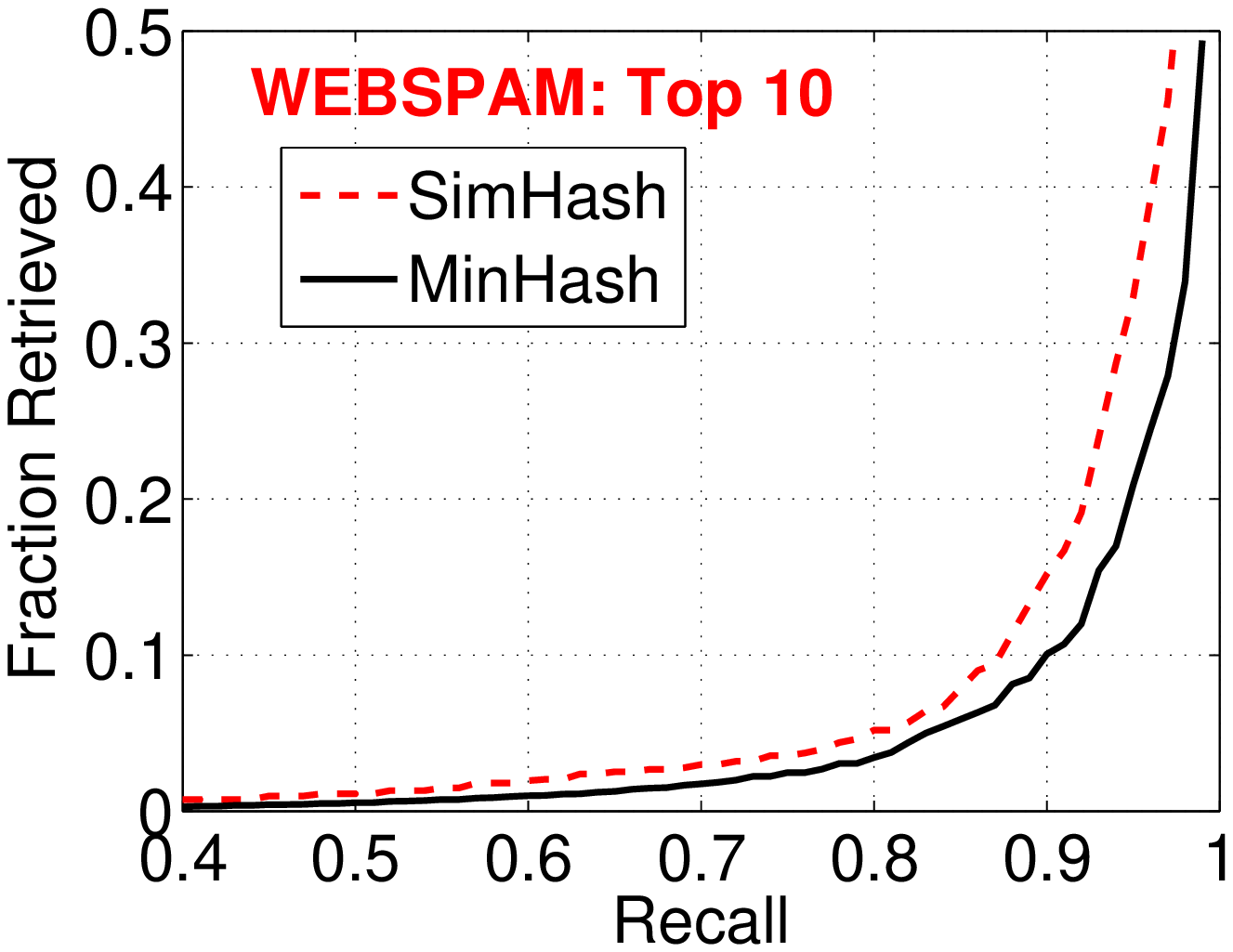}\hspace{-0.13in}
\includegraphics[width=1.7in]{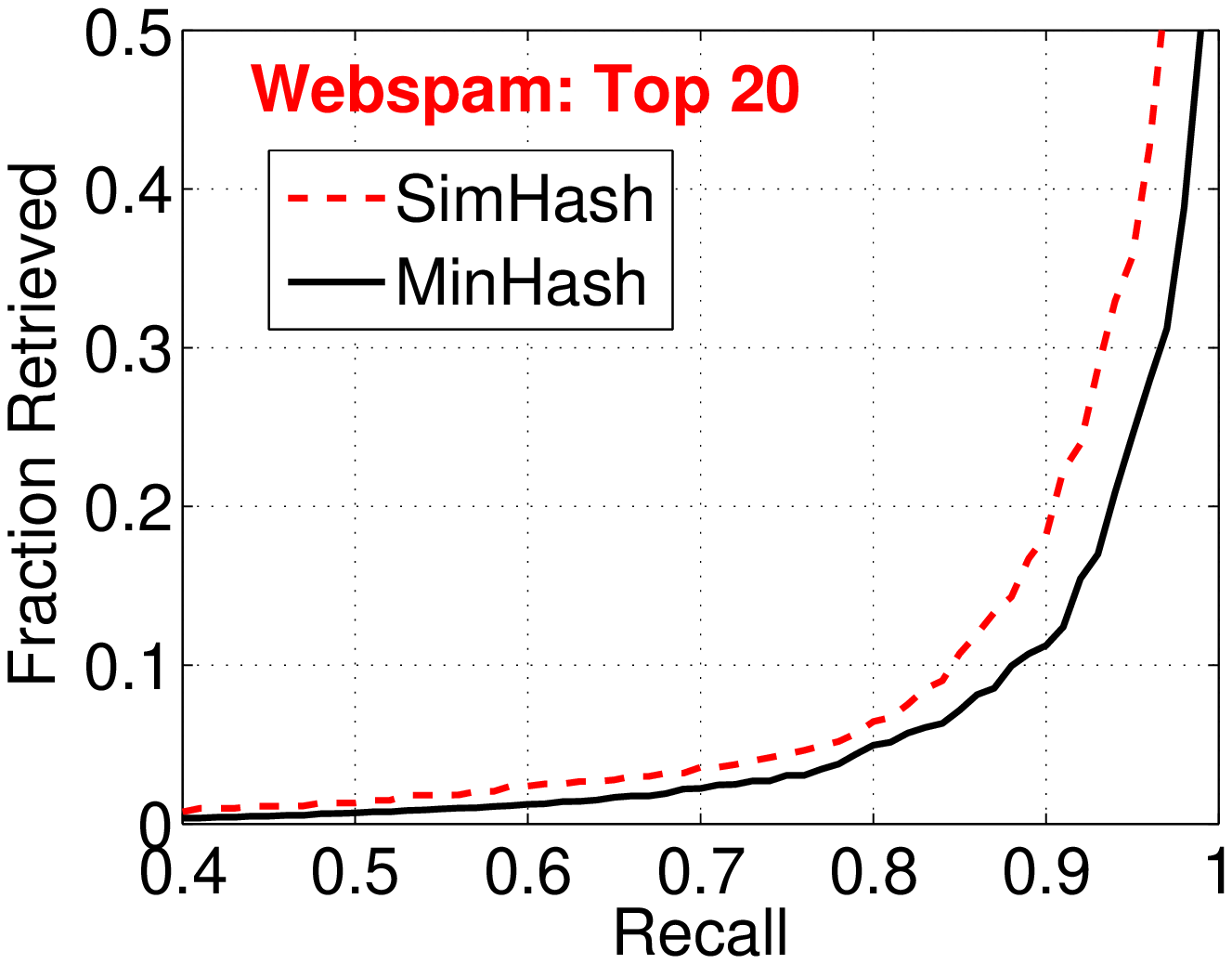}\hspace{-0.13in}
\includegraphics[width=1.7in]{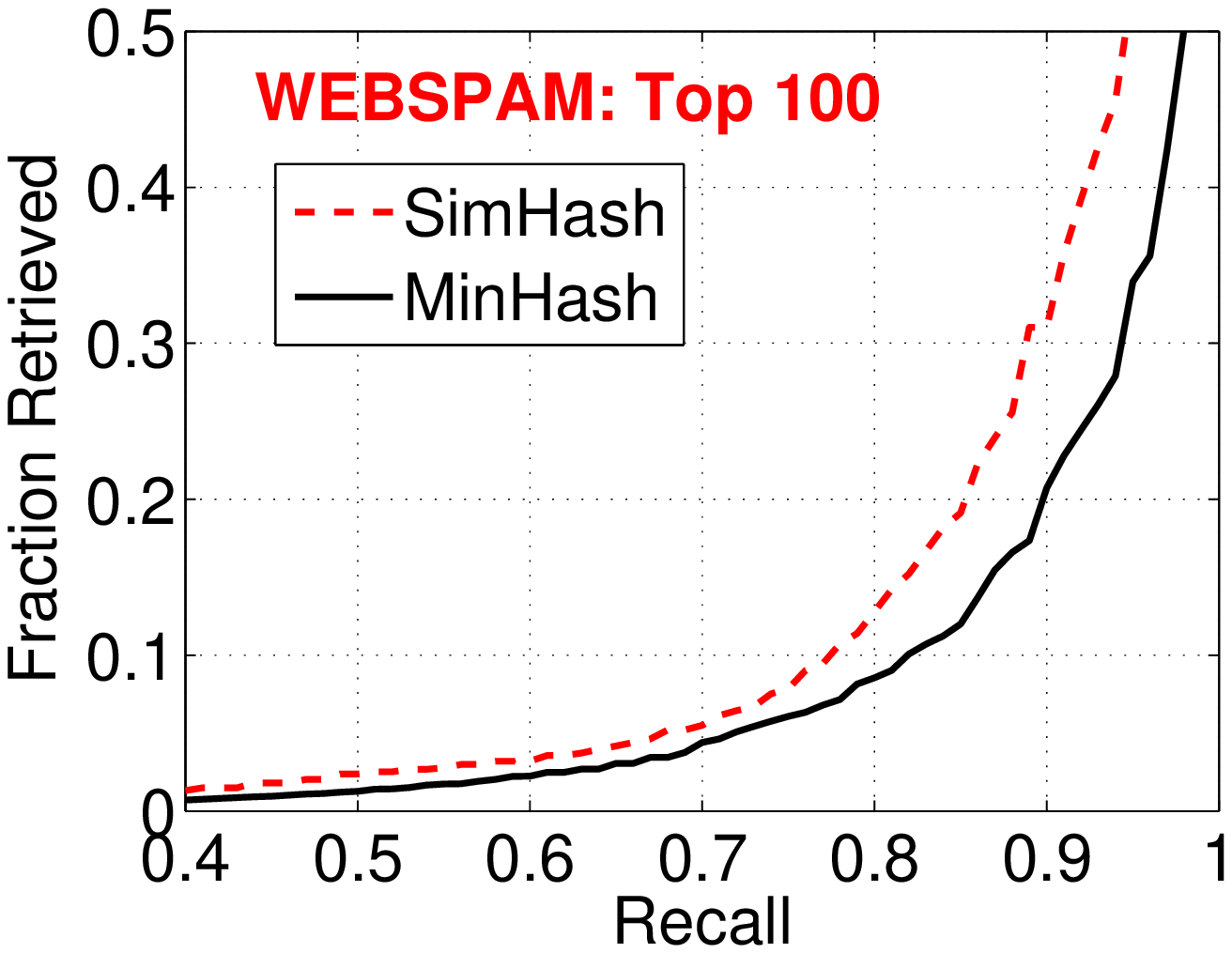}
}

\end{center}
\vspace{-0.2in}
\caption{Fraction of data points retrieved (y-axis) in order to achieve a specified recall (x-axis), for comparing SimHash with MinHash. Lower is better. We use top-$k$ (cosine similarities) as the gold standard for $k=1$, 10, 20, 100. For all 6 binarized  datasets, MinHash significantly outperforms SimHash. For example, to achieve a $90\%$ recall for top-1 on MNIST, MinHash needs to scan, on average, $0.6\%$ of the data points while SimHash has to scan $5\%$. For fair comparisons, we present the optimum outcomes (i.e., smallest fraction of data points) separately for MinHash and SimHash, by searching a wide range of parameters $(K,L)$, where $K\in\{1,2,...,30\}$ is the number of hash functions per table and $L\in\{1,2,...,200\}$ is the number of tables.}\label{fig_Topk}
\end{figure*}

For each combination of $(K,L)$ and for both of the hash functions, we computed the mean recall of the top-$k$ gold standard neighbors along with the average number of points reported per query.
We then compute the least number of points needed, by each of the two hash functions, to achieve a given percentage of recall of the gold standard top-$k$, where the least was computed over the choices of $K$ and $L$. We are therefore ensuring the best over all the choices of $K$ and $L$ for each hash function independently. This eliminates the effect of $K$ and $L$, if any, in the evaluations. The plots of the fraction of points retrieved  at different recall levels, for $k={1,\ 10,\ 20,\ 100}$, are  in Figure~\ref{fig_Topk}.

A good hash function, at a given recall should retrieve less number of points. MinHash needs to evaluate significantly less fraction of the total data points to achieve a given recall compared to SimHash. MinHash is  consistently better than SimHash, in most cases very significantly, irrespective of the choices of dataset and $k$. It should be noted that our gold standard measure for computing top-$k$ neighbors is cosine similarity. This should favor SimHash because it was the only known LSH for cosine similarity. Despite this ``disadvantage'', MinHash still outperforms SimHash in top near neighbor search with cosine similarity. This nicely confirms our theoretical gap analysis.

\clearpage\newpage

\begin{figure}[h!]
\begin{center}

\mbox{
\includegraphics[width=1.7in]{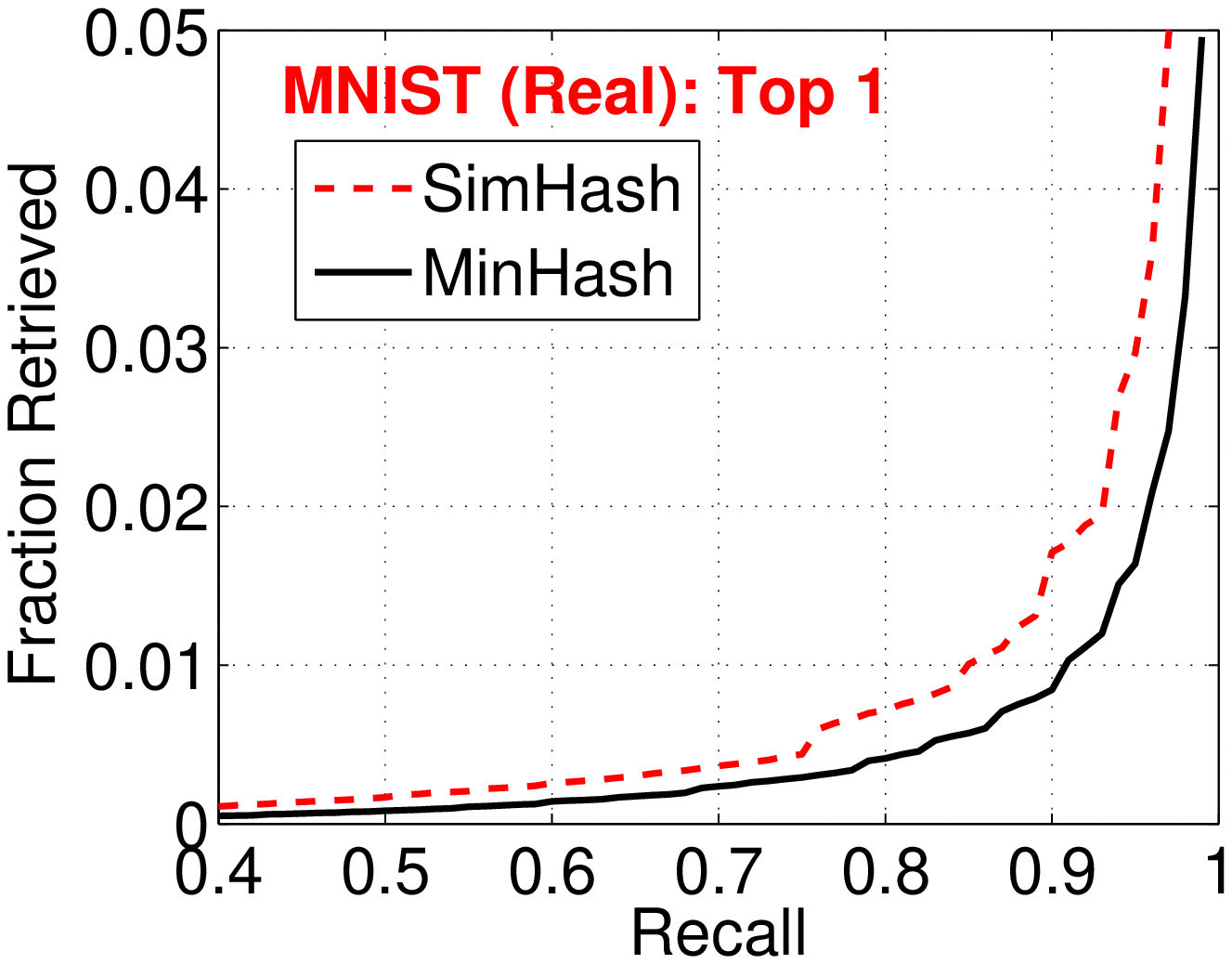}\hspace{-0.13in}
\includegraphics[width=1.7in]{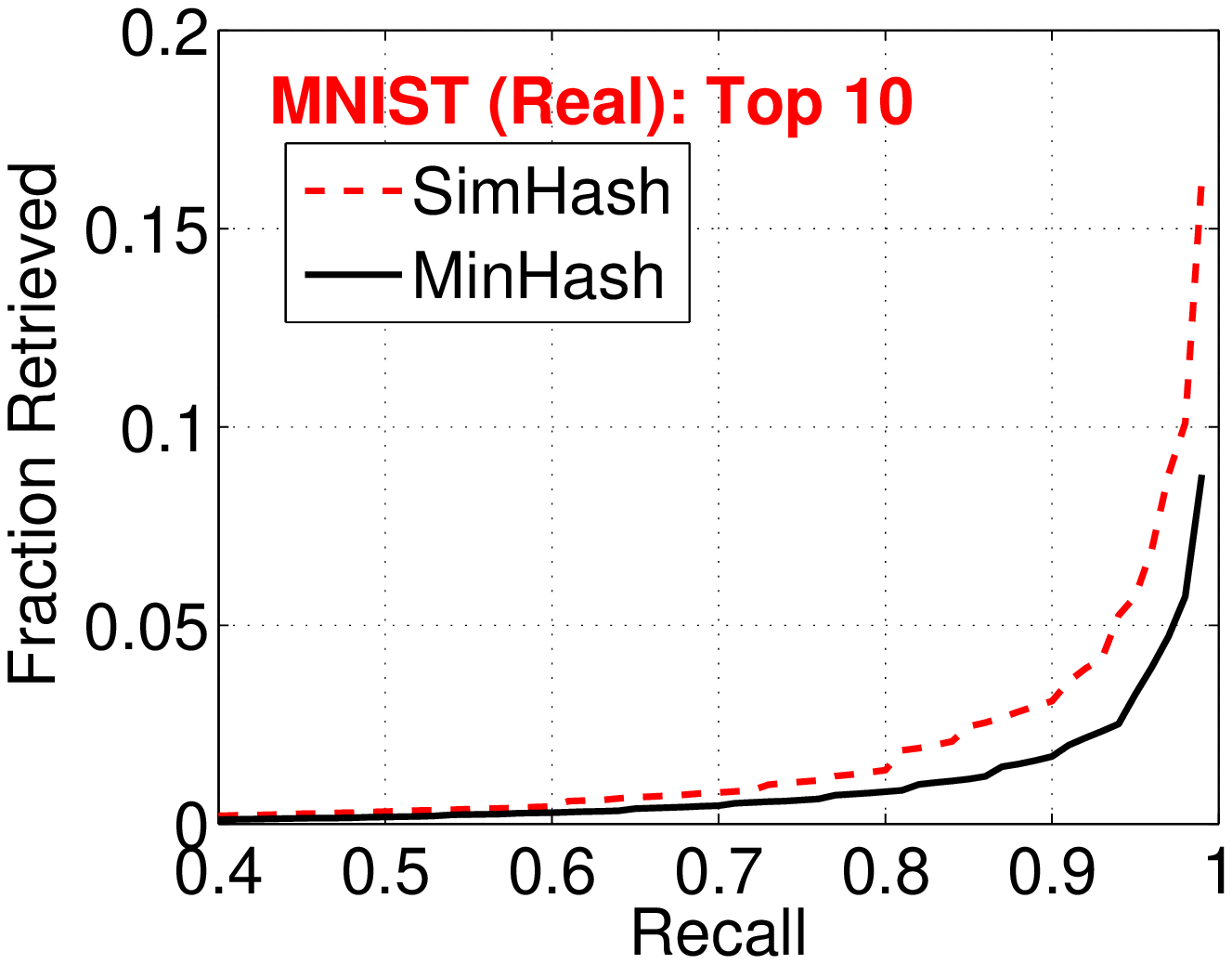}
}

\mbox{
\includegraphics[width=1.7in]{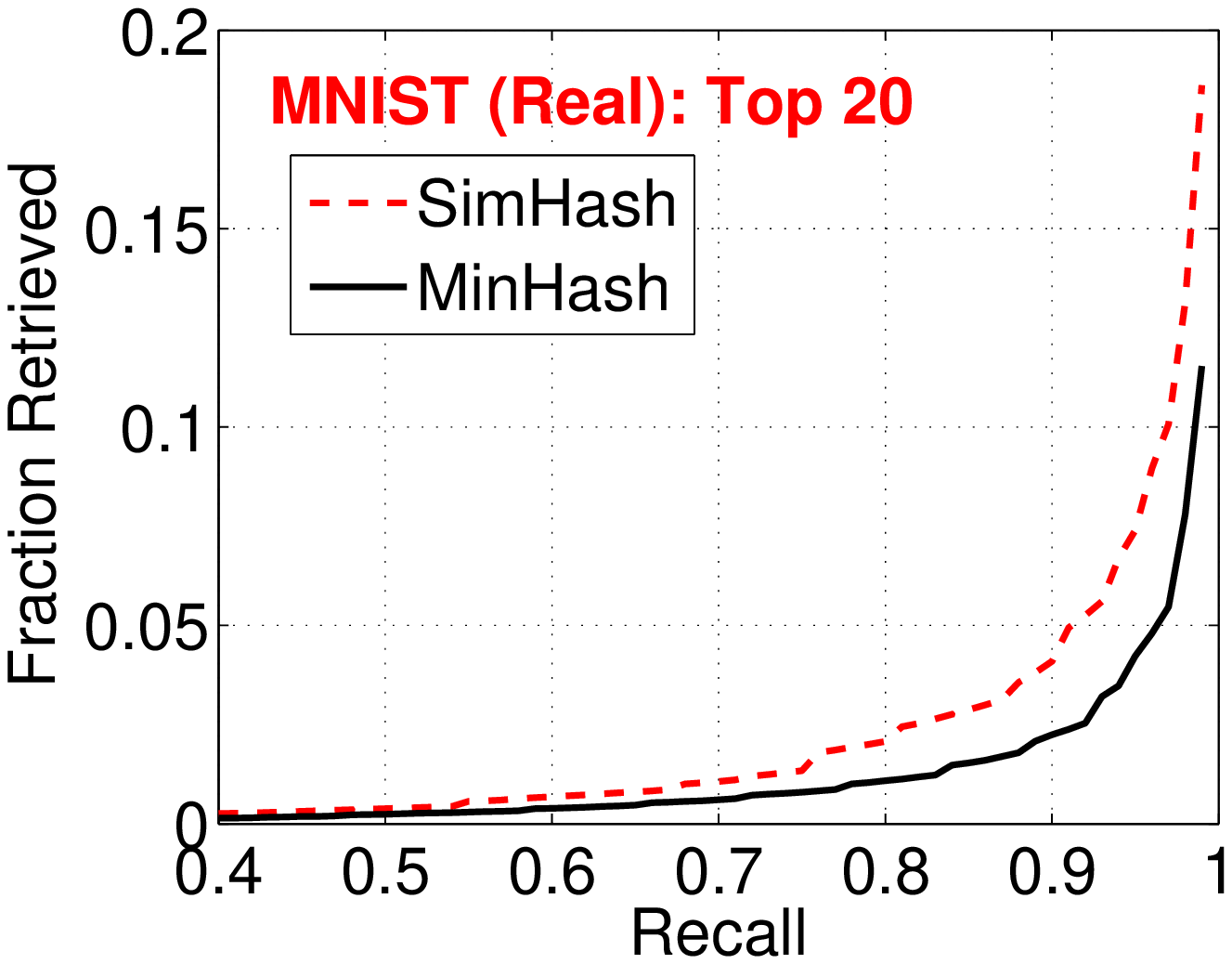}\hspace{-0.13in}
\includegraphics[width=1.7in]{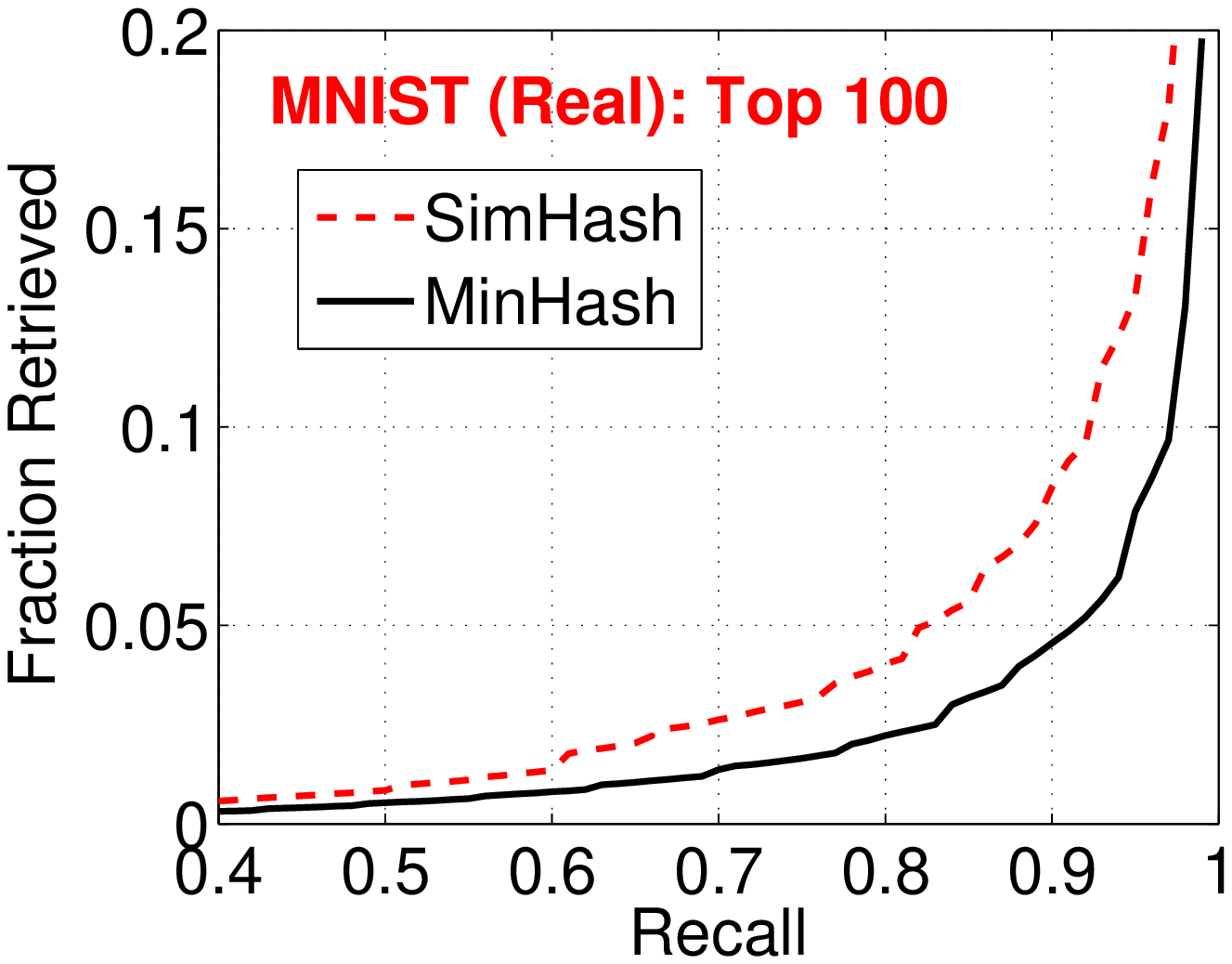}
}

\mbox{
\includegraphics[width=1.7in]{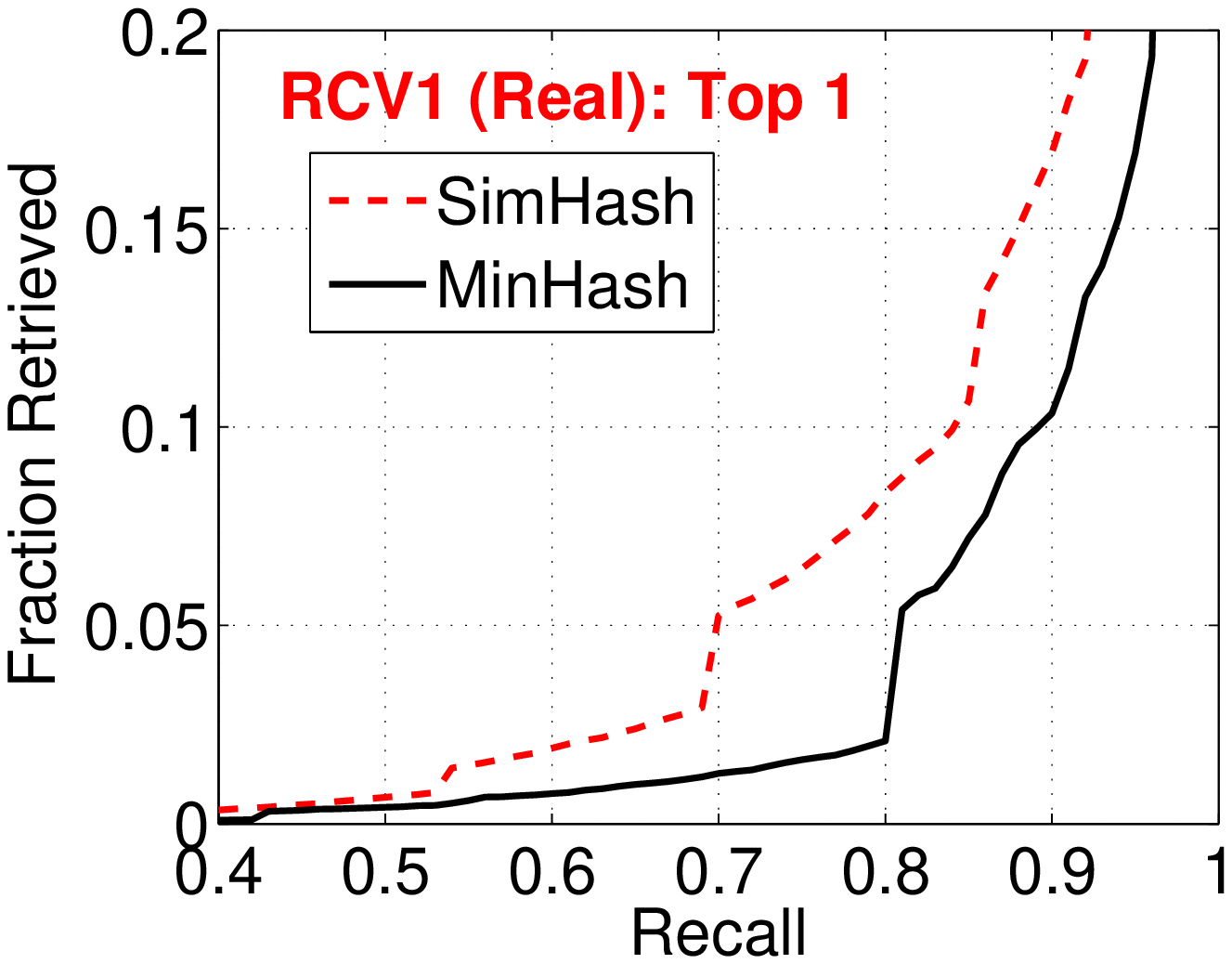}\hspace{-0.13in}
\includegraphics[width=1.7in]{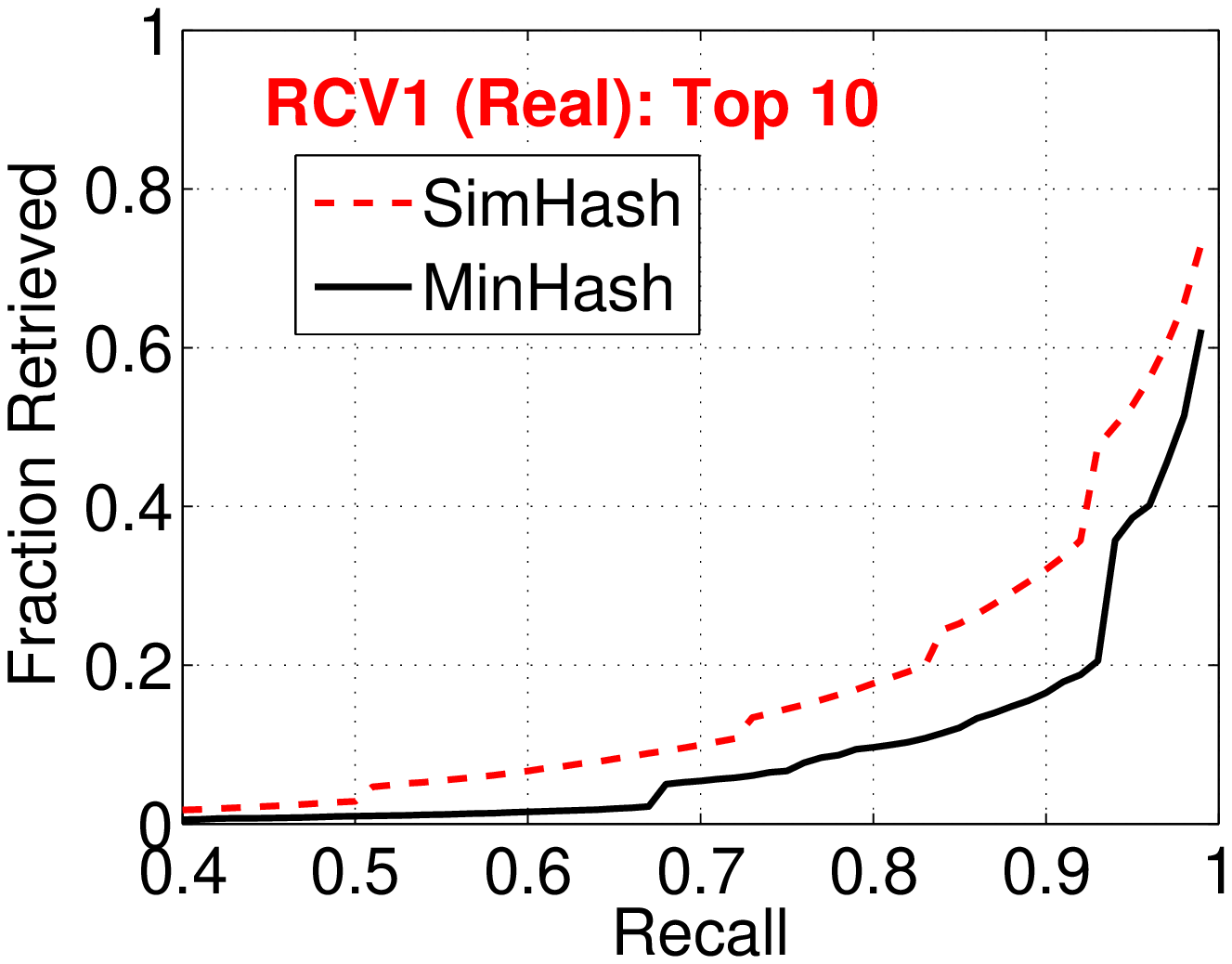}}
\mbox{
\includegraphics[width=1.7in]{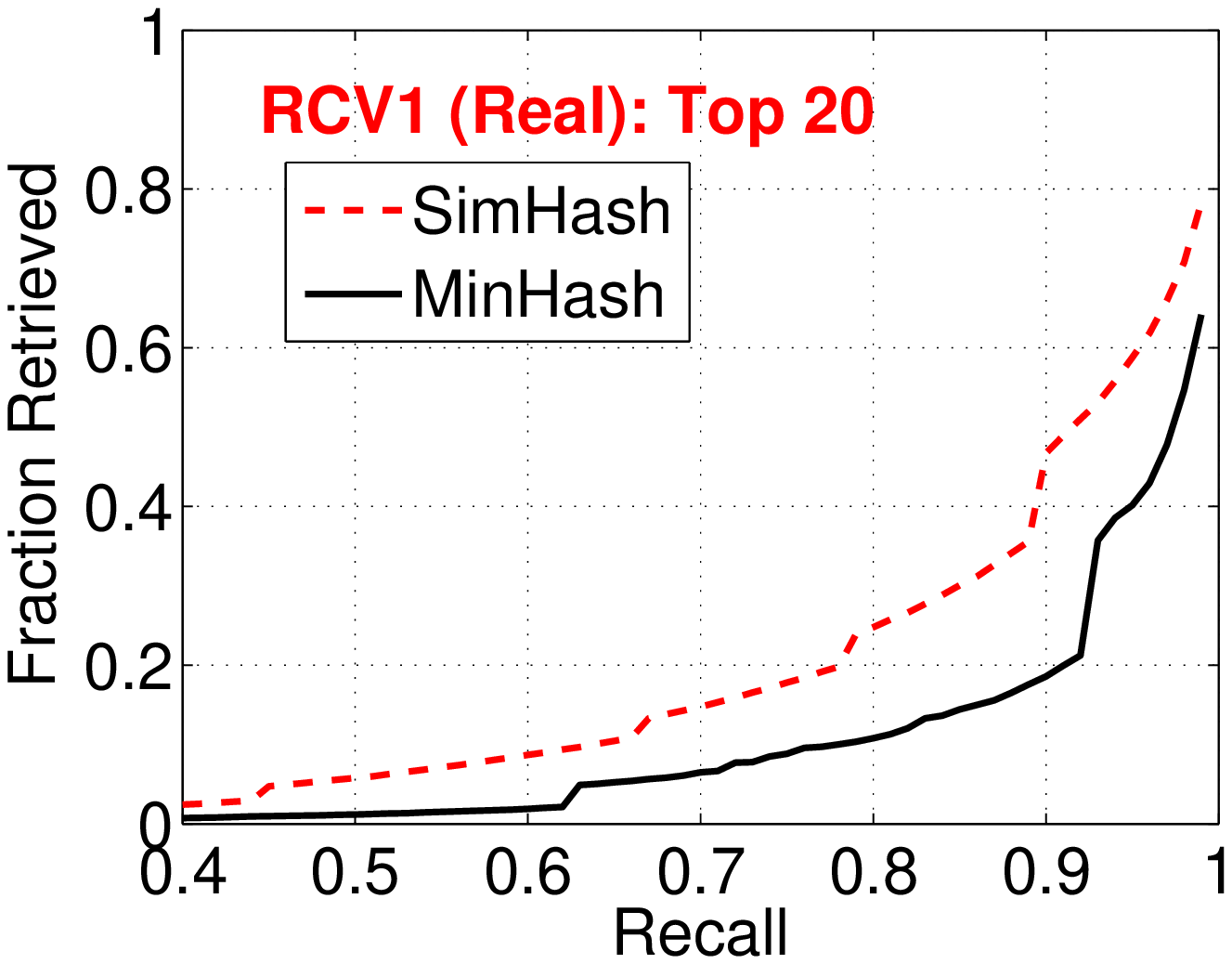}\hspace{-0.13in}
\includegraphics[width=1.7in]{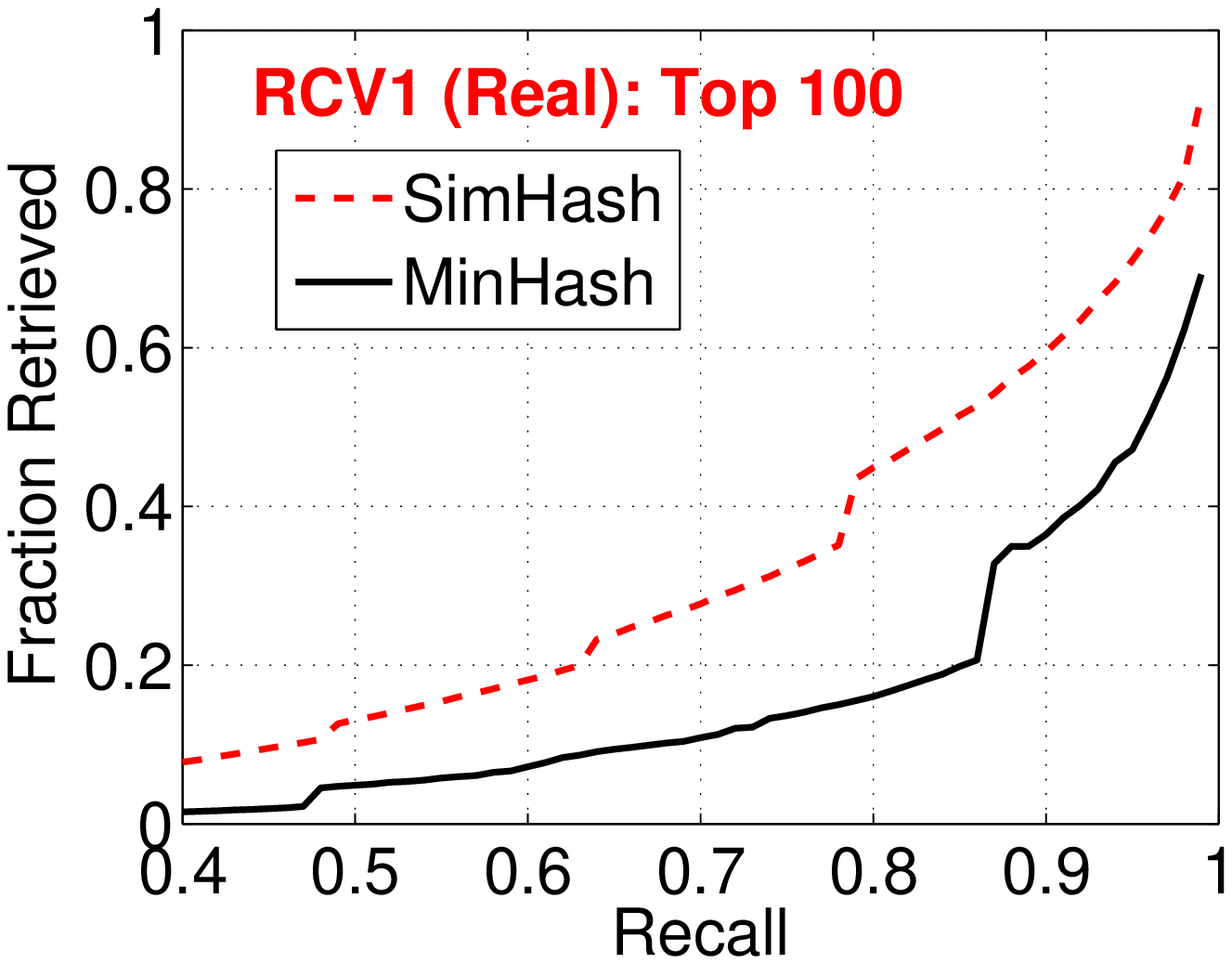}
}

\end{center}
\vspace{-0.2in}
\caption{Retrieval experiments on the original real-valued data. We apply SimHash on the original data and MinHash on the binarized data, and we evaluate the retrieval results based on the cosine similarity of the original data. MinHash still outperforms SimHash.}\label{fig_TopkReal}
\end{figure}

To conclude this section, we also add a set of experiments using the original (real-valued) data, for MNIST and RCV1. We apply SimHash on the original data and MinHash on the binarized data. We also evaluate the retrieval results based on the cosine similarities of the original data. This set-up places MinHash in a very disadvantageous place compared to SimHash. Nevertheless, we can see from Figure~\ref{fig_TopkReal} that MinHash still noticeably outperforms SimHash, although the improvements are not as significant, compared to the experiments on binarized data (Figure~\ref{fig_Topk}).

\vspace{-0.1in}
\section{Conclusion}
\vspace{-0.1in}

Minwise hashing (MinHash), originally designed for detecting duplicate web pages~\cite{Proc:Broder,Proc:Fetterly_WWW03,Proc:Henzinger_SIGIR06},  has been widely adopted in the search industry, with numerous  applications, for example, large-sale machine learning systems~\cite{Proc:HashLearning_NIPS11,Proc:Li_Owen_Zhang_NIPS12}, Web spam~\cite{Article:Urvoy08, Proc:Nitin_WSDM08}, content matching for online advertising~\cite{Proc:Pandey_WWW09},  compressing social networks~\cite{Proc:Chierichetti_KDD09},
advertising diversification~\cite{Proc:Gollapudi_WWW09}, graph sampling~\cite{Proc:Najork_WSDM09},  Web graph compression
\cite{Proc:Buehrer_WSDM08}, etc. Furthermore, the recent development of {\em one permutation hashing}~\cite{Proc:Li_Owen_Zhang_NIPS12,Proc:OneHashLSH_ICML14} has substantially reduced the preprocessing costs of MinHash, making the method  more practical.

In machine learning research literature, however, it appears that SimHash  is more popular for approximate near neighbor search. We believe part of the reason is that  researchers tend to use the cosine similarity, for which SimHash can be directly applied.

It is usually taken for granted that MinHash and SimHash are theoretically incomparable and the choice between them is decided based on whether the desired notion of similarity is cosine similarity or resemblance. This paper has shown that  MinHash is provably a better LSH than SimHash even for cosine similarity.  Our analysis provides a first provable way of comparing two LSHs devised for different similarity measures.  Theoretical and experimental evidence indicates significant computational advantage of using MinHash in place of SimHash. Since LSH is a concept studied by a wide variety of researchers and practitioners, we believe that the results shown in this paper will be useful from both theoretical as well as practical point of view.


\textbf{Acknowledgements}:  Anshumali Shrivastava is a Ph.D. student  supported by  NSF (DMS0808864, SES1131848, III1249316) and ONR (N00014-13-1-0764). Ping Li is partially supported by AFOSR (FA9550-13-1-0137), ONR (N00014-13-1-0764), and NSF (III1360971, BIGDATA1419210).

\appendix

\vspace{-0.1in}
\section{Proof of Theorem~\ref{thm_ineq}}
\vspace{-0.1in}

The only less obvious step is the  \emph{\bf Proof of tightness:}   Let a continuous function  $f(\mathcal{S})$ be a sharper upper bound i.e., $\mathcal{R} \le f(\mathcal{S}) \le \frac{\mathcal{S}}{2-\mathcal{S}}$.  For any rational $\mathcal{S} = \frac{p}{q}$, with $p,q  \in \mathbb{N}$ and $p \le q$, choose $f_1 = f_2 = q$ and $a = p$.  Note that $f_1, f_2 \text{ and } a$ are positive integers. This choice leads to $\frac{\mathcal{S}}{2-\mathcal{S}} = \mathcal{R} = \frac{p}{2q -p}$. Thus, the upper bound is achievable for all rational $\mathcal{S}$.
Hence, it must be the case that  $f(\mathcal{S}) = \frac{\mathcal{S}}{2-\mathcal{S}} = \mathcal{R}$ for all rational values of $\mathcal{S}$.  For any real number $c \in [0,1]$, there exists a Cauchy sequence of rational numbers $\{r_1,r_2, ...r_n, ...\}$ such that $r_n \in \mathbb{Q}$ and $\lim_{n \to \infty} r_n= c$. Since all $r_n$'s are rational,  $f(r_n) = \frac{r_n}{2-r_n}$. From the continuity of both $f$ and $\frac{\mathcal{S}}{2-\mathcal{S}}$, we have $f(\lim_{n \to \infty } r_n)=\lim_{n \to \infty } \frac{r_n}{2-{r_n}}$ which implies $f(c)= \frac{c}{2-c}$  implying $\forall c \in [0,1]$.

For tightness of  $\mathcal{S}^2$, let $\mathcal{S} = \sqrt{\frac{p}{q}}$, choosing $f_2 = a = p$ and $f_1 = q$ gives an infinite set of points having $\mathcal{R} = \mathcal{S}^2$. We now   use similar arguments in the proof  tightness of upper bound. All we need is the existence of a Cauchy sequence of  square root of rational numbers converging to any real $c$.
$\hfill{\square}$

{\small

\begin{thebibliography}{10}

\bibitem{Report:TeraLarning11}
Alekh Agarwal, Olivier Chapelle, Miroslav Dudik, and John Langford.
\newblock A reliable effective terascale linear learning system.
\newblock Technical report, arXiv:1110.4198, 2011.

\bibitem{Report:E2LSH}
Alexandr Andoni and Piotr Indyk.
\newblock E2lsh: Exact euclidean locality sensitive hashing.
\newblock Technical report, 2004.

\bibitem{Proc:Broder}
Andrei~Z. Broder.
\newblock On the resemblance and containment of documents.
\newblock In {\em the Compression and Complexity of Sequences}, pages 21--29,
  Positano, Italy, 1997.

\bibitem{Proc:Broder_STOC98}
Andrei~Z. Broder, Moses Charikar, Alan~M. Frieze, and Michael Mitzenmacher.
\newblock Min-wise independent permutations.
\newblock In {\em STOC}, pages 327--336, Dallas, TX, 1998.

\bibitem{Proc:Buehrer_WSDM08}
Gregory Buehrer and Kumar Chellapilla.
\newblock A scalable pattern mining approach to web graph compression with
  communities.
\newblock In {\em WSDM}, pages 95--106, Stanford, CA, 2008.

\bibitem{Report:Sibyl}
Tushar Chandra, Eugene Ie, Kenneth Goldman, Tomas~Lloret Llinares, Jim
  McFadden, Fernando Pereira, Joshua Redstone, Tal Shaked, and Yoram Singer.
\newblock Sibyl: a system for large scale machine learning.
\newblock Technical report, 2010.

\bibitem{Article:Chapelle_99}
Olivier Chapelle, Patrick Haffner, and Vladimir~N. Vapnik.
\newblock Support vector machines for histogram-based image classification.
\newblock 10(5):1055--1064, 1999.

\bibitem{Proc:Charikar}
Moses~S. Charikar.
\newblock Similarity estimation techniques from rounding algorithms.
\newblock In {\em STOC}, pages 380--388, Montreal, Quebec, Canada, 2002.

\bibitem{Proc:Chierichetti_KDD09}
Flavio Chierichetti, Ravi Kumar, Silvio Lattanzi, Michael Mitzenmacher,
  Alessandro Panconesi, and Prabhakar Raghavan.
\newblock On compressing social networks.
\newblock In {\em KDD}, pages 219--228, Paris, France, 2009.

\bibitem{Proc:Fetterly_WWW03}
Dennis Fetterly, Mark Manasse, Marc Najork, and Janet~L. Wiener.
\newblock A large-scale study of the evolution of web pages.
\newblock In {\em WWW}, pages 669--678, Budapest, Hungary, 2003.

\bibitem{Article:Friedman_75}
Jerome~H. Friedman, F.~Baskett, and L.~Shustek.
\newblock An algorithm for finding nearest neighbors.
\newblock {\em IEEE Transactions on Computers}, 24:1000--1006, 1975.

\bibitem{Article:Goemans}
Michel~X. Goemans and David~P. Williamson.
\newblock Improved approximation algorithms for maximum cut and satisfiability
  problems using semidefinite programming.
\newblock {\em Journal of ACM}, 42(6):1115--1145, 1995.

\bibitem{Proc:Gollapudi_WWW09}
Sreenivas Gollapudi and Aneesh Sharma.
\newblock An axiomatic approach for result diversification.
\newblock In {\em WWW}, pages 381--390, Madrid, Spain, 2009.

\bibitem{Proc:Hein_AISTATS05}
Matthias Hein and Olivier Bousquet.
\newblock Hilbertian metrics and positive definite kernels on probability
  measures.
\newblock In {\em AISTATS}, pages 136--143, Barbados, 2005.

\bibitem{Proc:Henzinger_SIGIR06}
Monika~Rauch Henzinger.
\newblock Finding near-duplicate web pages: a large-scale evaluation of
  algorithms.
\newblock In {\em SIGIR}, pages 284--291, 2006.

\bibitem{Proc:Indyk_STOC98}
Piotr Indyk and Rajeev Motwani.
\newblock Approximate nearest neighbors: Towards removing the curse of
  dimensionality.
\newblock In {\em STOC}, pages 604--613, Dallas, TX, 1998.

\bibitem{Proc:Jiang_CIVR07}
Yugang Jiang, Chongwah Ngo, and Jun Yang.
\newblock Towards optimal bag-of-features for object categorization and
  semantic video retrieval.
\newblock In {\em CIVR}, pages 494--501, Amsterdam, Netherlands, 2007.

\bibitem{Proc:Nitin_WSDM08}
Nitin Jindal and Bing Liu.
\newblock Opinion spam and analysis.
\newblock In {\em WSDM}, pages 219--230, Palo Alto, California, USA, 2008.

\bibitem{Proc:Li_Church_Hastie_NIPS06}
Ping Li, Kenneth~W. Church, and Trevor~J. Hastie.
\newblock Conditional random sampling: A sketch-based sampling technique for
  sparse data.
\newblock In {\em NIPS}, pages 873--880, Vancouver, BC, Canada, 2006.

\bibitem{Proc:Li_Konig_WWW10}
Ping Li and Arnd~Christian {K\"{o}nig}.
\newblock b-bit minwise hashing.
\newblock In {\em Proceedings of the 19th International Conference on World
  Wide Web}, pages 671--680, Raleigh, NC, 2010.

\bibitem{Proc:Li_Owen_Zhang_NIPS12}
Ping Li, Art~B Owen, and Cun-Hui Zhang.
\newblock One permutation hashing.
\newblock In {\em NIPS}, Lake Tahoe, NV, 2012.

\bibitem{Proc:Li_Internetware13}
Ping Li, Anshumali Shrivastava, and Arnd~Christian {K\"onig}.
\newblock b-bit minwise hashing in practice.
\newblock In {\em Internetware}, Changsha, China, 2013.

\bibitem{Proc:HashLearning_NIPS11}
Ping Li, Anshumali Shrivastava, Joshua Moore, and Arnd~Christian K\"onig.
\newblock Hashing algorithms for large-scale learning.
\newblock In {\em NIPS}, Granada, Spain, 2011.

\bibitem{Proc:Najork_WSDM09}
Marc Najork, Sreenivas Gollapudi, and Rina Panigrahy.
\newblock Less is more: sampling the neighborhood graph makes salsa better and
  faster.
\newblock In {\em WSDM}, pages 242--251, Barcelona, Spain, 2009.

\bibitem{Proc:Pandey_WWW09}
Sandeep Pandey, Andrei Broder, Flavio Chierichetti, Vanja Josifovski, Ravi
  Kumar, and Sergei Vassilvitskii.
\newblock Nearest-neighbor caching for content-match applications.
\newblock In {\em WWW}, pages 441--450, Madrid, Spain, 2009.

\bibitem{Proc:Shrivastava_ECML12}
Anshumali Shrivastava and Ping Li.
\newblock Fast near neighbor search in high-dimensional binary data.
\newblock In {\em ECML}, Bristol, UK, 2012.

\bibitem{Proc:OneHashLSH_ICML14}
Anshumali Shrivastava and Ping Li.
\newblock Densifying one permutation hashing via rotation for fast near
  neighbor search.
\newblock In {\em ICML}, 2014.

\bibitem{GoogleBlog}
Simon Tong.
\newblock Lessons learned developing a practical large scale machine learning
  system.
\newblock
  http://googleresearch.blogspot.com/2010/04/lessons-learned-developing-practical.html,
  2008.

\bibitem{Article:Urvoy08}
Tanguy Urvoy, Emmanuel Chauveau, Pascal Filoche, and Thomas Lavergne.
\newblock Tracking web spam with html style similarities.
\newblock {\em ACM Trans. Web}, 2(1):1--28, 2008.

\bibitem{Proc:Weiss_NIPS08}
Yair Weiss, Antonio Torralba, and Robert Fergus.
\newblock Spectral hashing.
\newblock In {\em NIPS}, 2008.

\end{thebibliography}

}

\end{document}